\documentclass[11pt]{article}
\usepackage{latexsym}
\usepackage{amsmath,amsfonts,amsbsy,amssymb,hyperref}
\usepackage[dvips]{graphicx,epsfig}

 \csname
@addtoreset\endcsname{equation}{section}

\def\mso{\mathfrak{so}}

\def\msp{\mathfrak{sp}}

\def\mhs{\mathfrak{hs}}

\def\mD{\mathfrak{D}}

\def\Real{{\mathbb R}}
\def\Comp{{\mathbb C}}

\def\bec{\begin{center}}
\def\ec{\end{center}}
\def\a{\alpha} \def\ad{\dot{\a}} 

\def\b{\beta}  \def\bd{\dot{\b}} 
\def\c{\gamma} \def\cd{\dot{\c}}
\def\C{\Gamma}
\def\d{\delta} \def\dd{\dot{\d}}
\def\D{\Delta}
\def\e{\epsilon} 
\def\ve{\varepsilon}

\def\k{\kappa}
\def\vark{\varkappa}
\def\l{\lambda}
\def\L{\Lambda}
\def\m{\mu}
\def\n{\nu}
\def\r{\rho}
\def\s{\sigma}

\def\th{\theta}

\def\y{\eta}
\def\z{\zeta}
\def\O{\Omega}
\def\o{\omega}

\def\sb{{\bar\s}}

\def\cA{{\cal A}}

\def\cP{{\cal P}}

\def\cA{{\cal A}}

\def\yb{{\bar y}}
\def\zb{{\bar z}}

\def\tn{\widetilde{\n}}
\def\tV{\widetilde{V}}
\def\ty{\widetilde{y}}
\def\tcP{\widetilde{\cP}}

\def\nn{\nonumber}
\newcommand{\eq}[1]{(\ref{#1})}

\def\be{\begin{equation}}
\def\ee{\end{equation}}
\def\bea{\begin{eqnarray}}
\def\eea{\end{eqnarray}}
\def\ba{\begin{array}}
\def\ea{\end{array}}

\def\ft#1#2{{\textstyle{{\scriptstyle #1}
\over {\scriptstyle #2}}}}

\def\ket#1{|#1\rangle}
\def\bra#1{\langle#1|}
\def\scs#1{\section{\bf{\sc \large #1}}}
\def\scss#1{\subsection{\bf{\sc  #1}}}
\def\scsss#1{\subsubsection{\bf{\sc \small #1}}}

\def\ad{\dot\alpha}
\def\bd{\dot\beta}
\def\sb{\bar\sigma}

\def\wV{\widehat{V}}
\def\wS{\widehat{S}}

\def\wPhi{\widehat{\Phi}}

\begin{document}



\vspace{8pt}

\begin{center}


{\huge\sc 4D Higher Spin Black Holes\\[10pt] 
with Nonlinear Scalar Fluctuations}


\vspace{35pt}

C a r l o~~~~I a z e o l l a \\[15pt]
{\it NSR Physics Department \\ G. Marconi University \\
via Plinio 44, Rome, Italy}\\[20pt]

P e r ~~~~ S u n d e l l\\[15pt]
{\it Departamento de Ciencias F\'isicas \\ Universidad Andres Bello\\ 
Republica 220, Santiago de Chile}\\[20pt]


\vspace{10pt} {\sc\large Abstract}\end{center}

We construct an infinite-dimensional space of solutions to Vasiliev's 
equations in four dimensions that are asymptotic to AdS spacetime 
and superpose massless scalar particle modes over static higher 
spin black holes. 
Each solution is obtained by a large gauge transformation of an
all-order perturbatively defined particular solution given in a 
simple gauge, in which the spacetime connection vanishes, the 
twistor space connection is holomorphic, and all local degrees of 
freedom are encoded into the residual twistor space dependence of 
the spacetime zero-forms.
The latter are expanded over two dual spaces of Fock space operators, 
corresponding to scalar particle and static black hole modes, equipped 
with positive definite sesquilinear and bilinear forms, respectively.
Switching on an AdS vacuum gauge function, the twistor space connection 
becomes analytic at generic spacetime points, which makes it possible
to reach Vasiliev's gauge, in which Fronsdal fields arise asymptotically, 
by another large transformation given here at first order.
The particle and black hole modes are related 
by a twistor space Fourier transform, resulting 
in a black hole backreaction already at the second order
of classical perturbation theory.
We speculate on the existence of a fine-tuned branch of 
moduli space that is free from black hole modes and directly 
related to the quasi-local deformed Fronsdal theory.
Finally, we comment on a possible interpretation
of the higher spin black hole solutions as black-hole microstates.

 \setcounter{page}{1}

\pagebreak

\tableofcontents

\vspace{2cm}


\scs{Introduction}\label{sec:intro}


\scss{Motivations}

Higher spin gravities are extensions of ordinary gravity theories
by massless fields with spin greater than two, based on the 
gauge principle. 
Remarkably, these assumptions lead to highly constrained
interactions governed by non-abelian higher spin symmetry 
algebras, whose consistency requires special matter sectors 
and non-vanishing cosmological constant.
The resulting framework thus provides a natural platform for 
studying tensionless string theory in anti-de Sitter spacetime \cite{sundborg,perjohan,minwalla} 
(for recent advances, see \cite{Gaberdiel:2014cha,Gaberdiel:2015mra,Gaberdiel:2015wpo})
and holography \cite{sundborg},\cite{Sezgin:2002rt}--\cite{Gaberdiel:2017dbk}
(for a different approach, see also \cite{Vasiliev:2012vf}),
without relying on any \emph{a priori} assumptions on 
dualities between strong and weak coupling on
the worldsheet or the conformal field theory side.

One of the outstanding problems is to connect two different approaches to higher spin gravity that are presently pursued:
Vasiliev's theory and the quasi-local deformed Fronsdal theory. The latter approach
(see \cite{Bengtsson:1983pd}--\cite{Francia:2016weg} and references therein; 
see also \cite{Bekaert:2010hw} for a review and more references, and  
\cite{Bekaert:2014cea}--\cite{Sleight:2017fpc} for more recent works 
on the holographic reconstruction of bulk vertices) provides a 
perturbatively defined deformation of the free Fronsdal action
following the Noether procedure supplemented by 
weak locality conditions --- as to ensure well-defined amplitudes
built using Green's functions in anti-de Sitter spacetime with 
proper boundary conditions.
Up to quartic order, this approach has so far been yielding 
holographic correlation functions corresponding to unitary
free conformal field theories.
As for exploring the moduli space of the theory, including
more nontrivial field theory duals, the deformed Fronsdal theory has 
so far been of limited use, however, in the absence of any 
fully nonlinear completion of the Noether procedure.

Vasiliev's equations 
\cite{vasiliev}--\cite{more} (for reviews, focusing on various aspects of the subject, see \cite{Vasiliev:1999ba}--\cite{Rahman:2015pzl}),
on the other hand,
provide a fully nonlinear classical theory based on 
differential graded algebras defined
on noncommutative Poisson manifolds, given locally
by direct products of a commuting spacetime manifold
and internal symplectic manifolds.
This formalism encodes the deformation of the Fronsdal 
equations on-shell into manifestly Cartan-integrable curvature 
constraints, at the expense,
however, of blurring the spacetime quasi-locality, and of introducing
additional moduli entering via the internal connection. 

It is therefore highly non-trivial to establish whether and how the deformed Fronsdal 
theory and Vasiliev's formulation are connected beyond the 
linearized approximation.
To this end, one may envisage two quite different routes,
depending on whether one expects to make contact
directly at the level of the spacetime action, or, 
more indirectly, at the level of on-shell 
actions subject to boundary conditions.
The former path, whereby
the internal degrees of freedom of Vasiliev's 
integrable system are viewed essentially as auxiliary fields,
without any proper dynamics of their own, involves a reduction down to a set of 
deformed Fronsdal equations on the spacetime 
submanifold
\cite{vasiliev,Vasiliev:1990vu,more,Vasiliev:1999ba,Sezgin:2002ru,Didenko:2014dwa,Giombi:2009wh,Vasiliev:2015wma,Boulanger:2015ova,Skvortsov:2015lja,Vasiliev:2016xui}, to be directly compared with those of the 
quasi-local deforrmed Fronsdal theory. 
Although this procedure introduces ambiguities
concerning the embedding of spacetime into the 
full noncommutative geometry and the internal
gauge fixing, one may entertain the idea 
that the quasi-locality requirement provides a guiding 
principle; for recent progress, see \cite{Vasiliev:2016xui,Sezgin:2017jgm,Didenko:2017lsn}.
Following the second route, one instead seeks to 
implement the action principle by deriving Vasiliev's 
equations from a globally defined Lagrangian density 
on the full noncommutative manifold \cite{Boulanger:2011dd,Boulanger:2012bj,Bonezzi:2015igv,Vasiliev:2015mka}, 
which requires a choice of topology
and a suitable class of functions forming a differential 
graded algebra equipped with a trace operation and 
a hermitian conjugation.
Making contact between the two approaches then consists of seeking up a point 
in moduli space leading to an on-shell action that agrees with 
that of the deformed Fronsdal theory. Note that this does not 
require any equivalence at the level of vertices
and Green's functions.

To pursue the comparison, the access to 
explicit classical solution spaces of Vasiliev's 
equations may provide useful guidance, whether to 
extract a quasi-local Fronsdal branch of moduli 
space, or to compute on-shell actions on the full 
noncommutative geometry.
So far, the classical moduli space has been explored 
in the direction of asymptotically anti-de Sitter solutions
with special higher spin Weyl curvatures: instanton-like solutions, higher spin black holes and other
generalized Type-D solutions 
\cite{Sezgin:2005pv,Iazeolla:2007wt,Didenko:2009td,us,us2,Bourdier:2014lya,Sundell:2016mxc};
see also \cite{Prokushkin:1998bq,Didenko:2006zd,Iazeolla:2015tca}
for exact solutions in three spacetime dimensions, and \cite{Gubser:2014ysa} 
for solutions obtained in axial gauge in twistor space.

The study of exact solutions of higher-spin gravity is relevant for addressing other important open questions: for instance, as the theory is supposed to describe an 
``extremely stringy'' extension of (super)gravity theories, a natural 
question to investigate is whether it admits black hole solutions, 
and how the main problems that their physics poses are framed in its language, and, 
possibly, resolved by the coupling of the gravitational fields with an 
infinite tower of massless fields of higher spin. 
Black-hole-like solutions of the Vasiliev equations have been constructed in \cite{Didenko:2009td,us,us2,Sundell:2016mxc}, 
but the present lack of a stringy generalization of geometry
has so far limited our capability to assess their precise nature --- for instance by establishing whether they possess
a horizon, whether their curvature singularity is physical or not, etc. 
Moreover, whereas the Type-D solution spaces in ordinary gravity
are finite-dimensional, it was observed in \cite{us,us2} 
that their higher-spin counterparts are infinite-dimensional. 
We are inclined to interpret this fact, together with other properties, as an indication that such solutions are
candidates for the microscopic description of 
hairy quantum black holes, rather then higher-spin black holes, as we shall comment on 
in more detail throughout this paper.
Clearly, one expects that the study of fluctuations over such black-hole-like solutions should help in answering some of the key physical questions above. 

In this paper, we shall extend the higher spin black hole 
systems of \cite{us,us2} by massless 
scalar particle and anti-particle modes\footnote{
A corollary of our analysis is that it is possible to 
map Weyl-ordered master fields on-shell to normal-ordered 
dittos obeying the central on-mass-shell theorem in the 
scalar field sector, as required for the embedding of the 
Vasiliev system into the Frobenius--Chern--Simons model 
on-shell \cite{FCS}.}.
The resulting configurations describe nonlinear propagating 
scalar fluctuation fields on static spherically-symmetric
higher spin black hole backgrounds.
Although these restrictions on the boundary conditions lead to
technical simplifications, they do  
not imply any loss of generality concerning our main
conclusions, as the solution method  
developed in this paper facilitates the \emph{systematic} fully nonlinear 
completion of linear combinations of  black-hole-like and particle modes, and therefore lends itself equally
well to the construction of solution spaces consisting 
of general time-dependent black hole modes superposed with 
general massless particle modes of any spin, 
which we leave for a future work.

Moreover, as we shall comment more in detail later on, the study in this paper already offers some results that may be relevant to the comparison of the two approaches to
higher spin gravity. Indeed, the quasi-local deformed Fronsdal theory
by construction builds a perturbation theory 
in which the fundamental massless particles 
are stable to any order --- which is indeed a necessary
condition for it to be a candidate bulk dual 
of free conformal theories.
At the classical level, this requires sufficiently 
local vertices; in particular, it requires that the result of two 
massless particles modes entering a cubic vertex 
can be expanded in terms of the same type of mode functions.
On the other hand, in the Vasiliev system, which is a set of zero-curvature 
constraints on a higher-dimensional fibered space, the 
particle modes are encoded into specific fiber functions, 
and the classical perturbation theory is generated
by applying homotopy contractors to initial data 
given by such elements.
Thus, the issue of whether the system admits any perturbative 
expansion scheme describing self-interacting particles 
requires a careful analysis as in the fully non-linear theory there exists a vertex 
that interchanges particle and black hole modes.
Indeed, the exact solutions of the present paper will show that, at least in certain gauges, particle modes 
interact to form black holes modes already
at second order of classical perturbation theory -- which can possibly be seen as a manifestation of the non-locality of the Vasiliev equations at each perturbative order. 
The special internal
gauge we start from is connected by a large gauge transformation to 
Vasiliev's (internal) gauge, in which Fronsdal fields arise at 
the linearized level.
Thus, aiming a direct correspondence between the Vasiliev
theory and the quasi-local deformed Fronsdal theory, one may ask
whether the ambiguity residing in Vasiliev's gauge beyond 
the linearized level can be exploited to cancel out the black 
hole modes at higher orders in perturbation theory, as to 
fine-tune to a quasi-local branch of Vasiliev's theory.
We shall not pursue this refined boundary value problem
in this paper, nor its analog in the alternative perturbative 
scheme employed by Vasiliev in his original work, though
we shall remark on these topics towards the end of the paper.
%

\scss{Black holes or black-hole microstates?}

The solutions to four-dimensional higher spin 
gravity that are referred to as higher spin black holes
 are distinct by a tower of electric or magnetic Weyl tensors of all spins that include and generalize the spin-2 Weyl tensor of an AdS Schwarzschild black hole. In particular, all Weyl tensors share the Killing symmetries, Petrov type D 
and principal spinors of the latter\footnote{The spin-two Weyl tensor is of Petrov Type D, that is, 
$C_{\a\b\c\d} = f u_{(\a}u_\b v_\c v_{\d)}$
where $(u_\a,v_\b)$ is a spin-frame, normalized such that 
$u^\a v_\a=1$, and $f$ is a complex function (which remains
invariant under redefinitions $u_\a\rightarrow \l u_\a$
and $v_\a\rightarrow \l^{-1}v_\a$, with $\l \in \Comp'$). See Appendix \ref{App:conv}.
In particular, black-hole Weyl tensors are distinguished, in gravity, by 
the fact that $u_\a$ and $v_\a$ are eigenspinors of the (self-dual part of the) Killing two-form 
of a time-like Killing vector. 
In the case of a static and spherically-symmetric
Type D solution, the function $f\sim {\cal M}_2 r^{-3}$
in the asymptotic region, where ${\cal M}_2$ is the 
spin-two charge, which is real in the electric case
and purely imaginary in the magnetic case.}; see 
\cite{Didenko:2009td} for the first instance 
of such a solution, and \cite{us,us2,Sundell:2016mxc} for 
developments. In gravity, such a form of the Weyl tensor is a local hallmark of a black-hole solution \cite{Mars}.
However, in higher spin gravity this identification is subtler: so far there is no known higher-spin invariant quantity ensuring that the singularity of the individual Weyl tensors is physical, and whether there 
exists any invariant notion of an event 
horizon  -- and whether an entropy could be attached 
to it\footnote{The conservative notion of an entropy, namely 
as a density of states at the saddle point of a path integral, 
requires the further notions of an on-shell action \cite{Boulanger:2011dd,Colombo:2012jx,Vasiliev:2015mka,FCS} 
and either a conserved set of macroscopic charges at infinity, or, 
alternatively, a conjugate (temperature) variable related to 
thermal fluctuations and Hawking radiation.
Alternatively, and more formally, one may think of the the entropy and 
the on-shell action as elements in the on-shell de Rham cohomology 
in spacetime form degree two and four; indeed, in \cite{Sezgin:2011hq} 
a structure group was found that yields a unique complex cohomology 
element for every strictly positive even degree (and vanishing
cohomology for every positive odd degree).
In \cite{Vasiliev:2015mka,Didenko:2015pjo}, cohomologies
in degrees two and four have been constructed using
different methods. See also \cite{Didenko:2015pjo} for the definition of generalized asymptotic charges.} -- remains an open problem.
Although there exist a large number of formally defined metrics
in higher spin gravity, as (non-abelian) higher spin symmetries 
mix different spins as well as numbers of derivatives (in units 
of the cosmological constant) none of them is \emph{a priori} preferred in the absence
of any metric-like action principle (though calibrations 
of areas may select special metrics related to brane
actions \cite{Sezgin:2011hq}).

On the other hand, as we shall describe more in detail below, the solutions found in \cite{us,us2} form an infinite-dimensional unitarizable higher-spin module of states (the norm being given by the ordinary supertrace), each one giving rise to a solution that has identical black-hole asymptotics but that is possibly non-singular and horizon-free, which would rather suggest an interpretation in terms of black-hole microstates.  In fact, viewing higher spin gravity as
gravity coupled to to higher spin gauge fields, it is tempting to view such solutions as higher-spin analogs of fuzzballs \cite{Mathur1}--\cite{Skenderis},  that we think of as solutions that are degenerate 
with the AdS--Schwarzschild black hole to a (macroscopic) 
observer in the asymptotic region, but that contain 
``hairs'' that modify the strongly coupled region
of the black hole geometry as to create a smooth 
solution without a horizon. We hope to investigate the details of this proposal in a future work.

Reasoning physically, we find it plausible to 
entertain the idea that entropic horizons emerge in broken 
phases of higher spin gravity, whereas the microstates are 
directly visible in the unbroken phase as a spectrum of 
regular solutions.
Indeed, the breaking of higher spin symmetries would
introduce a mass scale that could give rise to Yukawa-like
potentials that screen the higher spin hairs for
the asymptotic observer, while higher spin gauge fields 
extend into the asymptotic region in the unbroken 
phase.
Moreover, the black-hole-like solutions in Vasiliev's theory 
can be thought of as being smooth at the origin despite the fact
that each individual Weyl tensor blows up at this
point. Indeed, all Weyl tensors and auxiliary fields enter the Vasiliev equations packed into master fields
living on an extension of spacetime by a noncommutative twistor space and satisfying a deformed oscillator algebra: in these terms, a sign that the origin is not a special point is that the deformed oscillator algebra remains
well-defined there, even though the deformation is given by a distribution in twistor space\footnote{The global $AdS$ radial coordinate $r$ appears in the black-hole solutions as the parameter of
a delta sequence: away from the origin one has smooth Gaussian
functions, approaching a Dirac delta function as $r$
goes to zero \cite{us}.
But unlike the delta function on a commutative space,
which is singular thought of as an element in a ring 
of sections, the delta function in noncommutative 
twistors space is smooth thought of as a symbol
for an element of a star product algebra. Indeed, one can show \cite{us} that by changing ordering prescription (from Weyl ordering to normal ordering) one can map the delta function to a regular element, solve the equations in this basis and then move back to the original ordering -- generating the same solution that one would have obtained by solving directly the deformed oscillator algebra with the distributional deformation. In this sense we can say that the black-hole-like solutions are actually smooth.
}.

In order to stress this point, let us recall a few basic facts 
of the Vasiliev equations (leaving further details for Section 2).
Conceptually speaking, the equations are analogs of the 
constraints on the super-torsion and super-Riemann
tensor in supergravity, in the sense that, rather than 
working directly with Fronsdal fields, the formalism
employs a frame field, a Lorentz connection and an 
infinite tower of higher spin analogues.
These are introduced together with a corresponding tower 
of zero-forms, altogether forming a Cartan integrable system,
\emph{i.e.} a set of curvature constraints that can be used
to express all fields on shell in terms of (large) gauge 
functions and zero-form integration constants.
Although it shares plenty of features with 
topological field theory, the formalism makes use
of infinitely many fields, making it capable of 
describing systems with local degrees of freedom, which
enter via infinite-dimensional spaces of integration 
constants, constituting a module for the higher spin 
Lie algebra, known as the Weyl zero-form 
module, or as the twisted-adjoint representation.
Its Lorentz covariant basis define the generalized Weyl 
tensors (including spin-$0$ and spin-$1/2$ matter fields),
and all their covariant derivatives on-shell at some 
point of the base manifold, while its compact
basis describes particles, black holes, solitons 
and other states of the theory
\cite{Iazeolla:2008ix,us}.

The key to Vasiliev's fully nonlinear theory is the fact that 
both the higher spin algebra and its zero-form module 
arise as subspaces within one and the same associative 
algebra, given by sets of functions on a non-commutative 
twistor fiber space.
In order to couple the connection to the Weyl zero-form, 
Vasiliev extended spacetime with an additional 
non-commutative twistor base space, supporting 
remarkable closed and twisted-central two-forms 
whose star products with the Weyl zero-form 
serve as nontrivial sources for the two-form 
curvature in twistor space.
The resulting system thus describes a flat connection
and a covariantly constant zero-form in spacetime 
coupled to a set of deformed oscillators in twistor space.
This system can be treated in two dual fashions,
by either reducing it perturbatively to deformed set 
of Fronsdal fields on spacetime \cite{vasiliev,Vasiliev:1990vu,Vasiliev:1999ba,Sezgin:2002ru,Boulanger:2015ova}, or expressing
the fields using a gauge function (\emph{i.e.} a
large gauge transformation) and a set of deformed 
oscillators constructed using algebraic methods \cite{Vasiliev:1990bu,Vasiliev:1999ba,Sezgin:2005pv,Iazeolla:2007wt,Iazeolla:2008ix,us,Sundell:2016mxc}. 

In a previous work \cite{us} (see also \cite{us2,Sundell:2016mxc}), we have 
used the latter method to construct families of exact 
generalized Petrov Type D solutions,
encompassed by an Ansatz based on factorization of
noncommutative fiber and base twistor coordinates,
absorbed into generalized projectors and deformed oscillators,
respectively. 
The various families are distinguished by different 
Killing symmetries, while all share the Kerr-Schild 
property, \emph{i.e.}, the full Weyl zero-form coincides 
with the linearized one (in a special gauge).
Thus, the Weyl zero-form can be assigned a linear space, 
while the twistor space connection contains 
interference terms (that can be removed as well 
in certain gauges \cite{Didenko:2009td} though they
must be switched on eventually in going to
Vasiliev's gauge). 

In particular, the spherically-symmetric solutions 
are expanded over ``skew-diagonal'' projectors 
$\tcP_n \sim \ket{n/2}\bra{-n/2}$, that 
we shall refer to as \emph{twisted projectors}, with real or imaginary expansion coefficients $\n_n$,
where $n$ is a non-zero integer, and $\ket{n/2}$ 
denotes the supersingleton states with AdS energy $n/2$. 
These solution spaces form real higher-spin 
representations with positive definite bilinear
forms \cite{us,us2}.
Thus, for each distinct $n$, the classical solution 
lends itself to be interpreted as a microstate of a 
Euclidean theory, rather than a Lorentzian one as 
is the case for the particle states; see also the Conclusions. 

Looking more carefully at the classical solution
for a given $n$, its Lorentz covariant field content 
contains an infinite tower of Weyl tensors of all spins
that are spherically symmetric, static and of generalized 
Petrov Type D, each of which has a curvature singularity 
at its center; in particular, in the spin-two sector one recognizes the AdS-Schwarzschild 
Weyl tensor. The Weyl tensors carry asymptotic charges, that can be electric or magnetic depending on $n$, all proportional to the single deformation parameter $\n_n$, ${\cal M}_s(\nu)={\cal M}_s^n \nu_n$ (no sum on $n$).
On the other hand, starting from a given Fronsdal field of a fixed Lorentz spin, 
the corresponding linearized Type-D sector\footnote{As found in \cite{Didenko:2008va}, for every spin $s$ it is possible to construct a solution of the spin-$s$ free Fronsdal equations as $\varphi_{\m_1...\m_s} = \frac{M}{r}k_{\m_1}...k_{\m_s}$, where $k_\m$ is a Kerr-Schild vector. As stressed in \cite{Didenko:2009td}, the physical higher spin gauge fields corresponding to the higher spin black hole Weyl tensors described above indeed admit such a form in a suitable gauge. It is for this reason that we refer to such fields as to the linearized Type-D sector of spin $s$.} is finite-dimensional, and does not lend itself 
to any sensible interpretation as a quantum mechanical state space. 
However, the interactions of the Vasiliev system create 
``coherent'' states, consisting of all Lorentz spins, making 
up twisted projectors and thus facilitating a physical interpretation 
in terms of states in a real vector space with a Euclidean norm.

Conversely, one may consider deforming a whole tower of asymptotically 
defined (free) Fronsdal fields, with each spin-$s$ field carrying 
a charge ${\cal M}_s$.
In order to use Vasiliev's theory for this purpose, one has to 
tune the charges so that ${\cal M}_s={\cal M}_s(\nu)=\sum_n{\cal M}_s^n \nu_n$ 
for some ensemble $\{\nu_n\}$ \cite{us,us2}; in particular, one
may choose $\nu_{n}=\delta_{n,n_0} \nu_{n_0}$ as to activate
a single microstate.
In this sense, one may think of the asymptotic spin-$s$ charges 
${\cal M}_s$ for $s\geqslant 3$ as ``higher spin hairs''
that needs to be fine-tuned to form a microstate.
Moreover, it appears that the matrix ${\cal M}_s^n$ 
is non-invertible, which makes sense as the Lorentz spin 
is not expected to be a good observable in the strongly coupled 
region.

Clearly, to settle the issues of whether the singularity 
of the Weyl tensors is physical (not a gauge artifact), and 
whether or not these solutions possess an event horizon,
a more detailed study of the propagation of small 
fluctuations over the black-hole-like solutions is required.
As already mentioned, such a study is quite challenging, 
due to the non-locality of the interactions induced by 
higher spin symmetry, which requires
a proper generalization of the standard geometric tools used 
in the case of gravity. 

The aforementioned questions are tied to the issue of in
which classes of functions on twistor space
the master fields are allowed to take value.
On general grounds, the admissible classes must form 
differential graded associative algebras with well-defined trace operations   
\cite{Vasiliev:2015wma,Vasiliev:2015mka,Vasiliev:2016xui,Boulanger:2015ova,Skvortsov:2015lja}. 
We would like to stress that the class of functions making up
the twistor fiber space has an impact on the boundary behaviour 
of the fields on the base manifold, which we expect to be dual
to a set of observables, thereby determining a superselection 
sector of the theory; for examples, see \cite{Iazeolla:2008ix,Colombo:2010fu}. 
Conversely, certain boundary conditions in spacetime induce classes of 
functions in twistor fiber space whose star-product compositions
must be regularized.
This raises the issue of how to handle field configurations obtained
by superimposing different sectors equipped with separately well-defined
regular presentations. 
Indeed, a compatibility problem may arise upon dressing   
linear combinations of such linearized solutions into 
full solutions. 
This problem may have either only the trivial outcome, 
namely that each field configuration can only be dressed separately; 
or various non-trivial ones consisting of compatible combinations 
of regular presentations. 
In particular, as found in \cite{us}, the black hole solutions 
to the \emph{minimal} bosonic model require a specific regular 
presentation for the projectors, which ensures associativity and traceability.

In what follows, we shall examine how to superpose particle modes 
over black hole modes, while we leave the issue of a proper interpretation of the latter to a future publication.  Adhering to the current use in the literature, we shall therefore keep referring to them as higher spin black holes in this paper.

\scss{Summary of our new results}

In this work, we shall extend the spherically-symmetric, static 
black hole solutions found in \cite{us} by adding nonlinear
time-dependent scalar field modes.
To this end, we use mathematical methods that at first sight 
may appear to be distant from those used in ordinary treatments 
of gravity, but that are, on the other hand, natural from the 
point of view of the unfolded formulation.
As mentioned above, we thus absorb the spacetime dependence into
a gauge function, and construct a deformed oscillator algebra
from the Weyl zero-form and twistor space connection\footnote{
The gauge function method in four dimensions
differ from that employed in three-dimensional higher spin gravity 
\cite{Kraus:2012uf}.
In the latter context, the analysis of the scalar 
propagator on higher spin black hole backgrounds is simplified 
by the fact that any black hole (in fact, any vacuum solution) 
in three spacetime dimensions is locally equivalent to $AdS_3$,
which can be described by a vacuum gauge function. 
In four dimensions, however, the Weyl tensors deform the
gauge function, in order for canonical Fronsdal fields to appear.}.
The superposition of scalar particle and black hole modes in
the Weyl zero-form amounts to an expansion of its twistor fiber 
space dependence in terms of rotationally-invariant 
rank-$n$ projectors\footnote{
More generally, by the Flato-Fronsdal theorem, the operators contained in 
$\ket{j+\ft{1}2;\left(j\right)}\bra{j'+\ft{1}2;\left(j'\right)}$, with 
energy $j+\ft12$ and spin $j=0,\ft12,\dots$, have compact spins in $\left(j\right)\otimes
\left(j'\right)$, and contribute to the harmonic expansions in terms 
of particle modes in fields with Lorentz spins 
$s=j+j'$.} $\cP_n\sim \ket{n/2}\bra{n/2}$ and twisted 
projectors $\tcP_n\sim \ket{n/2}\bra{-n/2}$, respectively,
where $\ket{n/2}$, $n=\pm1,\pm2,\dots$, belong to supersingleton 
and anti-supersingleton weight spaces for positive and negative
$n$, respectively, and we have suppressed the spin degrees of freedom.
These elements form an indecomposable subalgebra of the star-product algebra:
evidently, the twisted projectors are not idempotent, but 
rather close under star products onto the projectors,
which in their turn close onto themselves.
Physically, this implies that, differently from 
black hole modes, the scalar particle modes only solve the linearized 
equations, and not the full equations as well. 
Therefore, once scalar particle modes are injected 
into the Weyl zero-form, the nonlinear corrections 
dress them into full solutions with black hole modes
formed as a back-reaction.

The technical detail responsible for the aforementioned phenomenon
is the presence of a Klein operator in a linear vertex in 
Vasiliev's equations, which has the effect of converting
particle modes into black holes modes by means of a Fourier
transformation in twistor space.
Schematically, the massless particle modes are first mapped, in 
accordance with the Flato-Fronsdal theorem \cite{Flato:1978qz}, 
to the aforementioned twistor-space realizations of operators 
$\ket{\ft{n}2}\bra{\ft{n'}2}$ ($n,n'>0$) in supersingleton state space \cite{Iazeolla:2008ix}. 
These operators are then mapped via the aforementioned linear vertex 
to an outgoing operator of the form $\ket{\ft{n}2}\bra{-\ft{n'}2}$ ,
where the bra is an anti-supersingleton state.
At the second order, the interactions produce an 
admixture of supersingleton-supersingleton and 
supersingleton-anti-supersingleton operators,
the latter corresponding to black-hole modes \cite{Didenko:2009td,us,us2}. 
Thus, without any further fine-tuning of the initial data,
the backreaction from particles produces black hole modes 
already at the second order of the classical perturbation 
theory, which one may view as a manifestation 
of the innate spacetime non-locality of the Vasiliev system.
In the bulk of the paper, we shall fill in the details
of the above sketchy derivation of this interesting
backreaction mechanism. 

The above results lend themselves to a holographic interpretation as follows:
Let us start by injecting into the bulk the states conjectured to be dual 
to the free vector model with conserved higher spin currents, by expanding the 
Weyl zero-form over operators corresponding to massless particle modes 
of arbitrary (integer) spin.
Following these modes into the bulk, the Weyl tensors become 
strongly coupled, and the non-local vertices bring forth higher 
spin black hole modes, that blow up at the origin.
In this sense, the projectors can be thought of as 
\emph{boundary states} -- since at the boundary all
Weyl tensors fall off and the theory linearizes, so that
one can think of preparing single Fronsdal fields -- while
the twisted projectors can be thought of as \emph{horizon states}, 
as they blow up in the interior, where the fields 
coalesce into collective degrees of freedom that can no longer be
assigned distinct Lorentz spins.

At first order, though, the black hole modes can be switched off,
and it is possible to make contact with the usual perturbative 
scheme in Vasiliev gauge\footnote{Our choice of nomenclature
reflects the fact that this gauge was used by Vasiliev in 
his original works on the classical perturbation theory
of his equations \cite{more}.}, as we shall see in Section \ref{Sec:V}.   
We defer to a future study the interesting problem of whether one can, 
order by order, remove the Type D modes from the twistor-space 
connection and the spacetime gauge fields, thereby obtaining field 
configurations expandable over only particle modes and defining a 
quasi-local Fronsdal branch of moduli space, that can be a proper 
candidate dual of the free large-$N$ vector model. 

We would also like to remark that while the internal connection does
not belong to any differential graded associative algebra, the
deformed oscillator variables formed from it belong to 
an associative subalgebra of the star 
product algebra for a general combined particle and black hole state.
Moreover, the effect of the vacuum $AdS$ gauge function on the
deformed oscillators is to make them real analytic on the twistor 
base manifold, as necessary for the existence of yet another field-dependent
gauge function, bringing the solution to Vasiliev's gauge, which 
we construct at the linearized level.
We consider the fact that the solution space exists in the Vasiliev 
gauge as being nontrivial.
This raises the prospect of using this requirement as a superselection
mechanism for admissible twistor space initial data\footnote{ 
Configurations that are singular on the twistor base manifold 
may instead have a role to play in a more general physical context, for example
in describing nonlinear density matrices.}.

\scss{Plan of the paper}

In Section 2, we spell out the properties of Vasiliev's equations in 
four dimensions that are relevant for constructing and interpreting our 
exact solutions.
Section 3 provides the solution Ansatz in a specific 
\emph{holomorphic gauge}. 
In Section 4, we apply the Ansatz to the aforementioned 
generalized projector algebra in a specific \emph{regular
presentation}, and demonstrate the backreaction mechanism
and the associativity of the deformed oscillator algebra
in twistor space.
Section 5 contains the analysis of the spacetime 
behaviour of the Weyl zero-form and the internal 
connection in the gauge reached from the holomorphic
gauge by first switching on the \emph{vacuum gauge
function}, and then a compensating field dependent
gauge function leading to the \emph{Vasiliev gauge} 
at the linearized level; the latter gauge permits 
a perturbative description in terms of self-interacting 
Fronsdal fields (see for example 
\cite{more,Vasiliev:1999ba,Sezgin:2002ru,Iazeolla:2008bp,Didenko:2014dwa}). 
Our conclusions and an outlook are in Section 6. 
The paper contains three appendices: Our conventions 
for spinors and the gravitational background fields 
in anti-de Sitter spacetime are given in Appendix \ref{App:conv}; 
useful properties of the Klein operator are contained in 
Appendix \ref{App:A}; finally, Appendix \ref{App:defosc} 
describes the method used for constructing the deformed oscillators.

\scs{Vasiliev's Four-Dimensional Bosonic Models}\label{sec:in}

%
In this section, we describe the basic algebraic structure of Vasiliev's
equations for bosonic gauge fields of all integer spins in four spacetime 
dimensions; in particular, the original form of Vasiliev's equations is given 
in Eqs. \eq{MC}--\eq{INT3}.
We would like to stress that, while our solutions only rely on the locally
defined Vasiliev system, we shall here also address some global issues that may be useful 
in providing them with a physical interpretation, as well as in extracting Fronsdal fields 
from the full system. 
To this end, we shall introduce a related additional set of boundary conditions, 
including ordering prescriptions, that we propose will lead to a well-defined 
perturbative expansion around asymptotically anti-de Sitter spacetime in the 
presence of particle and black-hole modes. 
%

\scss{Correspondence space}

Vasiliev's higher spin gravity is formulated in terms of a finite set 
of master fields.
Geometrically, these are horizontal differential forms
on a bundle space, that we shall refer to as the correspondence 
space, whose fibers as well as base manifold are noncommutative.
We shall denote the correspondence space, its base and its fiber space
by ${\cal C}$, ${\cal B}$ and ${\cal Y}$, respectively.
The fibers are symplectic, and the higher spin algebra arises naturally
as symplectomorphisms, while the base can in general be a differential 
Poisson manifold (equipped with an integration measure).
Further below, we shall largely trivialize these structures, though 
the requirement that higher spin gravity admits a global 
formulation within this category of quantum geometries may prove fruitful
in constraining the theory and constructing new models.

The noncommutative structure of ${\cal C}$ is carried by a differential 
associative algebra $\O({\cal C})$ consisting of symbols, which is
a class of differential forms on ${\cal C}$, thought of as a classical
manifold, equipped with a star product, differential and trace operation.
Formally, one may obtain these structures by deforming the wedge product,
de Rham differential and classical integration measure for smooth and 
bounded forms on ${\cal C}$ (thought of as a compact space) along a 
semi-classical differential Poisson structure \cite{Chu,Beggs,Tagliaferro,Zumino}.
For the application to higher spin gravity, however, it is crucial
that these deformations are \emph{finite}, as Vasiliev's theory makes
explicit usage of non-trivial roots of the 
unity, that have no classical limit, in order to construct
closed and central elements in positive form degree that can be 
used to build cocycles gluing sections to connections.
We shall use the term master field to refer to sections and connections alike,
though, more precisely, the elements in $\O({\cal C})$ are sections, which means 
that the zero-forms must be bounded, while the connections are assumed
to act faithfully on $\O({\cal C})$ and have curvatures in $\O({\cal C})$.

The choice of $\O({\cal C})$ is dictated by the boundary 
conditions of the theory.
We shall consider solution spaces in which the master fields 
can be expanded in terms of a basis of functions on ${\cal Y}$ 
star-multiplied with component forms on ${\cal B}$, which are
determined on-shell by differential constraints.
As for the section components, it is assumed that they have finite 
integrals over ${\cal B}$ in order to yield finite star products 
and traces; thus they should belong to $L^1({\cal B})$\footnote{On $\Real^2$, the integral
version of the star product is equivalent to the twisting 
of the convolution product by a cocycle given by a phase factor,
which does not affect the convergence of the integration;
thus, as the convolution product closes for functions
in $L^1(\Real^2)$, so does the star product; we thank
Giuseppe de Nittis for pointing out this fact to us.}.
As for connection components, it is instead assumed that
they give rise to well-defined open Wilson lines,
which can be implemented in noncommutative geometry
using deformed oscillator algebras; for example, see \cite{Ambjorn:2000cs,Gross:2000ba,Bonezzi:2017vha}. 

As for the space of functions on ${\cal Y}$, which one may
think of as the infinite-dimensional representation matrices
of the theory, we shall consider modules of the underlying 
higher spin algebra that lend themselves to the unfolded 
framework for harmonic expansion on $AdS_4$ \cite{Iazeolla:2008ix}, \emph{i.e.}
the elements must be real-analytic in order for the Lorentz tensorial 
component fields to be well-defined.
In the case of the Weyl zero-form, which contain the local
degrees of freedom of the theory, we shall furthermore
assume that its function space admits a positive definite 
sequi- or bilinear form in bases where compact higher spin 
generators are diagonalized, in order to attempt to provide
the theory will a quantum-mechanical interpretation; assuming 
that thar these inner products are induced via the trace 
operation, the functions on ${\cal Y}$ in the Weyl zero-form 
must be elements in $L^2({\cal Y})$.

\paragraph{Noncommutative structure.} 
The noncommutative structure thus amounts to a differential 
$\widehat d:\O({\cal C})\rightarrow \O({\cal C})$ and 
a compatible associative binary composition rule $(\cdot)\star(\cdot):
\O({\cal C})\otimes \O({\cal C})\rightarrow \O({\cal C})$,
that are (finite) deformations of the de Rham differential and the
wedge product, respectively, such that if $\widehat f,\widehat g,\widehat h\in\O({\cal C})$
then\footnote{The hat on $\widehat d$ indicates that the it is in general
a nontrivial deformation of the de Rham differential, whereas the
hats on the elements in $\O({\cal C})$ are used to distinguish them
from elements on various subbundles that will be introduced below.}
\be \widehat d^{\,\,2}\widehat f~=~0\ ,\qquad 
\widehat d\left(\widehat f\star\widehat g\right)~=~
\left(\widehat d\,\widehat f\right)\star\widehat g+
(-1)^{{\rm deg}(\widehat f)}\widehat f\star\left(\widehat d\,\widehat g\right)\ ,\ee
\be \widehat f\star (\widehat g\star \widehat h)~=~(\widehat f\star \widehat g)\star \widehat h\ .\ee
These operations are in addition assumed to be compatible with an 
hermitian conjugation operation $\dagger$, \emph{viz.}
\be \left(\widehat f\star\widehat g\right)^\dagger~=~
(-1)^{{\rm deg}(\widehat f) {\rm deg}(\widehat g)}\big(\widehat g\big)^\dagger\star \big(\widehat f\,\big)^\dagger\ ,
\qquad \left(\widehat d\,\widehat f\right)^\dagger~=~\widehat d\left(\big(\widehat f\,\big)^\dagger\right)\ ,\qquad ((\widehat f)^\dagger)^\dagger=\widehat f\ ,\ee
for all $\widehat f,\widehat g\in\O({\cal C})$.
We shall assume that there exists a cover of ${\cal C}$ 
by charts and corresponding bases for local symbol calculus, 
which one may refer to as a \emph{quantum atlas}. 
The corresponding similarity transformations (including 
transitions between charts) are given by Kontsevich gauge 
transformations \cite{Kontsevich1997}, which combine 
bundle isomorphims (including higher spin gauge 
transformations) with re-orderings of symbols generated
by (symmetric) polyvector fields.
The resulting ambiguities are factored out at the level of
classical observables given by functionals that are invariant
under (small) Kontsevich gauge transformations.

\paragraph{Topology of correspondence space.} 
As for the bundle structure of the correspondence space, \emph{viz.}
\be {\cal Y} \hookrightarrow {\cal C} \longrightarrow {\cal B}\ ,\ee
we take the fiber space ${\cal Y}\cong \Real^4$ to be a symplectic
manifold with global canonical coordinates; the structure group is 
thus the group of symplectomorphisms of ${\cal Y}$, playing the 
role of higher spin gauge group\footnote{
The elevation of the space of horizontal forms on ${\cal C}$,\emph{i.e.} the kernel of inner derivation 
along vertical vector fields, to a differential graded subalgebra of $\Omega({\cal C})$ 
compatible with the quantum geometry, requires a differential Poisson structure with special (abelian) Killing 
vectors \cite{Arias:2016agc} along ${\cal Y}$ in order to project the star product
and a special vertical top-form to project the quantum differential.}.
As has already been remarked above, although ${\cal Y}$ is not compact, 
the class of functions on ${\cal Y}$ used to expand the sections will 
in effect be assumed to provide ${\cal Y}$ with a compact topology 
(at the level of the trace operation); further below, we shall
achieve this by taking the functions on ${\cal Y}$ to be symbols 
of (super)traceable operators in Fock spaces (defined using Weyl order).

We furthermore assume ${\cal B}$ to be closed (hence compact) while 
allowing the (higher spin) connection to blow up on a submanifold 
\be {\cal S}\subset{\cal B}\ ,\ee
modulo gauge and ordering artifacts, that is, we assume
that there exists a quantum atlas such that the connection is 
smooth away from ${\cal S}$; conversely, we assume that the 
singularities in the connection on ${\cal S}$ are not
removable by going to a ``finer'' quantum atlas.
We remark that, in this sense, \emph{large} Kontsevich gauge transformations
can be used to generate physical singularities in the connection
whose structure define boundary states, such as, for example, conformal
classes of higher spin gauge fields corresponding to
sources for conserved currents on the dual conformal
field theory side.

In order to describe asymptotically anti-de Sitter higher spin geometries,
we shall take the differential Poisson structure on ${\cal C}$ to 
be flat and torsion free \cite{Chu,Beggs}, and assume a 
the bundle structure of ${\cal C}$ to be trivial with
\be {\cal B}\cong {\cal X}\times {\cal Z}\ ,\ee
where 
\be {\cal X}\cong S^1\times S^3\ee
is commutative, and 
\be {\cal Z}\cong S^2\times\overline S^2\ee
is a non-commutative space obtained by adding (commuting) points at infinity 
to ${\cal Z}'\cong \Real^4$ \cite{FCS}.
In other words, we have 
\be {\cal C}={\cal X}\times {\cal T}\ ,\qquad 
{\cal T}:={\cal Y}\times {\cal Z} \cong \Real^4\times S^2\times\overline S^2\ .\ee
where ${\cal T}$ is thus obtained by adding points at infinity to 
\be {\cal T}':={\cal Y}\times {\cal Z}' \cong \Real^8\ ,\ee
equipped with global canonical coordinates.

\paragraph{Ordering schemes and boundary conditions.}

In providing the (perturbatively) exact solutions,
we shall start from an Ansatz in which the sections belong to 
the algebra
\be \Omega({\cal C})=\Omega({\cal X})\otimes \Omega({\cal T})\ ,\ee
where $\Omega({\cal T})$ consists of symbols defined
using \emph{Weyl order} that are forms in $L^1({\cal Z})$
valued in an extension ${\cal W}$ of the Weyl algebra on ${\cal Y}$
by inner Klein operators \cite{FCS}, as we shall describe further below.
The trace operation on $\Omega({\cal T})$ is given by integration 
over ${\cal Z}$ combined with a trace operation on ${\cal W}$ given
by a regularized integral over ${\cal Y}$; see \cite{FCS} for the definition. 
We shall then apply a large gauge transformation, and convert the
sections to elements in the algebra $\Omega({\cal T}')$ consisting 
of symbols defined using \emph{normal order} that are real-analytic 
at the origin of ${\cal T}'$, that is, they can be expanded in the 
basis of monomials in the canonical coordinates on ${\cal T}'$,
at generic spacetime points.
As for ${\cal X}$, we shall keep the time-direction periodic 
by working with a compact spectrum of states realized in ${\cal W}$.
Furthermore, we decompactify $S^3$ by allowing the connection
to blow up at its south and north poles, $p_{\rm S}$ and $p_{\rm N}$, 
respectively, corresponding to the boundary and origin of the anti-de 
Sitter spacetime background, \emph{i.e.} we take
\be {\cal S}= S^1\times p_{\rm S}\times {\cal P}_{\rm S} \,\cup\, 
S^1\times p_{\rm N}\times {\cal P}_{\rm N}\ ,\ee
where ${\cal P}_{\rm S,N}$ are planes passing through the origin 
of ${\cal T}$.
We propose that requiring the connection at $p_{\rm S}$ to 
consist of asymptotically defined Fronsdal fields, defined order by order 
in perturbation theory, yields an irreducible gauge equivalence class
of connections; the implementation of this boundary condition 
requires the aforementioned large gauge transformation,
as will be discussed further below. 
We shall furthermore propose that the quasi-local Fronsdal branch arises 
upon further requiring smooth connections at $p_{\rm N}$.

As for the status of the boundary conditions on the connection, we 
shall implement them only to the linearized order.
It remains to be seen whether they can be reached at higher orders,
and, if so, whether they are non-trivial (in the sense that
they select an irreducible gauge equivalence class) in view of the 
ambiguities residing in the form of large Kontsevich gauge 
transformations.
We shall also leave for future work the issue of whether the 
sections in $\Omega({\cal T}')$ can actually be mapped back to 
$\Omega({\cal T})$ (by going from normal back to Weyl order).
We remark that this would imply the Weyl zero-form is actually
expandable over the basis of ${\cal W}$ in terms of bounded 
component fields, which may seem surprising in the case of the 
black-hole-like solutions, whose separate linearized Weyl tensors 
have singularities at $p_{\rm N}$.
However, as we shall see, these can be summed up and converted to regular
elements in the extended Weyl algebra (given by delta functions 
on ${\cal Y}$), at least at the linearized order.
Another problem that we shall not address in any detail is whether 
the quasi-local Fronsdal branch can be equivalently described  
using an entirely normal-ordered scheme.
To this end, it would be natural to equip $\Omega({\cal T}')$ with 
the standard trace given by the integral over ${\cal T}'$, which 
would require the symbols to be elements of $L^1({\cal T}')$; 
for a brief discussion, see Section 5.3. 

\paragraph{Higher spin gauge fields and deformed oscillator algebras.}

Large gauge transformations are required in order to construct 
connections that obey the dual boundary conditions in normal 
and Weyl order.
In normal order, the corresponding gauge condition in 
$\Omega({\cal T}')$, namely, the Vasiliev gauge condition 
to be spelled out below, ensures that the restriction of the 
connection on ${\cal C}$ from ${\cal B}$ to ${\cal X}$,
or more precisely, projecting onto the elements in $\Omega({\cal T}')$ 
that are constant on ${\cal Z}$, defines a higher spin connection
on ${\cal X}$ valued in a higher spin Lie algebra, to be defined below.
The Vasiliev gauge condition implies that the gauging of the higher
spin algebra, or unfolding procedure, yields a set of Fronsdal fields 
with particular self-interaction upon taking the 
spin-two subsector to describe a Lorentzian spacetime.
Dually, for a generic point $p\in{\cal X}$, the restriction 
of the connection on ${\cal C}$ to $\{p\}\times {\cal Z}'$ is assumed 
to equip ${\cal T}'$ with a \emph{deformed oscillator algebra} 
$\widehat {\cal A}|_p$ that is a subalgebra of $\Omega({\cal T}')$
and that in its turn is assumed to admit a Weyl ordered description 
in $\Omega({\cal T})$.

We would like to stress the fact that Vasiliev's procedure is
designed to describe interactions in spacetime
that are dual to deformed oscillator algebras in twistor space
rather than to obey any quasi-locality conditions in spacetime. 
Whether the resulting conversion of singularities in spacetime 
to distributions in twistor space provides a physically acceptable 
model for black holes remains to be spelled out in more detail.
A promising fact, noted already in \cite{us}, is that the relevant
Type D sector, which is infinite-dimensional in higher spin
gravity already in the metric-like formulation, is mapped
by means of a simple $\mathbb Z_2$-transformation to 
the ordinary massless particle spectrum consisting of 
lowest-weight spaces, as we shall recall below.

\paragraph{Local coordinates.}
We coordinatize ${\cal C}$ using ($\underline\a=(\a,\ad)$; $\a,\ad=1,2$)
\footnote{Our spinor conventions
are collected in Appendix \ref{App:conv}.}
\be \Xi^{\underline M}~=~ (X^M;Y^{\underline\a})\ ,\qquad X^M~=~ (x^\mu; Z^{\underline\a})\ ,\qquad (Y^{\underline\a};Z^{\underline\a})~=~(y^\a,\yb^{\ad};z^\a,-\zb^{\ad})\ ,\ee
with reality properties
\be (x^\mu)^\dagger~=~ x^\mu\ ,\qquad 
(y^\a)^\dagger~=~\yb^{\ad}\ ,\qquad 
(z^\a)^\dagger~=~\zb^{\ad}\ ,\ee
and canonical commutation rules
\be [y^\a,y^\b]_\star ~=~2i\e^{\a\b}\ ,\qquad 
[z^\a,z^\b]_\star~=~-2i\e^{\a\b}\ , \qquad [y^\a,z^\b]_\star ~=~0 \ ,\label{oscillators}\ee
\emph{idem} their complex conjugates.
We also have 
\be \left[ \widehat d \Xi^{\underline M}, \widehat f\,\right]_\star =0 \ ,\ee
where the bracket is graded, and we choose a symbol calculus such that 
\be \widehat d \Xi^{\underline M} \star \widehat f= \widehat d \Xi^{\underline M} \wedge \widehat f\ ;\ee
in what follows we shall suppress the wedge product when ambiguities cannot arise.
The local representatives of horizontal forms and the differential acting on them are given, respectively, by 
\be \widehat f|_{\rm hor} \stackrel{\rm loc}{=} \widehat f|_{dY^{\underline\a}=0}~=~\widehat f(X,dX;Y)\ ,\qquad
\widehat d|_{\rm hor}  \stackrel{\rm loc}{=}  \widehat d|_{dY^{\underline\a}=0}~=~d X^M \partial_M\ .\ee

\paragraph{Star product formula.}
On ${\cal T}'$, we shall use Vasiliev's normal ordered star product, 
given by the following twisted convolution formula:
\begin{eqnarray}
&&\widehat f_{1}\left( y,\bar{y},z,\bar{z}\right) \star 
\widehat f_{2}\left( y,\bar{y},z,\bar{z
}\right)  \notag \\[5pt]
&=&\int_{{\cal R}_{\Real}} \frac{d^{2}ud^{2}\bar{u}d^{2}vd^{2}\bar{v}}{\left( 2\pi \right) ^{4}}
e^{i\left( v^{\alpha }u_{\alpha }+\bar{v}^{\dot{\alpha}}\bar{u}_{\dot{\alpha}
}\right) }\ \widehat f_{1}\left( y+u,\bar{y}+\bar{u};z+u,\bar{z}-\bar{u}\right)
\widehat f_{2}\left( y+v,\bar{y}+\bar{v};z-v,\bar{z}+\bar{v}\right) \text{ ,}  \notag
\\
&&  \label{star prod def}
\end{eqnarray}
where the integration domain 
\be  {\cal R}_{\mathbb R}\ =\ \{~(u_\a,\bar u_{\ad};v_\a,\bar v_{\ad})~:~ 
(u_{\a})^\dagger=u_\alpha\ , \quad 
(\bar u_{\ad})^\dagger=\bar u_{\ad}\ , \quad
(v_{\a})^\dagger=v_\alpha\ , \quad 
(\bar v_{\ad})^\dagger=\bar v_{\ad}\ \}\ ,\ee
that is, all auxiliary variables are integrated over the real line.
In particular, it follows that 
\begin{equation}
y_{\alpha }\star y_{\beta }=y_{\alpha }y_{\beta }+i\varepsilon _{\alpha
\beta }\text{ , \qquad }y_{\alpha }\star z_{\beta }=y_{\alpha }z_{\beta
}-i\varepsilon _{\alpha \beta }\text{ ,}\ee
\be z_{\alpha }\star y_{\beta
}=z_{\alpha }y_{\beta }+i\varepsilon _{\alpha \beta }\text{ , \qquad }z_{\alpha
}\star z_{\beta }=z_{\alpha }z_{\beta }-i\varepsilon _{\alpha \beta }\text{ ,
}  \label{yz comm}
\end{equation}
\emph{idem} the anti-holomorphic variables.
Equivalently, in terms of the creation (+) and annihilation (-) operators
\begin{equation}
a_{\alpha }^{\pm }\ := \ \frac{1}{2}\left( y_{\alpha }\pm z_{\alpha }\right) \text{
,}
\end{equation}
one has 
\be a_{\alpha }^{-} \star a_{\beta }^{+}\ = \ a_{\alpha }^{-} a_{\beta }^{+}+
i \varepsilon _{\alpha \beta }\ ,\qquad
a_{\alpha }^{+}\star a_{\beta }^{-}\ = \ a_{\alpha }^{+} a_{\beta }^{-}\text{ , } \ee
\be a_{\alpha }^{+}\star a_{\beta }^{+}\ = \ a_{\alpha }^{+} a_{\beta }^{+}\ ,\qquad
a_{\alpha }^{-}\star a_{\beta }^{-}\ = \ a_{\alpha }^{-}a_{\beta }^{-}\text{ .}\ee
Thus, the star-product formula provides a realization of the operator
product in terms of symbols defined in the above \emph{normal order}.
To be more precise, letting ${\cal O}_{\rm Normal}$ denote the Wigner map 
that sends a classical function $\widehat f(y,z)$ to the operator 
${\cal O}_{\rm Normal}(\widehat f(y,z))$ with symbol $\widehat f(y,z)$ 
defined in normal order, we have
\be {\cal O}_{\rm Normal}(\widehat f_{1}\left( y,z\right) \star \widehat f_{2}\left(
y,z\right))=
{\cal O}_{\rm Normal}(\widehat f_{1}\left( y,z\right)) {\cal O}_{\rm Normal}(\widehat f_{2}\left(
y,z\right))
\text{ .}\label{Onormal}\ee
Letting ${\cal O}_{\rm Weyl}$ denote the corresponding map defined using Weyl order,
we have  
\be {\cal O}_{\rm Weyl}(f(y))={\cal O}_{\rm Normal}(f(y))\ ,\qquad
{\cal O}_{\rm Weyl}(f(z))={\cal O}_{\rm Normal}(f(z))\ .\ee
It follows that 
\begin{eqnarray}
{\cal O}_{\rm Weyl}(f_{1}\left( y\right) \star f_{2}\left( y\right)) &=&
{\cal O}_{\rm Weyl}(f_{1}\left( y\right) ) {\cal O}_{\rm Weyl}( f_{2}\left( y\right)) \text{ ,} \\
{\cal O}_{\rm Weyl}(f_{1}\left( z\right) \star f_{2}\left( z\right)) &=&
{\cal O}_{\rm Weyl}(f_{1}\left( z\right) ) {\cal O}_{\rm Weyl}(f_{2}\left( z\right))\text{ ,}
\end{eqnarray}
and that 
\be
{\cal O}_{\rm Normal}( f_{1}\left( y\right) \star f_{2}\left( z\right))
\ =\ {\cal O}_{\rm Weyl}( f_{1}\left( y\right) ) {\cal O}_{\rm Weyl}( f_{2}\left( z\right))\ =\
{\cal O}_{\rm Weyl}( f_{1}\left( y\right) f_{2}\left( z\right)) \text{ ,}  \label{f1f2 Weyl}\ee
where the fact that the $Y$ and $Z$ oscillators are mutually commuting
has been used in the last step. 
In other words, if the symbol $\widehat f(y,z)$ of an operator defined using normal
order can be factorized as $\widehat f(y,z)=\sum_\lambda f_\l(y)\star f^\l(z)$ (over
some classes of functions or distributions on ${\cal Y}$ and ${\cal Z}$),
then its symbol defined using Weyl order is given by $\sum_\lambda f_\l(y)f^\l(z)$.
We would like to stress, however, that, in what follows, all symbols will always
be given using the normal order.

\paragraph{Regularization of star products.} The twisted convolution 
formula extends the Moyal product from the space of real analytic 
functions to the space of Fourier transformable functions, including 
delta function distributions (including their derivatives).
When applied to group elements and projectors, this may result 
in auxiliary Gaussian integrals that involve indefinite diagonalizable 
bilinear forms; resorting to the original Moyal-product shows that 
these integrals must be performed by means of analytical continuation 
in the eigenvalues of the bilinear forms.

\scss{Master fields}

The connection on ${\cal C}$ is a one-form $\widehat A$ with curvature
\be \widehat F:= \widehat d\widehat A+\widehat A\star \widehat A\ ,\ee
transforming in the adjoint representation of the structure group,
and obeying the Bianchi identity
\be \widehat D\widehat F:=\widehat d\widehat F+[\widehat A, \widehat F]_\star \equiv 0\ .\ee
Higher spin gravity also makes use of twisted-adjoint \cite{Vasiliev:1990vu,more,Vasiliev:1999ba}, 
and, more generally, bi-fundamental \cite{FCS} master fields, that can 
be introduced geometrically by replacing $\Omega({\cal C})$ by 
$\Omega({\cal C}) \times {\cal K} \times {\cal F}$, where ${\cal K}$ 
is generated by outer Klein operators \cite{vasiliev,Vasiliev:1990vu,Vasiliev:1999ba} and ${\cal F}$ 
is an internal graded Frobenius algebra
\cite{FCS}.
Consistent truncation \cite{FCS} leads to a twisted
adjoint zero-form $\widehat \Phi$ and a pair of closed
and twisted-central two-forms $(\widehat J,\,\,\widehat{\!\!\bar J})$.
The resulting fields obey the reality conditions
\be (\widehat \Phi,\widehat A,\widehat J,\widehat{\bar J})^\dagger~=~
(\pi(\widehat \Phi),-\widehat A,-\widehat{\bar J},-\widehat J)\ ,\label{reality}\ee
where $\pi$ and $\bar\pi$ are the involutive automorphisms defined by 
$\widehat d\,\pi=\pi\,\widehat d$, $\widehat d\,\bar\pi=\bar\pi\,\widehat d$ and
\be \pi(x^\mu;y^\a,\yb^{\ad};z^\a,\zb^{\ad})~=~
(x^\mu;-y^\a,\yb^{\ad};-z^\a,\zb^{\ad})\ ,\qquad \pi(\widehat f\star\widehat g)~=~
\pi(\widehat f)\star \pi(\widehat g)\ ,\ee
\be \bar\pi(x^\mu;y^\a,\yb^{\ad};z^\a,\zb^{\ad})~=~
(x^\mu;y^\a,-\yb^{\ad};z^\a,-\zb^{\ad})\ ,\qquad \bar\pi(\widehat f\star\widehat g)~=~
\bar\pi(\widehat f)\star \bar\pi(\widehat g)\ .\ee
In order to define the twisted-adjoint representation, we introduce 
the graded bracket
\be [\widehat f,\,\widehat g\,]_\pi:=
\widehat f\star\widehat g-(-1)^{{\rm deg}(\widehat f){\rm deg}(\widehat g)}\widehat g\star\pi(\widehat f)\ ,\qquad 
\widehat f,\widehat g\in\O({\cal C})\ .\ee
The twisted adjoint covariant covariant derivative 
\be \widehat D\widehat \Phi:=\widehat d\,\widehat \Phi+[\widehat A,\widehat\Phi]_\pi\ ,\ee
obeys the Bianchi identity
\be \widehat D^2\widehat \Phi\equiv [\widehat F,\widehat\Phi]_\pi\ .\ee
The conditions on the two-form making it closed and twisted-central read
\be \widehat d\widehat J=0\ ,\qquad [\widehat J,\widehat f]_\pi=0\ ,\label{e2}\ee
for all $\widehat f\in \Omega({\cal C})$.
The bosonic models require the integer-spin projection
\be \pi\bar\pi(\widehat \Phi,\widehat A)~=~(\widehat \Phi,\widehat A)\ ,\label{piofj}\ee
which together with the reality conditions lead to real 
Fronsdal fields with integer rank, each rank occurring once.
The twisted-central elements obey the stronger conditions
\be \pi(\widehat J,\,\,\widehat{\!\!\bar J})~=~\bar\pi(\widehat J,\,\,\widehat{\!\!\bar J})~=~
(\widehat J,\,\,\widehat{\!\!\bar J})\ .\ee
In the minimal models, the odd-spin Fronsdal fields are
removed by the stronger even-spin projection
\be \tau(\widehat \Phi,\widehat A,\widehat J,\widehat{\bar J})~=~(\pi(\widehat \Phi),-\widehat A,-\widehat J,-\widehat{\bar J})\ ,\label{min}\ee
where $\tau$ is the graded anti-automorphism defined by $\widehat d\,\tau ~=~\tau\, \widehat d$ and
\be \tau(x^\mu;Y^{\underline\a};Z^{\underline\a})~=~(x^\m;iY^{\underline\a};-iZ^{\underline\a})\ ,\qquad \tau(\widehat f\star\widehat g)~=~(-1)^{\widehat f\widehat g} \tau(\widehat g)\star \tau(\widehat f)\ ;\ee
from $\tau^2=\pi\bar\pi$ it follows that \eq{min} implies \eq{piofj}.

\paragraph{Twisted-central and closed elements.} 

Eq. \eq{e2} admits the following non-trivial solution \cite{Didenko:2009td}:
\begin{equation}
\widehat J=j_z\star \kappa_y\ ,\qquad j_z:=-\tfrac{i}{4}dz^\alpha \wedge dz^\b \varepsilon _{\alpha \beta }\kappa_z 
\text{ , \qquad }\kappa_{y}:=2\pi \delta
^{2}\left( y\right) \text{ , \qquad }\kappa_{z}:=2\pi \delta ^{2}\left( z\right)
\text{ ,}\label{factorized}
\end{equation}
where $\kappa_y$ is an inner Klein operator obeying
\begin{equation}
\kappa_y \star f(y)\star \kappa_y= f(-y) \text{ , \qquad }\kappa _{y}\star
\kappa _{y}\ =\ 1\ , \label{kyprop}
\end{equation}
idem $\kappa_z$.
Thus, one may write \cite{Didenko:2009td,berezin}
\be \widehat J=-\tfrac{i}{4}dz^\alpha \wedge dz^\b 
\varepsilon _{\alpha \beta }\widehat \kappa\ ,\qquad
\widehat \kappa:=\kappa _{y}\star \kappa _{z}=\exp(i y^\a z_\a)\ ,\label{kappa}\ee
where thus
\be \widehat \kappa\star \widehat f(y,z)=\widehat \kappa \widehat f(z,y)\ ,\qquad 
\widehat f(y,z)\star \widehat \kappa=\widehat \kappa \widehat f(-z,-y)\ ,\ee
\be \widehat \kappa\star
\widehat f(y,z)\star \kappa=\pi(\widehat f(y,z))\ ,\qquad \widehat \kappa\star\widehat \kappa=1\ .\ee
By hermitian conjugation one obtains
\begin{equation}\widehat{\overline {J}}=-(\widehat J) ^{\dag }=-
\tfrac{i}{4}d\bar z^{\dot\alpha} \wedge d\bar z^{\dot\beta}
\varepsilon _{\dot{\alpha}\dot{\beta}}\widehat {\bar{\kappa}}\ .\end{equation}
%

\scss{Equations of motion}\label{fieldequations}

\paragraph{Master field equations.}
Vasiliev's equations of motion are given by
\be  \widehat D\widehat \Phi~=~0\ ,\qquad   
 \widehat F+{\cal F}(\widehat \Phi)\star \widehat J+
 \overline{\!\cal F}(\widehat \Phi)\star \,\,\widehat{\!\!\bar J}~=~0\ ,\label{e1}\ee
where the interaction ambiguities\footnote{The functionals ${\cal F}$
cannot be fixed by any \emph{a priori} considerations on-shell.
In the off-shell formulation proposed in \cite{FCS}, the
parameters in ${\cal F}$ represent degrees of freedom entering via 
a dynamical two-form.}.
\be {\cal F}~=~\sum_{n=0}^\infty f_{2n+1}(\widehat \Phi) \left(\widehat\Phi\star\pi(\widehat\Phi)\right)^{\star n}\star \widehat \Phi\ ,\ee
and $\overline{\!\cal F}=({\cal F})^\dagger$, where $f_{2n+1}$ are 
complex-valued zero-form charges, which are functionals of
$\widehat \Phi$ obeying
\be \widehat d f_{2n+1}~=~0\ ,\label{f2n+1constraints}\ee
on-shell.
Factoring out perturbatively defined redefinitions of 
$\widehat \Phi$, one has \cite{Vasiliev:1999ba,Sezgin:2011hq}
\be {\cal F}~=~{\cal B}\star\widehat\Phi\ ,\qquad {\cal B}~=~
\exp_\star\left(i\theta(\widehat\Phi)\right)\ ,\label{calB}\ee\be \theta~=~ 
\sum_{n=0}^\infty \theta_{2n}(\widehat\Phi)\,\left(\widehat\Phi\star\pi(\widehat\Phi)\right)^{\star n}\ ,\label{Theta}\ee
where $\theta_{2n}$ are real-valued zero-form charges.
Introducing the parity operation $P$ given by the  
automorphism of $\O({\cal C})$ defined by
\be P(x^\mu;y^\a,\yb^{\ad};z^\a,\zb^{\ad})~=~(x^\mu;\yb^{\ad},y^\a;-\zb^{\ad},-z^\a)\ ,\qquad \widehat  d P~=~P \widehat d\ ,\label{Posc}\ee
and by a linear action on the expansion coefficients of the master fields,
it follows that 
\be P(\widehat J, \widehat{\bar J})=(\widehat{\bar J},\widehat J)\ .\ee
Hence ${\cal B}$ breaks parity except in the following two 
cases \cite{Sezgin:2003pt}:
\bea \mbox{Type A model (scalar)}&:& \theta~=~0\ ,\qquad P(\widehat \Phi,\widehat A)~=~(\widehat \Phi,\widehat A)\ ,\\[5pt]
\mbox{Type B model (pseudo-scalar)}&:& \theta~=~\frac{\pi}2\ ,\qquad P(\widehat \Phi,\widehat A)~=~(-\widehat \Phi,\widehat A)\ .\eea
The equations of motion are Cartan integrable, and hence
admit the following on-shell Cartan gauge transformations:
\be \delta_{\widehat\e}\widehat \Phi~=~-[\widehat \e,\widehat\Phi]_\pi\ ,\qquad \delta_{\widehat\e}\widehat A~=~\widehat D\widehat\e:=\widehat d\widehat\e+[\widehat A,\widehat\e]_\star\ ,\ee
where the parameters obey the same kinematic conditions as $\widehat A$,
and the two-form is treated as a background in the sense that $\delta_{\widehat\e}\widehat J~=~0$
idem $\widehat{\bar J}$. 

\paragraph{Flat connection and deformed oscillators.}

In order to exhibit more explicitly the deformation of the
curvature induced by the closed and twisted-central elements, we decompose 
\be \widehat A~=~\widehat U+\widehat V\ ,\ee
where
\be \widehat U~:=~ dx^\mu\widehat U_\mu(x;Z,Y)\ ,\ee\be  \widehat V~:=~ dZ^{\underline\a}\widehat V_{\underline\a}(x;Z,Y)~=~dz^\a \widehat V_\a(x;Z,Y)+dz^{\ad} \widehat V_{\ad}(x;Z,Y)\ ,\ee
and introduce the deformed oscillators
\be \widehat S_{\underline\a}~:=~( \widehat S_{\a}, -\widehat{\bar S}_{\ad})~:=~Z_{\underline\a} -2i\widehat V_{\underline\a}~=~(z_\a-2i \widehat V_\a,-\zb_{\ad}+2i \widehat{\bar V}_{\ad})\ .\label{SV}\ee
Letting $d=dx^\mu\partial_\mu$, the master field equations can be rewritten as
\be d \widehat U+\widehat U\star \widehat U~=~0\ ,\qquad d\widehat \Phi+\widehat U\star\widehat \Phi-\widehat \Phi\star \pi(\widehat U)~=~0\ ,\label{MC}\ee
\be d\widehat S_{\underline\a}+[\widehat U,\widehat S_{\underline\a}]_\star~=~0\ ,\label{dSa}\ee
\be \widehat S_\a\star\widehat\Phi+\widehat\Phi\star\pi(\widehat S_\a)~=~0\ ,\quad
\widehat {\bar S}_{\ad}\star\widehat\Phi+\widehat\Phi\star\bar\pi(\widehat {\bar S}_{\ad})~=~ 0  \ ,\label{INT1}\ee\be
[\widehat S_\a,\widehat S_\b]_\star~=~ -2i\e_{\a\b}(1-{\cal B}\star\widehat \Phi\star\widehat \kappa)\ ,\quad [\widehat {\bar S}_{\ad},\widehat {\bar S}_{\bd}]_\star~=~ -2i\e_{\ad\bd}(1- \overline{\cal B}\star  \widehat\Phi\star\widehat {\bar \kappa}) \ ,\label{INT2}\ee\be
[\widehat S_\a,\widehat{\bar S}_{\ad}]_\star~=~ 0\ .\label{INT3}\ee
The gauge transformations now read
\be \delta_{\widehat \e} \,\widehat \Phi~=~-[\widehat\e,\widehat\Phi]_\pi\ ,\qquad 
\delta_{\widehat \e}\,\widehat S_{\underline\a}\ =\ -[\widehat \e,\widehat V_{\underline\a}]_\star\ ,\qquad 
\delta_{\widehat \e}\, \widehat U~=~ d\widehat \e+[\widehat U,\widehat \e\,]_\star\ .\ee
Thus, the master field equations describe a flat connection on ${\cal X}$ and
a covariantly constant deformed oscillator algebra 
\be  \widehat{\cal A}:= \left\{ (\widehat\Phi\star \widehat \kappa)^{\star k} \star (\Phi\star\widehat{\bar\kappa} )^{\star \bar k}\star \widehat S_{(\underline \a_1}\star\cdots \star\widehat S_{\underline \a_m)}\right\}\ ,\ee
generated by the internal master fields $(\widehat \Phi,\widehat S_{\underline\a})$.
These two objects belong to subalgebras of $\Omega({\cal T}')$ that 
we expect to be determined by the boundary conditions, as we shall
examine in more detail in the case of higher spin black holes and
fluctuation fields.
%

\paragraph{Lorentz covariantization.}

To obtain a manifestly locally Lorentz covariant formulation 
one introduces a canonical Lorentz connection $(\o^{\a\b},{\bar \o}^{\ad\bd})$ 
by means of the field redefinition \cite{Vasiliev:1999ba,Sezgin:2002ru,Sezgin:2011hq}
\be \widehat W~:=~ \widehat U-\widehat K\ ,\qquad \widehat K~:=~ \frac1{4i} \left(\o^{\a\b} \widehat M_{\a\b}+\bar \o^{\ad\bd} \widehat {\overline M}_{\ad\bd}\right)\ ,\label{fieldredef}\ee
where 
\bea \widehat M_{\a\b}&:=& \widehat M^{(0)}_{\a\b}+\widehat M^{(S)}_{\a\b}\ ,\qquad
\widehat {\overline M}_{\ad\bd}\: =\ \widehat{\overline M}^{(0)}_{\ad\bd}+\widehat{\overline{M}}^{(\bar S)}_{\ad\bd}\ ,\label{fullM}\eea
are the full Lorentz generators, consisting of the internal part
\bea \widehat M^{(0)}_{\a\b}&:=& y_{(\a} \star y_{\b)}-z_{(\a} \star z_{\b)}\ ,\qquad
\widehat{\overline M}^{(0)}_{\ad\bd}\ :=\ \yb_{(\ad}\star \yb_{\bd)}- \zb_{(\ad} \star \zb_{\bd)}\ ,\label{M(0)}\eea
rotating the $Y$ and $Z$ oscillators, and the external part
\bea \widehat M^{(S)}_{\a\b}&:=& \widehat S_{(\a}\star \widehat S_{\b)}\ ,\qquad
\widehat{\overline M}^{(\bar S)}_{\ad\bd}\ :=\  \widehat {\bar S}_{(\ad}\star \widehat {\bar S }_{\bd)}\ ,\label{M(S)}\eea
rotating the spinor indices carried by $(\widehat S_\a,\widehat{\bar S}_{\ad})$. 
As a result, the master equations read
\bea &\nabla\widehat W+\widehat W\star \widehat W + \frac1{4i} \left(r^{\a\b} \widehat M_{\a\b}+\bar r^{\ad\bd} \widehat {\overline M}_{\ad\bd}\right)\ =\ 0\ ,\quad \nabla\widehat \Phi+\widehat W\star\widehat \Phi-\widehat \Phi\star\pi(\widehat W)\ =\ 0\ ,\label{2.55}&\\[5pt]
&\nabla\widehat S_\a+\widehat W\star\widehat S_\a-\widehat S_\a\star \widehat W\ =\ 0\ ,\quad \nabla\widehat {\bar S}_{\ad}+\widehat W\star\widehat {\bar S}_{\ad}-\widehat {\bar S}_{\ad}\star \widehat W\ =\ 0&\\[5pt]
&
\widehat S_\a\star\widehat\Phi+\widehat\Phi\star\pi(\widehat S_\a)\ =\ 0\ ,\quad
\widehat {\bar S}_{\ad}\star\widehat\Phi+\widehat\Phi\star\bar\pi(\widehat {\bar S}_{\ad})\ =\ 0 &\\[5pt]
&
[\widehat S_\a,\widehat S_\b]_\star\ =\ -2i\e_{\a\b}(1-{\cal B}\star\widehat \Phi\star\widehat\kappa)\ ,\quad [\widehat {\bar S}_{\ad},\widehat {\bar S}_{\bd}]_\star\ =\ -2i\e_{\ad\bd}(1- \overline{\cal B}\star \widehat\Phi\star\widehat{\bar \kappa})&\\[5pt]
&
[\widehat S_\a,\widehat{\bar S}_{\ad}]_\star\ =\ 0\ ,&\label{2.56}\eea
where $r^{\a\b}:=d\o^{\a\b}+\o^{\a\c}\o^{\b}{}_\c$ and $\bar r^{\ad\bd}:=d\bar \o^{\ad\bd}+\o^{\ad\cd}\o^{\bd}{}_{\cd}$,
and
\bea \nabla \widehat W&:=& d\widehat W+\frac1{4i} \left[\o^{\a\b} \widehat M^{(0)}_{\a\b}+\bar \o^{\ad\bd} \widehat {\overline M}^{(0)}_{\ad\bd}~,~\widehat W\right]_\star\ ,\label{lorcovfirst}\\[5pt]
\nabla \widehat \Phi&:=& d\widehat \Phi+\frac1{4i} \left[\o^{\a\b} \widehat M^{(0)}_{\a\b}+\bar \o^{\ad\bd} \widehat {\overline M}^{(0)}_{\ad\bd}~,~\widehat \Phi\right]_\star\ ,\\[5pt]
\nabla \widehat S_\a&:=& d\widehat S_\a+\o_\a{}^\b\widehat S_\b+\frac1{4i} \left[\o^{\b\c} \widehat M^{(0)}_{\b\c}+\bar \o^{\bd\cd} \widehat {\overline M}^{(0)}_{\bd\cd}~,~\widehat S_\a\right]_\star\ ,\\[5pt]
\nabla \widehat S_{\ad}&:=& d\widehat S_{\ad}+\bar \o_{\ad}{}^{\bd}\widehat {\bar S}_{\bd}+\frac1{4i} \left[\o^{\b\c} \widehat M^{(0)}_{\b\c}+\bar \o^{\bd\cd} \widehat {\overline M}^{(0)}_{\bd\cd}~,~\widehat {\bar S}_{\ad}\right]_\star\ .\label{lorcovlast}\eea
The field redefinition implies a local shift-symmetry 
with parameter $(\varsigma^{\a\b},{\bar \varsigma}^{\ad\bd})=
dX^M(\varsigma_M{}^{\a\b},{\bar \varsigma}_M{}^{\ad\bd})$ defined by
\be \delta_{\varsigma} (\widehat U,\widehat \Phi,\widehat S_\a,\widehat{\bar S}_{\ad})~=~0\ ,\qquad  \delta_{\varsigma} (\o^{\a\b},{\bar \o}^{\ad\bd})~=~ (\varsigma^{\a\b},{\bar \varsigma}^{\ad\bd})\quad\Rightarrow\quad \delta_{\varsigma}\widehat W~=~ -\frac1{4i} \left(\varsigma^{\a\b} \widehat M_{\a\b}+\bar \varsigma^{\ad\bd} \widehat {\bar M}_{\ad\bd}\right)\ ,\ee
which can be used to embed the canonical Lorentz connection 
into the full master fields by imposing
\be\left.{\partial^2\over \partial y^\a \partial y^\b} \widehat W\right|_{Y=Z=0}~=~0\ ,\qquad \left.{\partial^2\over \partial \bar y^{\ad} \partial {\bar y}^{\bd}} \widehat W\right|_{Y=Z=0}~=~0\ .\label{shiftgauge}\ee
%

\paragraph{Perturbatively defined Fronsdal fields.}

To obtain a perturbative expansion in terms of self-interacting Fronsdal fields
that is locally Lorentz covariant and diffeomorphism invariant
on ${\cal X}$, one proceeds by imposing the initial conditions
\be \widehat \Phi|_{Z=0} \ = \ \Phi\ ,\qquad \widehat W|_{Z=0} \ = \ W\ ,\label{2.61}\ee
where $W$ is valued in the bosonic higher spin algebra $\mhs_1(4)$ or in
its minimal projection $\mhs(4)$ and $\Phi$ in the corresponding 
twisted-adjoint representation \cite{vasiliev,Vasiliev:1990vu,more,Vasiliev:1999ba,Sezgin:2002ru}, 
and the \emph{Vasiliev gauge condition}
\be z^\a\widehat{V}_\a+\bar z^{\ad}\widehat{\bar V}_{\ad} \ = \ 0\ ,\label{vgc}\ee
in normal order, after which it is possible to solve the constraints
involving $Z$-derivatives in a perturbative expansion in terms of $\Phi$
while maintaining real-analyticity on ${\cal T}$.
The remaining constraints can then be shown to hold on ${\cal X}\times {\cal Z}$
provided they hold at ${\cal X}\times \{Z=0\}$, where they define a
non-linear free differential algebra on ${\cal X}$ generated 
by $\Phi$ and $W$ \cite{vasiliev,more,Sezgin:2002ru}\footnote{Concerning the 
Vasiliev gauge condition \eq{vgc}, prior to Lorentz covariantization, it 
ensures that terms involving $\partial_\mu \Phi$ drop out from the 
perturbative expression for $\widehat A_\mu$; for details, see \cite{Sezgin:2000hr}.
Thus, a modification of  \eq{vgc} to $z^\a\widehat{V}_\a+\bar z^{\ad}\widehat{\bar V}_{\ad}=\widehat \Upsilon$,
where $\widehat\Upsilon$ is a Lorentz singlet with a parturbative expansion starting at the second order, 
leads to non-canonical higher order corrections to the Cartan curvature 
of $W$ of involving $\widehat \Upsilon$ and its twistor space derivatives
at $Z=0$.
Eliminating these terms perturbatively, using the constraint 
on $\Phi$, yields a canonical Cartan curvature for $W$.}.
Assuming furthermore that the Vasiliev frame field
\be e_{\a\ad}:= \left.\frac{\partial^{2}}{\partial y^{\a}\partial \yb^{\ad}}W\right|_{Y=0}\ee
is invertible, the free differential algebra yields i) algebraic
constraints that express $(\Phi,W)$ in terms of a remaining 
set of dynamical fields, namely $e^{\a\ad}$, the scalar\footnote{Under the parity operation $P$, one 
has $P(\phi)=\phi$ in the type A model, and $P(\phi)=-\phi$ in the type B model.}
\be \phi~:=~\Phi|_{Y=0}\ ,\ee
and the Fronsdal tensors ($s\geqslant 1$, $s\neq 2$)
\be \phi_{\mu(s)}~:=~\left.2i e^{\a_1\ad_1}_{(\mu_1} \cdots e^{\a_{s-1}\ad_{s-1}}_{\mu_{s-1}} \frac{\partial^{2s-2}}{\partial y^{\a_1}\cdots\partial y^{\a_{s-1}}\bar\partial \yb^{\ad_1}\cdots \bar\partial \yb^{\ad_{s-1}}} W_{\mu_s)}\right|_{Y=0}\ ,\ee
modulo auxiliary gauge symmetries; ii) 
dynamical metric-like equations of motion; and iii) Bianchi identities; for details, 
see Appendix D in \cite{us}. 

As for the auxiliary fields, of particular interest 
are the pure Weyl tensors $C_{\a(2s)}$ ($s\geqslant 1$) 
that make up the generating function ($s\geqslant 0$)
\be C~:=~\Phi|_{\yb =0}\ ,\qquad C_{\a(2s)}~:=~
\left.\frac{\partial^{2s}}{\partial y^{\a_1}\cdots
\partial y^{\a_{2s}}}C\right|_{y=0}\ .\label{2.60}\ee
The remaining components of $\Phi$ are given by all
possible derivatives of the Weyl tensors that are
non-vanishing on-shell.

The following remarks are in order:

The Vasiliev gauge is required at the linearized level 
in order to obtain a Lorentz covariant description of
free Fronsdal fields on the mass shell.
Beyond this order, we observe that:
i) Eq. \eq{vgc} is manifestly Lorentz covariant and conveniently removes 
all gauge artifacts up to residual $\mhs_1(4)$ gauge transformations on 
${\cal X}$;
ii) It also yields perturbatively well-defined master fields 
in regular sub-classes of $\Omega({\cal T}')$ for generic 
points in ${\cal X}$ as introduced in \cite{Vasiliev:1990vu,Prokushkin:1998bq}
and further developed in \cite{Vasiliev:2015wma};
iii) Provided that the asymptotic spacetime gauge fields defined in normal order
are Fronsdal fields order-by-order in classical perturbation theory,
and that the master fields can be mapped to $\Omega({\cal T})$, 
which ensures finite invariants given by integrals ${\cal B}$
and traces over ${\cal W}$, then any modification of \eq{vgc} 
preserving the topology of $\Omega({\cal T})$ will not affect 
the invariants, and hence lead to a physically equivalent 
description of higher spin dynamics.
In what follows, however, we shall impose the Vasiliev gauge only 
at the leading order pending further consolidation of the above 
approach.

Secondly, Vasiliev's procedure yields Fronsdal field interactions
in the Vasiliev gauge that are highly non-local in the sense that 
the field redefinition \cite{Vasiliev:2016xui} mapping them to the 
quasi-local Fronsdal theory \cite{Bekaert:2014cea,Bekaert:2015tva,Sleight:2016dba} 
is large in the metric-like sense \cite{Taronna:2016xrm}.
Moreover, as found in \cite{Boulanger:2015ova,Skvortsov:2015lja},
this procedure yields classical fields that do \emph{not} lend 
themselves to on-shell amplitude computations using the prescription 
of \cite{GKP,W}, which assumes the existence of a quasi-local action 
principle with canonical self-adjoint metric-like kinetic terms 
in anti-de Sitter spacetime.
In response to these subtleties, criteria for classifying the 
locality properties of interactions, field redefinitions and 
gauge transformations given in twistor space have been proposed in 
\cite{Vasiliev:2015wma}, though it remains to adapt them to
the non-polynomial classes of initial data describing black holes 
and massless particles, and to furthermore verify the proposal at the level of the 
prescription of \cite{GKP,W}.
The appropriate interpretation of the quasi-local
Fronsdal theory may thus be as a quantum effective
theory \cite{Sleight:2017pcz}.

As for the underlying microscopic action for Vasiliev's full 
equations, an alternative proposal formulated directly in terms 
of the master fields in correspondence space, has been given in 
\cite{Boulanger:2011dd,FCS}, whereby the amplitudes 
are to be obtained by evaluating the on-shell action on classical master fields 
(containing asymptotically anti-de Sitter regions) obtained by dressing free 
Fronsdal fields expanded over suitable higher spin representations.
Thus, the findings of the present paper may serve as a starting point
for corresponding amplitude computations, once an appropriate on-shell action
has been found, a problem which is currently under investigation \cite{Boulanger:2011dd,Vasiliev:2015mka,FCS}.

\scs{Solution Method}

In this Section, we give the method based on
gauge functions and separation of variables
in twistor space that we shall use to solve
Vasiliev's equations.
In particular, by going to a convenient gauge,
we shall provide a perturbatively defined solution 
for general zero-form initial data.

\scss{Gauge functions}\label{gaugefmethod}

The basic idea is to build families of exact solutions 
in gauges in which the spacetime connection vanishes.
The latter can then be switched on and the solutions brought
to the Vasiliev gauge by means of large gauge transformations.
To this end, one makes use of the fact that the commutative nature 
of ${\cal X}$ implies that, locally on ${\cal X}$,
the general solution to Eqs. \eq{MC} and \eq{dSa} is given by 
\be \widehat U~=~\widehat g^{-1}\star d\widehat g\ ,\qquad 
\widehat \Phi~=~\widehat g^{-1}\star \widehat \Phi'\star \pi(\widehat g)
\ ,\qquad
\widehat S_{\underline\a}~=~ \widehat g^{-1}\star \widehat S'_{\underline\a}\star \widehat g\ ,\label{Lrot}\ee
where $\widehat g(x,Y,Z)$ is a gauge function and  
\be \partial_\mu\widehat\Phi'=\partial_\mu\widehat S'_{\underline\a}=0\ .\ee
By a choice of coordinates, we may assume that
\be \widehat g|_{x=0}~=~1\ ,\qquad  (\widehat\Phi',\widehat S'_{\underline\a})=(\widehat\Phi,\widehat S_{\underline\a})|_{x=0}\ .\ee
The remaining master field equations read
\bea
&\widehat S'_\a\star\widehat\Phi'+\widehat\Phi'\star\pi(\widehat S'_\a)\ =\ 0\ ,\quad
\widehat {\bar S}{}'_{\ad}\star\widehat\Phi'+\widehat\Phi'\star\bar\pi(\widehat {\bar S}{}'_{\ad})\ =\ 0 &\label{prime1}\\[5pt]
&
[\widehat S'_\a,\widehat S'_\b]_\star\ =\ -2i\e_{\a\b}(1-{\cal B}'\star\widehat \Phi'\star\kappa)\ ,\quad [\widehat {\bar S}{}'_{\ad},\widehat {\bar S}{}'_{\bd}]_\star\ =\ -2i\e_{\ad\bd}(1- \overline{\cal B}'\star \widehat\Phi'\star\bar \kappa)&\label{prime2}\\[5pt]
&
[\widehat S'_\a,\widehat{\bar S}{}'_{\ad}]_\star\ =\ 0\ ,&\label{prime3}\eea
to be solved subject to boundary conditions on $(\widehat\Phi',\widehat S'_{\underline\a})$ on ${\cal T}$.
We shall focus on classical solutions that admit perturbative expansions
\be\widehat \Phi'~=~\sum_{n=1}^\infty \widehat \Phi^{\prime(n)}\ ,\qquad  \widehat S'_{\underline\a}~=~ \sum_{n=0}^\infty \widehat S^{\prime(n)}_{\underline\a}~\equiv~Z_{\underline\a}-2i\sum_{n=0}^\infty \widehat V^{\prime(n)}_{\underline\a} \ ,\label{solclass}\ee
where $(\widehat S^{\prime(n)}_{\underline\a},\widehat \Phi^{\prime(n)})$ 
is an $n$-linear functional in the integration constant
\be \Phi'(Y)~=~\widehat \Phi'(Y,Z)|_{Z=0}\ ,\label{ic}\ee
and $\widehat S^{\prime(0)}_{\underline\a}$ is a flat connection 
obeying
\bea [\widehat S^{\prime(0)}_{\underline\a},\widehat S^{\prime(0)}_{\underline\b}]_\star ~=~ -2iC_{\underline{\a\b}}  \ := \ -2i\left(\ba{cc}\e_{\a\b} & 0 \\ 0& \e_{\ad\bd}\ea\right) \ ,\label{fc}\eea
which admit non-trivial solutions obtained by activating Fock space 
projectors on ${\cal Y}\times {\cal Z}$; see Appendix \ref{App:defosc}
for details.

In order to obtain solutions containing asymptotically anti-de Sitter 
regions with free Fronsdal (gauge) fields, we use a gauge function 
$\widehat G$ subject to the Vasiliev gauge condition, \emph{viz.}
\be Z^{\underline\a} ( (\widehat G)^{-1} \star \widehat S'_{\underline\a} \star \widehat G )\ = \ 0\ ,\label{VGC}\ee
and a vacuum condition
\be \widehat G|_{\Phi'=0}=L\ ,\ee
where $L:{\cal X}\rightarrow \exp \mhs_1(4)$; for the sake
of simplicity, we shall choose
\be L:{\cal X}\rightarrow SO(2,3)/SO(1,3)\ ,\ee
such that $L^{-1}\star dL$ describes anti-de Sitter spacetime.

We recall from Section 3.2 that Eq. \eq{VGC} is necessary 
at the linearized level but optional at higher orders, though 
under extra assumptions on the topology of ${\cal Z}$ and the 
nature of the physical observables of the theory, one may argue 
that it can be imposed to all orders.
We would also like to stress that in order for $\widehat G$ to be of 
physical relevance it has to be a \emph{large gauge tranformation},
which affects the asymptotics of the fields on ${\cal X}\times {\cal Z}$, 
and some observables of the theory.
The space of such transformations consists of equivalence classes 
formed by factoring out the proper (or small) gauge transformations, 
which by their definition do not affect any observable,
but that may nonetheless be useful in order to remove
physically irrelevant coordinate singularities and other 
inconvenient gauge artifacts.

Having obtained $\widehat G$ (as a functional of $L$
and $\Phi'$), the generating functions $C$ 
and $W$ of the pure Weyl tensors, including the physical scalar field, and of the 
Fronsdal gauge fields, including the spin-two frame field, 
defined in Eqs. \eq{2.60} and \eq{2.61}, respectively, are 
given by 
\be C=\left.\left(\widehat G^{-1}\star \widehat \Phi'\star \pi(\widehat G)\right)\right|_{Z=0,\yb=0}\ ,\ee
\be W~=~ \left.\left[\widehat G^{-1}\star d \widehat G- \frac1{4i}\,\left(\o^{\a\b} \left( y_\a \star y_\b+\widehat G^{-1}\star \widehat S'_\a\star\widehat S'_\b\star\widehat G \right) + h.c.\right) \right]\right|_{Z=0}\ ;\label{Wmu}\ee
finally, the condition \eq{shiftgauge} is imposed as to determine 
$(\o_\m^{\a\b},\bar \o_\m^{\ad\bd})$.

We remark that at the classical level, the fields 
$(\widehat\Phi',\widehat S'_{\underline\a})$ and the gauge function, 
which must belong to an associative algebra, may be singular or given 
by distributions on ${\cal T}$, as long as 
$(\widehat W,\widehat \Phi,\widehat S_{\underline\a})$
are real-analytic on ${\cal T}$ for generic points in ${\cal X}$.
At the semi-classical level, stronger conditions arise from demanding that  
$(\widehat W,\widehat \Phi,\widehat S_{\underline\a})$ belong to an 
associative bundle over ${\cal X}$ with well-defined invariants, whose
construction is beyond the scope of this work.

In summary so far, the gauge function method gives rise to solution spaces 
that depend on the following classical moduli (see
also \cite{us} and references therein):  
\begin{itemize}
\item[(i)] the constants entering via the parametrization 
of $\Phi'(Y)$, which describe local degrees of freedom;
\item[(ii)] the degrees of freedom entering via the homogenous 
solutions to the Vasiliev gauge condition modulo proper gauge 
transformations, that is, the space of boundary gauge functions   
$\widehat G|_{\partial{\cal X}\times \{ Z=0\}}\sim L|_{\partial{\cal X}}$;
\item[(iii)] star-product projectors entering via the flat connections 
$\widehat V^{\prime(0)}$ on ${\cal Z}$;
\item[(iv)] windings contained in the transition functions entering via
the atlas for ${\cal X}$.
\end{itemize}
In what follows, we shall activate (i) and (iii).

\scss{Separation of variables in twistor space and holomorphicity}\label{genansatz}

The deformed oscillator problem \eq{prime1}--\eq{prime3} can be solved 
formally by factorizing the dependencies on $Y$ and $Z$, and assuming
that the internal connection $\widehat V'$ is the sum of a holomorphic 
and an anti-holomorphic one-form on ${\cal Z}$ that can be 
expanded perturbatively in the zero-form initial data $\Phi'$
defined in \eq{ansatz1}.
One can show that it is addition possible to assume that the
zero-form is uncorrected, leading to an Ansatz of the form
\bea \wPhi'(Y,Z)& = & \Phi'(Y) \ \equiv \ \Psi(Y)\star \kappa_y \ \equiv \ 
\overline \Psi(Y)\star \bar\kappa_{\bar y} \ , \label{ansatz1}\\[5pt]
\wV'_\a (Y,Z) & = & \sum_{n=0}^\infty(\Psi(Y))^{\star n}\star V^{(n)}_\a(z) 
\ ,\label{ansatz2} \\[5pt]
\widehat{\overline{V}}'_{\ad} (Y,Z) & = & 
\sum_{n=0}^\infty(\overline \Psi(Y))^{\star n}\star \overline{V}^{(n)}_{\ad}(\bar{z}) 
\label{ansatz3}\eea
where $\kappa_y$ and $\bar\kappa_{\bar y}$ are defined in \eq{factorized},
and we have implicitly defined 
\be \Psi \ := \ \Phi'\star \k_y \ ,\qquad \overline \Psi \ := \ \Phi'\star \bar\k_{\bar y}\ .\ee
The reality conditions imply that
\be \overline \Psi=\Psi^\dagger\ ,\qquad (V^{(n)}_\a(z))^\dagger\ =\ -\overline{V}^{(n)}_{\ad}(\bar{z})\ ,\ee
while the bosonic projection gives 
\be \pi\bar\pi(\Psi)= \Psi\ ,\qquad [\Psi,\overline \Psi]_\star=0\ .\label{psipsibar}\ee
As for the internal connection, the bosonic projection 
combined with the holomorphicity implies that
\be \pi(V^{(n)}_\a) \ = \ -V^{(n)}_\a \ , \qquad  
\bar{\pi}({\overline{V}}^{(n)}_{\ad}) \ = \ -
{\overline{V}}^{(n)}_{\ad} \ .\label{piV}\ee
We shall refer to the above Ansatz as the \emph{holomorphic gauge},
noting the residual symmetry under gauge transformations
with holomorphically factorized gauge functions 
\be \widehat N'=\widehat h'(\Psi,z) \star\widehat{\bar{h}}'(\overline\Psi,\bar{z})\ ,\qquad 
\pi(\widehat h')=\widehat h'\ ,\qquad \bar{\pi}(\widehat{\bar{h}}')= \widehat{\bar{h}}'\ ,\qquad 
(\widehat h')^\dagger=\widehat{\bar{h}}'\ .\ee
In this gauge, one has $\widehat F_{\a\ad}=0$ and $\widehat D'_\a \widehat \Phi'=0$,
\emph{idem} for the hermitian conjugate;
the former follows immediately from holomorphicity and \eq{psipsibar}, while 
the latter can be seen by rewriting 
\bea \widehat D'_\a \widehat \Phi'= \sum_{n=0}^\infty\Psi^{\star n+1}\star \{ V^{(n)}_\a,\kappa_z\}_\star \ = \ 0\ ,\eea
and using \eq{piV}.
The remaining constraints on $\widehat F'_{\a\b}$ and $\widehat{\bar F}'_{\ad\bd}$ 
(see Eqs. \eq{prime2}) now read
\bea \partial_{[\a} \wV'_{\b]}+\wV'_{[\a}\star \wV'_{\b]}  & = & -\frac{i}{4}\,\e_{\a\b} 
\,{\cal B}(\Psi\star \Psi) \star \Psi\star\k_z \ ,\label{ZcurvF1}\\[5pt]
 \partial_{[\ad} \widehat{\bar{V}}'_{\bd]}+\widehat{\bar{V}}'_{[\ad}\star \widehat{\bar{V}}'_{\bd]}  
 & = & -\frac{i}{4}\,\e_{\ad\bd} \,\overline{{\cal B}}(\overline\Psi\star 
 \overline\Psi)\star \overline{\Psi}\star\bar{\k}_{\zb}\ ,\label{ZcurvF2}\eea
which are formally equivalent to those solved in \cite{us}
using the extension of the non-perturbative method of \cite{Prokushkin:1998bq}
given in \cite{Sezgin:2005pv,us} as to include non-trivial flat connections 
$\widehat V^{\prime(0)}_{\underline\a}$; for details, see Appendix \ref{App:defosc}.
For the sake of simplicity, we shall henceforth take 
\be \widehat S^{\prime(0)}_{\underline\a}=Z_{\underline\a}\ ,\qquad {\cal B}=b:=e^{i\theta_0}\ ,\ee
stressing that there is no technical obstruction to generalize the results as to include
the full interaction ambiguity and non-trivial vacuum connections on ${\cal Z}$.

Next, we shall choose a particular solution for the internal connection.
To this end, we use a spin-frame $u^\pm_\a$ obeying 
\be u^{\a+}u^-_\a=1\ ,\ee
to define conjugate variables 
\be z^\pm:=u^{\pm\a}z_\a\ ,\qquad [z^-,z^+]_\star = -2i\ ,\ee
that allows us to represent $\kappa_z$, appearing in the
right-hand side of of \eq{prime2}, as the following 
delta function sequence \cite{us}\footnote{Following \cite{us}, letting $\k_z(\e):=\frac{1}{\e}\,e^{-\frac{i}{\e}w_z}$, 
one has 
$$\k_z(\e)\star z_\a=i(\e {\cal D}_\a{}^\b \partial_\b +\partial_\a) \k_z(\e)\ ,\qquad 
z_a\star \k_z(\e)=i(\e {\cal D}_\a{}^\b \partial_\b-\partial_\a) \k_z(\e)\ ,\qquad  
\k_z(\e)\star \k_z(\e)=(1+\e^2)^{-1} \exp\frac{-i\e}{1+\e^2} w_z\ .$$
Assuming that $\check \kappa_z:=\lim_{\e\to 0^+}\k_z(\e)$ exists as a
distribution, it follows that 
$$\check \kappa_z \star z_\a =- z_\a  \star\check \kappa_z\ ,\qquad \check \kappa_z\star\check \kappa_z=1\ ,$$
that is, it obeys the defining property of $\kappa_z$, \emph{viz.} $\check \kappa_z\star f(z)\star \check \kappa_z=\pi(f(z))$.
Thus, setting aside the subtler existence issue, we are led to \eq{deltasequence}.
}:
\be \k_z=\lim_{\e\to 0^+}\frac{1}{\e}\,e^{-\frac{i}{\e}w_z}\ ,\qquad 
w_z:=z^+z^-=\ft12 z^\a z^\b {\cal D}_{\a\b}\ ,\qquad {\cal D}_{\a\b}:=2u^+_{(\a} u^-_{\b)}\ 
.\label{deltasequence}\ee
Fixing the residual holomorphic gauge symmetries, we can construct the  
particular solution
\bea \wV'_{\a} \ = \ 2i \sum_{n=1}^\infty  {1/2 \choose n}\left(-\frac{b}{2}\right)^{n}\int_{-1}^1 \frac{dt}{(t+1)^2} \frac{(\log (1/t^2))^{n-1}}{(n-1)!}\,z_\a e^{i\frac{t-1}{t+1}w_z} \star \Psi^{\star n}\ . \label{wV'} \eea
Indeed, expansion of Eqs. \eq{ZcurvF1}-\eq{ZcurvF2} 
in $\star$-powers of $\Psi$ yields
\bea  & \partial_{[\a} V^{(1)}_{\b]}+\frac{i}{4}\,\e_{\a\b} \,b \,\k_z\ = \ 0 \ ,&\\[5pt]
& \partial_{[\a} V^{(n)}_{\b]} + \frac{1}{2} \sum_{p+q=n}\left[V^{(p)}_{\a}, V^{(q)}_{\b}\right]_\star \ = \ 0 \ ,\quad n\geq 2 &\ ,\eea
which are solved by the coefficient 
of $\Psi^{\star n}$ in \eq{wV'}, \emph{i.e.}
\bea V^{(n)}_{\a} \ = \ 2i {1/2 \choose n}\left(-\frac{b}{2}\right)^{n}\int_{-1}^1 \frac{dt}{(t+1)^2} \frac{(\log (1/t^2))^{n-1}}{(n-1)!}\,z_\a e^{i\frac{t-1}{t+1}w_z} \ .\eea
The above particular solution obeys $z^\alpha V^{(n)}_{\a}=0$,
and is given by a distribution that is not real-analytic at $z^\alpha =0$; thus, the symbol 
of $\widehat V'_\a$ in Weyl order has the same properties.

\paragraph{A twistor-space distribution.} At first order, the 
$z$-dependence of $\wV^{(1)\pm}$ is captured by the distribution
\be I^\pm(z) \ := \ 2z^\pm\int_{-1}^1 \frac{dt}{(t+1)^2}\, e^{i\frac{t-1}{t+1}z^+z^-}  \ = \ 
\frac{1}{iz^\mp}\ ,\label{I}\ee
where we have used $\lim_{\e\rightarrow 0^+} e^{-\frac{i}{\e}z^+z^-}=0$,
which follows from \eq{deltasequence}. 
It enjoys the following properties 
\be i z^\mp I^\pm=1\ ,\qquad \partial_\pm I^\pm=\kappa_z\ ,\label{propI}\ee  
of which the first one clearly follows from  \eq{I}, while
the second one requires more care, as $1/z^\mp$ is not differentiable at $z^\mp=0$.
Thus, in order to differentiate $I$ we must first rewrite it as a differentiable
distribution, for which we use 
\bea \partial_\pm I^\pm&=&
\partial_\pm \left(\int_0^{z^\pm} dz^{\prime \pm}\lim_{\e\to 0^+}\frac{1}{\e}e^{-\frac{i}{\e}z^{\prime \pm}z^\mp}\right) \nn\\[5pt]&=& 
2\pi\partial_\pm\left(\int_0^{z^\pm} dz^{\prime \pm}\d(z^{\prime \pm})\d(z^\mp)\right) \ = \ 2\pi\partial_\pm\theta(z^\pm)\d(z^\mp)\ ,\eea
which yields the second equation in \eq{propI}.

\paragraph{Importance of boundary conditions.} As we shall see, choosing the initial data to 
correspond to massless particles and black holes, it follows
that $\Psi$ belongs to an associative algebra in which $\Psi^{\star n}$ 
can be rigorously defined, and that the internal connection becomes 
real-analytic in ${\cal Z}$ for generic spacetime points 
once the gauge function $L$ is switched 
on and the star products in \eq{wV'} are performed (in normal order).
Thus, the solution space can be brought 
to the Vasiliev gauge, at least at the linearized level, where thus
Fronsdal fields arise in the asymptotic region, as we shall 
examine further below.
%

\scs{Internal Solution Space with Scalar Field and
Black Hole Modes}\label{Sec:new}

In this Section, we examine the exact twistor-space solution
in holomorphic gauge in the case when $\Phi'$ consists of 
operators with distinct eigenvalues under left and right
star-multiplication by the generators of the compact Cartan 
subalgebra of $\mso(2,3)$ \cite{us}, corresponding to linearized 
massless scalar modes and spherically symmetric black hole modes. 
Unlike in a local field theory, black hole modes arise as a 
back-reaction to the scalar particle modes already at second 
order in perturbation theory. 
We would like to stress that the exact solutions in twistor space
can be thought of as solution to the full set of equations over
a special base point in ${\cal X}$ where the gauge function is
equal to one; the inclusion of nontrivial gauge functions will 
be the topic of the next Section.

\scss{Particle and black hole modes}\label{ptandbh}

Extending the approach of \cite{Iazeolla:2008ix,us}, 
we shall take $\Phi'(Y)$ to belong to a reducible 
twisted-adjoint representation space consisting of 
eight unitarizable irreps as follows:
\be \mathfrak{H}^{(+)}\oplus \pi(\mathfrak{H}^{(+)})\oplus \mathfrak{H}^{(-)}\oplus \pi(\mathfrak{H}^{(-)})\ ,\label{spectrum}\ee
where 
\be \mathfrak{H}^{(\pm)}:=\mathfrak{D}^{(\pm)}\oplus\mathfrak{S}^{(\pm)}
\ ,\qquad \mathfrak{S}^{(\pm)}:=\mD^{(\pm)}\star \kappa_y\ ,\ee
and $\mathfrak{D}^{(\pm)}$ consist of operators 
obtained by acting with reflectors\footnote{The reflector 
sends elements in the direct product of two 
supersingletons to operators acting on the supersingleton Hilbert space. 
The twisted-adjoint action of the higher spin algebra on such an operator is therefore reflected  
to the left action of the higher spin algebra on the corresponding state in the tensor product of two 
supersingletons, that decomposes into a sum of massless particle states via the Flato-Fronsdal theorem 
\cite{Flato:1978qz}. 
In particular, we shall use projectors on supersingleton states 
that correspond to rotationally invariant massless scalar particle states \cite{Iazeolla:2008ix}.
As far as parity is concerned, the reflector preserves the
parity of the states in $\mD(1;(0))$, while it reverses the
parity of the states in $\mD(2;(0))$.
In particular, the reflector maps the rotationally invariant states, which are
parity even in $\mD(1;(0))$ and parity odd in $\mD(2;(0))$, to the aforementioned projectors, 
all of which are even under $P$.} \cite{Iazeolla:2008ix} 
on squares of the scalar and spinor singletons 
$\mathfrak{D}(\frac12;(0))$ and $\mathfrak{D}(1;(\frac12))$, 
respectively, \emph{viz.}
\be \mathfrak{D}^{(-)}\stackrel{\rm refl.}{\cong} 
[\mathfrak{D}(\frac12;(0))]^{\otimes 2}\cong  \bigoplus_{s=0,1,2,\dots}\mathfrak{D}(s+1;(s))\ ,\ee
\be \mathfrak{D}^{(+)}\stackrel{\rm refl.}{\cong}  
[\mathfrak{D}(1;(\frac12))]^{\otimes 2}\cong
\mathfrak{D}(2;(0))\oplus \bigoplus_{s=1,2,\dots}\mathfrak{D}(s+1;(s,1))\ .\ee
These are lowest-weight spaces with positive energies sent by
$\pi$ to corresponding highest-weight spaces with negative energies.
The resulting eight spaces in Eq. \eq{spectrum} form separate twisted-adjoint 
$\mathfrak{hs}_1(4)$-orbits. 
The superscript $(\pm)$ refers to the eigenvalue of 
the involution given by the composition of the twisted-adjoint action of 
$(-1)^{E+J'} $ and the parity map $P$, where $J'$ is a spatial angular 
momentum chosen as to make $e_\star^{i\pi J'}\star P_r\star e_\star^{-i\pi J'} = -P(P_r)$, 
\emph{idem} on $M_{0r}$. 
Thus, this involution commutes with the action of the full $\mhs_1(4)$ algebra, and 
as consequence it splits the module \eq{spectrum} into two submodules; for further details, 
see \cite{Iazeolla:2008ix}.

The reference states can be taken
to be the scalar ground states represented by the projectors on 
the scalar and spinor (anti-)singleton ground states, respectively, \emph{viz.}
\be {\cal P}_{\pm 1}(E)\equiv {\cal P}_{\pm 1;(0)}(E)=4\exp(\mp 4E)\cong \ket{\pm\frac12;(0)}\bra{\pm\frac12;(0)}\ ,\ee
\be {\cal P}_{\pm 2}(E)\equiv {\cal P}_{\pm 2;(0)}(E)=\mp 8\exp(\mp 4E)(1\mp 4E)\cong \ket{\pm 1;(\frac12)}^i\,{}_i\bra{\pm 1;(\frac12)}\ ,\ee
where $E$ is the anti-de Sitter energy operator and $i=1,2$
is the doublet index of the spin-$1/2$ representation of $\mso(3)$; 
for further details, see Section \ref{Sec:genproj} and \cite{Iazeolla:2008ix,us}. 
For simplicity, we shall limit the expansion of $\Phi'$ over $\mathfrak{D}^{(\pm)}\oplus \pi(\mathfrak{D}^{(\pm)})$
to the rank-$|n|$ supersingleton projectors ${\cal P}_n(E)$, $n=\pm 1,\pm 2,\dots$,
given by the sum of the projectors onto the (anti-)singleton 
states with energy $n/2$ and spin $|n|-1/2$, \emph{viz.}\footnote{
More precisely, in \cite{us}, the projectors ${\cal P}_n(E)$ 
were referred to as symmetry-enhanced, in order to distinguish 
them from the rank-$1$, biaxially-symmetric projectors $P_{n_1 n_2}(E,J)$, 
depending on $E$ as well as an angular momentum $J$.} 
\be {\cal P}_n(E)\equiv {\cal P}_{n;(0)}(E)\cong 
\ket{\frac{n}2;(\frac{|n|-1}2)}^{i(|n|-1)}{}_{i(|n|-1)}\bra{\frac{n}2;(\frac{|n|-1}2)}\ ,
\qquad n=\pm 1,\pm 2,\dots\ ,\label{calPnsum}\ee
where the notation $i(|n|-1)=i_1 i_2\ldots i_{|n|-1}$ denotes $|n|-1$ symmetrized doublet indices. These projectors obey
\be (E-\frac{n}2)\star {\cal P}_n(E)=0={\cal P}_n(E)\star (E-\frac{n}2)\ ,\qquad [M_{rs},\cP_n(E)]_\star  = 0\ ,\ee
from which it follows that they are rotationally invariant 
and that their twisted-adjoint energy eigenvalues are given 
by $n$, \emph{i.e.}
\be E\star\cP_n(E)-\cP_n(E)\star\pi(E) = \{E,\cP_n(E)\}_\star  = n\cP_n(E) \ .\label{twadj}\ee
Thus, under the twisted-adjoint action \eq{twadj}, $\cP_n(E)$ behave as 
enveloping algebra realizations of rotationally invariant modes of 
an $AdS_4$ massless scalar field \cite{Iazeolla:2008ix}.
In other words, the twisted-adjoint action of $\mso(2,3)$ on the 
$\cP_n(E)$ realizes the (left) $\mso(2,3)$-action on the tensor 
product of two super(anti-)singleton states, that gives rise to specific 
scalar modes, in accordance with the Flato-Fronsdal theorem 
\cite{Flato:1978qz}.
Indeed, via the linearization of the second equation in \eq{Lrot},
this projector provides the initial data in twistor space 
for the \emph{scalar-field mode in spacetime with energy $n$ and 
vanishing spatial angular momentum} found by Breitenlohner and 
Freedman \cite{Breitenlohner:1982jf}, as will be shown below 
in Section \ref{sec:LWSWeyl}. 

Turning to the spaces $\mathfrak{S}^{(\pm)}\oplus \pi(\mathfrak{S}^{(\pm)})$, these are required
for the non-linear completion of the particle states in our gauge, 
as the interactions involve $\star$-multiplications with $\k_y$ 
that map the projectors associated to particle states to 
\emph{twisted projectors} 
\be \widetilde{{\cal P}}_{n}:= {\cal P}_{n}(E)\star\k_y \cong 
\ket{\frac{n}2;(\frac{|n|-1}2)}^{i(|n|-1)}{}_{i(|n|-1)}\bra{-\frac{n}2;(\frac{|n|-1}2)}\ ,
\qquad n=\pm1,\pm2,\dots\ ,\ee
by transforming singleton states into anti-singleton states with opposite energy,
and vice-versa\footnote{As a consequence of the action of parity on the oscillators \eq{Posc}, and of the third of Eqs. \eq{Pkkb}, the twisted projectors $\tcP_n$ are alternatively odd or even under $P$, $P(\tcP_n) = (-1)^n\tcP_n$.}.
These elements are the initial data for \emph{static}, \emph{spherically symmetric}
higher spin generalizations of Schwarzschild black holes \cite{Didenko:2009td,us}. 
Indeed, following the same reasoning as above, one can show that 
\be E\star\tcP_n-\tcP_n\star\pi(E) = [E,\cP_n(E)]_\star \star \k_y  = 0\ ,\ee
implying that the twisted projectors give rise to static solutions.
Concerning the more detailed nature of the linearized
fluctation fields, we recall that while the lowest-weight space modes 
give rise to fields that are regular everywhere in spacetime, the black hole 
modes exhibit the typical singularity at $r=0$, as dictated by 
the spherical symmetry and the generalized Petrov Type D structure 
in each spin-$s$ Weyl zero-form component (see Appendix \ref{App:conv} for
a quick review of Petrov's classification).

Thus, starting from a linearized particle mode, the 
star-product interactions generate black hole modes 
in $(\wV'_\a, \widehat{\bar V}_{\ad}')$ already at 
second order, and hence in the spacetime gauge fields
once the gauge function is switched on.
We interpret the perturbative mixing of particle and black hole modes 
as a manifestation of the non-locality of the interactions in the 
Vasiliev equations.
This raises the issue of whether one can fine-tune $\Phi'$ and the
gauge function $L$ to create a branch of moduli space that is
smooth at the origin of spacetime, which could correspond
holographically, via suitable generating functions, to 
three-dimensional conformal field theories at zero temprerature; 
we shall comment on this open problem towards the conclusions.

\paragraph{Remark on more general Killing vectors.} 
We recall that, as studied in \cite{us,us2}, the solution space
based on $\mathfrak D^{(\pm)}\equiv \mathfrak D^{(\pm)}(E;J)$ 
admits a generalization to six different inequivalent solution 
spaces $\mathfrak D^{(\pm)}(K;\widetilde K)$, with different physical 
meaning, labelled by pairs of mutually commuting and normalized 
elements $K$ and $\widetilde K$ of the Cartan subalgebra of $\msp(4;\Comp)$ taken from the
set $\{ E, J, iB, iP \}$, where $J$ is an angular momentum, 
$B$ a (hermitian) boost and $P$ a (hermitian) translation.
Thus, restricting to symmetry-enhanced solutions, the construction 
sketched above can be repeated for the more general projectors
\be {\cal P}_n(K)\ ,\qquad K =E, J, iB, iP\ .\ee
The corresponding twisted-adjoint $\mhs(4)$ orbits give rise
to linearized solutions spaces given by expansions over 
lowest-weight spaces that in general not unitary. 
In the cases that $K=E,iP$, their fully non-linear completions,
which are soliton-like states, contain twisted projectors 
(that necessarily appear at higher orders).
In the cases that $K=J,iB$, however, the ${\cal P}_n(K)$ are eigenstates 
of one-sided star multiplication by $\k_y$, and hence the full
solutions can be given by expansions over the ${\cal P}_n(K)$ alone.

\scss{Regular presentation of generalized projectors}\label{Sec:genproj}

Next, as anticipated, we shall expand the initial datum $\Phi'$, and therefore 
$\Psi$ and $\wV'_{\a}$, in terms of projectors and twisted projectors,
which we shall refer to collectively as \emph{generalized projectors}.
To this end, let us start by recalling that the 
rotationally-invariant supersingleton projector ${\cal P}_n(E)$
given in \eq{calPnsum} can be realized as \cite{us} ($n=\pm 1, \pm 2, ...$)
\bea {\cal P}_{n}(E) &=&  4(-1)^{n-\ft{1+\ve}2} \,e^{-4E}L^{(1)}_{n-1}(8E) \ = \  
2(-1)^{n-\ft{1+\ve}2}\,\oint_{C(\ve)} \frac{d\eta}{2\pi i}\,\left(\frac{\eta+1}{\eta-1}\right)^{n}\,e^{-4\eta E}\ ,\label{projint}\eea
where $L^{(1)}_{n}(x)$ are generalized Laguerre polynomials 
and the contour integral is performed around a small contour $C(\ve)$  
encircling $\ve:=n/|n|$.  
As for the twisted projectors, the corresponding presentation is given by
\bea  \tcP_n \ := \ \cP_n(E)\star\k_y \ = \ 4\pi (-)^{n-\ft{1+\ve}2}\,\oint_{C(\ve)} \frac{d\eta}{2\pi i}\,\left(\frac{\eta+1}{\eta-1}\right)^{n}\,\d^2(y-i\eta \s_0\yb)\ .\label{tPE} \eea
The twisted projectors are not idempotent, but rather the generalized projectors 
form a subalgebra of the $\star$-product algebra containing an ideal spanned by the projectors, \emph{viz.}
\bea \cP_n\star \cP_m & = & \d_{nm}\cP_n \ , \label{genproj1}\\[5pt]
\cP_n\star \tcP_m & = & \d_{nm}\tcP_n \ , \label{genproj2}\\[5pt]
\tcP_n\star \cP_m & = & \d_{n,-m}\tcP_n \ , \label{genproj3}\\[5pt]
 \tcP_n\star \tcP_m & = & \d_{n,-m}\cP_n \ ,\label{genproj4}\eea
where we have used the property \eq{kyprop} and performed the 
star products before evaluating the contour integrals.
In this way, we achieve orthonormality between the $\cP_n$ 
with positive and negative $n$ in Eq. \eq{genproj1}, from which 
Eqs. \eq{genproj2}--\eq{genproj4} follow by associativity\footnote{\label{footnote2}
Performing the auxiliary contour integrals first, on the other hand, 
one encounters star products between two $\delta^2(y\mp i\sigma_0 \yb)$ which are divergent as $y_\a\mp i(\sigma_0)_\a{}^{\bd} \yb_{\bd}$
are abelian.
We shall comment on alternative perturbative schemes towards the conclusions.}; 
see Appendix F in \cite{us} for a proof and a precise prescription for the contour.
For this reason, we shall refer to Eqs.\eq{projint} and \eq{tPE} as the
\emph{regular presentations} of the generalized projectors.

As the nonlinear corrections are given by star-powers of the adjoint elements $\Psi$ and $\bar{\Psi}$ (see for example Eq. \eq{wV'}), we note that the fact that the product of two twisted projectors
is an ordinary projector manifests the fact that free scalar modes alone do not solve
the full equations, but rather their completion into full solutions 
includes spherically symmetric higher spin black hole modes. 
It is also interesting to note that the relation between free massless scalar 
and higher spin black hole modes is just a Fourier transform 
with respect to the $y$ variables, since that is what the star-multiplication 
with $\k_y$ amounts to \cite{Didenko:2009td}. 

\scss{Explicit form of internal solution in holomorphic gauge}

With the purpose of generalizing the black-hole-like solutions found in 
\cite{us} to solutions superposing particle and black hole modes, 
let us examine in more detailed the expansion of the initial datum 
$\Phi'$ over projectors and twisted projectors, for which we shall use
the following notation:
\bea \Phi' \ = \ \sum_{n=\pm1, \pm2,...}\left(\n_n \cP_n +\tn_n\tcP_n \right)\star\k_y \ = \  \sum_{n=\pm1, \pm2,...}\left(\n_n \tcP_n +\tn_n\cP_n \right)\ ,\label{Phiprproj}\eea
i.e.,
\bea \Psi \ = \ \sum_{n=\pm1, \pm2,...}\left(\n_n \cP_n +\tn_n\tcP_n \right) \ ,\label{F}\eea
where $\nu_n$ and $\tn_n$ are constant deformation parameters, the ${\cal P}_n$ and the $\tcP_n$ absorb 
all the $Y$-dependence and obey
\be  \pi\bar\pi({\cal P}_n)~=~{\cal P}_n~=~({\cal P}_n)^\dagger\ ,\qquad 
\pi({\cal P}_n)~=~{\cal P}_{-n}\ ,\qquad \k_y\bar{\k}_{\yb}\star \cP_n \ = \ (-1)^n\cP_n\ .\label{Pkkb}\ee
As a consequence, the reality conditions on $\Phi$ \eq{reality} impose the following 
restrictions on the deformation parameters:
\bea & \n^\ast_n \ = \ (-1)^n\n_n \ , \qquad \Rightarrow \qquad \n_n \ = \ i^n\m_n \ , \quad \m_n\,\in\,\Real& \\[5pt]\label{realonparameters1}
& \tn^\ast_n \ = \ \tn_{-n}  \ . & \label{realonparameters2} \eea
Thus, from \eq{realonparameters2} it follows that the sum over the particle states 
\eq{Phiprproj} must run symmetrically around $n=0$, such that every particle mode 
is accompanied by its counterpart with opposite energy. 

It is now possible to evaluate the product $\Psi^{\star k}$ that appears 
in $\wV'_\a$ (see Eq. \eq{wV'}) explicitly in terms of the deformation parameters. 
To this end, we use the generalized projector algebra \eq{genproj1}-\eq{genproj4}
to write 
\bea \Psi^{\star k} \ = \ \sum_{n=\pm1, \pm2,...}\left(\n^{(k)}_n \cP_n +\tn^{(k)}_n\tcP_n \right)  \ ,\label{Fk}\eea
where $\n^{(k)}_n$ and $\tn^{(k)}_n$ are homogeneous polynomials built out of $\n_n$ and $\tn_n$, 
defined by 
\be \n^{(1)}_n= \nu_n\ ,\qquad \tn^{(1)}_n=\tn_n\ ,\ee
and, for $k\geq 2$, by the recursive relations
\bea \n^{(k)}_n & = & \nu_n \n^{(k-1)}_n + \tn_{-n} \tn^{(k-1)}_n \ , \label{recursive1}\\[5pt]
\tn^{(k)}_n & = & \tn_n \n^{(k-1)}_n + \n_{-n} \tn^{(k-1)}_n\ .\label{recursive2}\eea
The solution to these relations is given by 
\bea \n^{(k)}_n & = & \frac{1}{2^{k+1}\C_n}\left[(\D\n_n+\C_n)(\n_n+\n_{-n}+\C_n)^k-(\D\n_n-\C_n)(\n_n+\n_{-n}-\C_n)^k\right]\ , \label{nuk}\\[5pt]
\tn^{(k)}_n & = & \frac{\tn_n}{2^{k}\C_n}\left[(\n_n+\n_{-n}+\C_n)^k-(\n_n+\n_{-n}-\C_n)^k\right]\ , \label{tnuk} \eea
where $\D\n_n:=\n_n-\n_{-n}$ and $\C_n := \sqrt{(\D\n_n)^2+4|\tn_n|^2}$.
One has 
\bea \n^{(k)}_n -\n^{(k)}_{-n}  \ = \ \frac{\tn^{(k)}_n }{\tn_n }\,\D\n_n \ .\label{Delnuk}\eea
We note that the zeroes in the denominator $\Gamma_n$ are
cancelled by zeroes in the numerators.
Moreover, for every $n$ such that $\tn_n=0$ it follows 
from \eq{recursive2} that $\tn^{(k)}_n=0$, \emph{i.e.}, the problem 
reduces to the one already solved in \cite{us, us2}, with 
$\n^{(k)}_n=\n_n^k$, $\tV_{n,\a}=0$, and the particle sector disappears.

Inserting \eq{Fk} into \eq{wV'}, the internal connction takes the form
\bea \wV'_\a \ = \ \sum_{n=\pm 1, \pm 2,...} \left(V_{n,\a}\star\cP_n + \tV_{n,\a}\star\tcP_n\right) \ , \label{wV'proj}\eea
and analogously for $\widehat{\bar V}_{\ad}$.
Thus, the coefficients $V_{n,\a}$ and $\tV_{n,\a}$ are given by
\bea V_{n,\a} & = & 2iz_\a\sum_{k=1}^\infty  {1/2 \choose k}\left(-\frac{b}{2}\right)^{k} \n^{(k)}_n \int_{-1}^1 \frac{dt}{(t+1)^2} \frac{(\log (1/t^2))^{k-1}}{(k-1)!}\,e^{i\frac{t-1}{t+1}w_z}  \ ,\label{Vna}\\[5pt]
\tV_{n,\a} & = & 2iz_\a\sum_{k=1}^\infty  {1/2 \choose k}\left(-\frac{b}{2}\right)^{k} \tn^{(k)}_n \int_{-1}^1 \frac{dt}{(t+1)^2} \frac{(\log (1/t^2))^{k-1}}{(k-1)!}\,e^{i\frac{t-1}{t+1}w_z}  \ ,\label{tVna} \eea
with $\n^{(k)}_n$ and $\tn^{(k)}_n$ given by \eq{nuk} and \eq{tnuk}. 
We note that the relation \eq{Delnuk} among the deformation parameters translates into
\bea V_{n,\a}-V_{-n,\a} \ = \ \frac{\D\n_n}{\tn_n }\,\tV_{n,\a} \ . \eea
This completes the solution in twistor space, \emph{i.e.}, in holomorphic gauge.

\paragraph{Alternative solution method.} Another way of obtaining the internal connection, 
which is closer to the procedure followed in \cite{us}, is to first expand
$\Phi'$ and $(\wV'_\a,\widehat{\bar V}_{\ad}')$ in generalized projectors, \emph{viz.}
\be \Phi' \ = \ \sum_{n =\pm1, \pm2, ...} \left(\n_n \tcP_n+\tn_n \cP_n\right) \ ,\ee
\be \wV'_\a \ = \ \sum_{n=\pm 1, \pm 2,...} \left(V_{n,\a}(Z)\star\cP_n + \tV_{n,\a}(Z)\star\tcP_n\right) \ , \ee
\be\widehat{\bar{V}}'_\a \ = \ \sum_{n=\pm 1, \pm 2,...} \left(\bar{V}_{n,\ad}(Z)\star\cP_n + \bar{\tV}_{n,\a}(Z)\star\tcP_n\right) \ee
and then insert these expansions into Eqs. \eq{prime1}-\eq{prime3}. 
Using the generalized projector algebra \eq{genproj1}-\eq{genproj4}, the
coefficients of $\cP_n$ and $\tcP_n$ in Eq. \eq{prime1} yields
the two conditions
\bea & \n_n \left(V_{n,\a}+\pi(V_{n,\a})\right)+\tn_{-n}\tV_{n,\a}+\tn_n\pi(\tV_{-n,\a}) \ = \ 0 \ ,& \\[5pt]
& \tn_n \left(V_{n,\a}+\pi(V_{-n,\a})\right)+\n_{-n}\tV_{n,\a}+\n_n\pi(\tV_{n,\a}) \ = \ 0 \ ,&\eea
respectively. 
They are solved by
\bea &\pi(V_{n,\a}) \ = \ -V_{n,\a} \ , \qquad \pi(\tV_{n,\a}) \ = \ -\tV_{n,\a} \ , &  \\[5pt]
& V_{n,\a}-V_{-n,\a} \ = \ \frac{\D\n_n}{\tn_n }\,\tV_{n,\a} \ ,&\label{V-n}\\[5pt]
&\tV_{-n,\a} \ = \ e^{-2i\varphi_n}\tV_{n,\a}\ ,&\label{tV-n}\eea
provided $\tn_n\neq 0$, and where, recalling the reality conditions \eq{realonparameters2}, 
we have defined the phase factor
\be e^{2i\varphi_n}~:=~\frac{\widetilde \nu_n}{(\widetilde \nu_n)^\ast}\ .\ee
Identical considerations hold for $\widehat{\bar V}_{\ad}$. 
Next, Eq. \eq{prime3} gives rise to the conditions
\bea & \partial_{\a}\bar{V}_{n,\ad} -\bar{\partial}_{\ad}V_{n,\a}+ [V_{n,\a}, \bar{V}_{n,\ad}]_\star+ [\tV_{n,\a}, \bar{\tV}_{-n,\ad}]_\star \ = \ 0  \ ,& \label{defVo}\\[5pt]
& \partial_{\a}\bar{\tV}_{n,\ad} -\bar{\partial}_{\ad}\tV_{n,\a}+ [V_{n,\a}, \bar{\tV}_{n,\ad}]_\star+ [\tV_{n,\a}, \bar{V}_{-n,\ad}]_\star \ = \ 0  \ .&   \label{defVto}\eea
Solving perturbatively in powers of $\n_n,\tn_n$, it is possible to use the
gauge freedom to set to zero any non-holomorphic terms in $V_{n,\a},\tV_{n,\a}$, 
leading to the holomorphic Ansatz satisfying
\bea & \partial_{\a}\bar{V}_{n,\ad} \ = \ 0 \ = \ \bar{\partial}_{\ad}V_{n,\a} \ , &\\
& \partial_{\a}\bar{\tV}_{n,\ad} \ = \ 0 \ = \ \bar{\partial}_{\ad}\tV_{n,\a} \ . &\eea
Finally, inserting the expansions into Eq. \eq{prime2}, which is equivalent to \eq{ZcurvF1} and \eq{ZcurvF2}, 
and requiring the coefficients of $\cP_n$ and $\tcP_n$ to vanish, one obtains
\bea & \partial_{[\a}V_{n,\b]} +V_{n,[\a}\star V_{n,\b]}+ \tV_{n,[\a}\star \tV_{-n,\b]} \ = \ -\frac{i}{4}\e_{\a\b}\,b \,\n_n\k_z  & \label{mixed1}\\[5pt]
& \partial_{[\a}\tV_{n,\b]} +\tV_{n,[\a}\star V_{-n,\b]}+ V_{n,[\a}\star \tV_{n,\b]} \ = \ -\frac{i}{4}\e_{\a\b}\,b \,\tn_n\k_z  & \ ,\label{mixed2}\eea
which can be rewritten using the relations \eq{V-n} and \eq{tV-n} as
\bea & \partial_{[\a}V_{n,\b]} +V_{n,[\a}\star V_{n,\b]}+ e^{-2i\varphi_n} \tV_{n,[\a}\star \tV_{n,\b]} \ = \ -\frac{i}{4}\e_{\a\b}\,b \,\n_n\k_z  \ ,& \label{defV}\\[5pt]
& \partial_{[\a}\tV_{n,\b]} +\tV_{n,[\a}\star V_{n,\b]}- \frac{\D\n_n}{\tn_n }\,\tV_{n,[\a}\star \tV_{n,\b]} + V_{n,[\a}\star \tV_{n,\b]} \ = \ -\frac{i}{4}\e_{\a\b}\,b \,\tn_n\k_z  \ .&   \label{defVt}\eea
It follows that the black hole modes of $V_{n,\a}$ form a closed subsector, 
whereas the particle modes do not. 
Indeed, setting $\tn_n=0$ reduces the deformed oscillator 
problem \eq{mixed1}-\eq{mixed2} to the one already solved 
in \cite{us,us2}, with $\tV_{n,\a}=0$ and $\nu^{(k)}_n=\n^k_n$;
as for the fact that black hole modes obey both the linear and the non-linear 
equations, see \cite{Didenko:2009td} and references therein.
Setting $\nu_n=0$, on the other hand, the black hole sector 
resurfaces through the nonlinearities: even though the first 
equation has no curvature deformation term on the right-hande 
side, the quadratic terms in $\tV_{n,\a}$ nonetheless provide 
a source for $V_{n,\a}$ as soon as one goes beyond the linearized 
approximation.
Indeed, setting $\nu_n=0$ does not imply that $\nu_n^{(k)}$ vanishes for $k>1$,
as we shall examine in detail further below.
Thus, introducing a particle mode necessarily turns on the 
black hole sector in the $Z$-space connection $\wV'_\a$.

We proceed by defining the normal modes
\be \Sigma^{[\pm]}_{n,\a}~:=~z_\a-i\left(V_{n,\a}+V_{-n,\a}\pm 
\frac{\C_n}{\tn_n}\,\widetilde V_{n,\a}\right)\ , \label{SigmaV}\ee
in terms which one obtains a set of two decoupled deformed-oscillator equations for every $n$ as follows:
\be \left[\Sigma^{[\pm]}_{n,\a},\Sigma^{[\pm]}_{n,\b}\right]_{\star}~=~-2i\e_{\a\b}\left(1- b M_n^{[\pm]}\kappa_z\right)\ ,\ee\be\pi(\Sigma^{[\pm]}_{n,\a})~=~-\Sigma^{[\pm]}_{n,\a}\ ,\quad \partial_{\ad}\Sigma^{[\pm]}_{n,\a}~=~0\ ,\ee
with complex deformation parameters
\be M_n^{[\pm]} \ := \ \frac{1}{2}\left(\n_n + \n_{-n} \pm \C_n\right)\ .\ee
These equations can be solved using the refined version of the method
found in \cite{Prokushkin:1998bq} given in \cite{Sezgin:2005pv,Iazeolla:2007wt,us}.
The solution reads
\be \Sigma_{n,\a}^{[\pm]} ~=~ 4z_{\a}\int_{-1}^1 \frac{dt}{(t+1)^2}\,f^{[\pm]}_n(t)  \,e^{ i\ft{t-1}{t+1} w_z}\ ,\label{WSigmaansatz}\ee
where $f^{n[\pm]}(t) $ obeys the integral equations
\be  (f^{[\pm]}_n\circ f^{[\pm]}_n)(t) ~=~ \d(t-1)-\frac{b M^{[\pm]}_{n}}{2} \ ,\label{ringeq}\ee
where
\bea  (h_1 \circ h_2)(u) ~:=~ \int_{-1}^1 dt\int_{-1}^1 dt'\,h_1(t)\,h_2(t')\,\d(tt'-u)\ .\label{ringo}\eea
Using the above integral representation, the commutator of two deformed 
oscillators \eq{WSigmaansatz} reproduces the singular source term using
the limit representation \eq{deltasequence} with parameter $\epsilon$
identified as $1+t$.
For each $n$, the $\circ$-product problem is solved by
\be f^{[\pm]}(t)~=~\delta(t-1)+j^{[\pm]}(t)\ ,\label{j2}\ee
\be j^{[\pm]}(t)~=~q^{[\pm]}(t)+\sum_{k=0}^\infty \l_{\s,k} p_k(t)  \ ,\ee
\be q^{[\pm]}(t)~=~-\frac{b M^{[\pm]}_{n}}{4}{}_1F_1\left[\frac{1}{2};2;\frac{bM^{[\pm]}_{n}}{2}\log(1/t^{2})\right] \ = \ \sum_{k=1}^\infty{1/2 \choose k}\left(-\frac{bM^{[\pm]}_n}{2}\right)^k\frac{[\log(1/t^2)]^{k-1}}{(k-1)!}\ ,\label{j}\ee
where $p_k(t):=\ft{(-1)^k}{k!}\,\d^{(k)}(t)$ act as projectors 
in the $\circ$-product algebra; fur further details as well as 
the explicit expression for the coefficients $\l_k$, see Appendix \ref{App:defosc}.  
Choosing $\l_k=0$, it follows from \eq{SigmaV} combined with \eq{V-n} that 
\bea V_{n,\a} & = & i z_\a\,\sum_{k=1}^\infty{1/2 \choose k}\left(-\frac{b}{2}\right)^k\int_{-1}^1\frac{dt}{(t+1)^2}\,\frac{[\log(1/t^2)]^{k-1}}{(k-1)!}\,\,e^{ i\ft{t-1}{t+1} w_z} \nn\\[5pt]
&& \hspace{3cm}\times \ \ \left[(M^{[+]}_n)^k\left(1+\frac{\D\n_n}{\C_n}\right)+(M^{[-]}_n)^k\left(1-\frac{\D\n_n}{\C_n}\right)\right]\\[5pt]
&= & 2iz_\a\sum_{k=1}^\infty  {1/2 \choose k}\left(-\frac{b}{2}\right)^{k} \n^{(k)}_n \int_{-1}^1 \frac{dt}{(t+1)^2} \frac{(\log (1/t^2))^{k-1}}{(k-1)!}\,e^{i\frac{t-1}{t+1}w_z}  \ ,\label{Vna2}
\eea
and
\bea \widetilde V_{n,\a}& = & 2 i z_\a\,\frac{\tn_n}{\C_n}\sum_{k=1}^\infty{1/2 \choose k}\left(-\frac{b}{2}\right)^k\int_{-1}^1\frac{dt}{(t+1)^2}\,\frac{[\log(1/t^2)]^{k-1}}{(k-1)!}\,\,e^{ i\ft{t-1}{t+1} w_z} \nn\\[5pt]
&& \hspace{3cm}\times \ \ \left[(M^{[+]}_n)^k-(M^{[-]}_n)^k\right]\\[5pt]
 & = & 2iz_\a\sum_{k=1}^\infty  {1/2 \choose k}\left(-\frac{b}{2}\right)^{k} \tn^{(k)}_n \int_{-1}^1 \frac{dt}{(t+1)^2} \frac{(\log (1/t^2))^{k-1}}{(k-1)!}\,e^{i\frac{t-1}{t+1}w_z}  \ , \label{tVna2}\eea
which coincides with the result that we had previously found in Eqs. \eq{Vna}-\eq{tVna}.
We remark that the reduced deformed oscillators 
$(\Sigma_\a^{n[\pm]},\bar{\Sigma}_{\ad}^{n[\pm]})$ are real 
analytic away from $Z=0$ \cite{us}, where singularities in
${\cal Z}$ arise from the singularities of the integrand at $t=-1$; see further comments below \eq{wV'}.
%


\scss{Black-hole backreaction from scalar particle modes}\label{Sec:particles}


In what follows, we give explicitly the solution in 
holomorphic gauge in the case in which  $\n_n=0 , \,\,\forall n$,
i.e., in which the Weyl zero-form contains only scalar modes.
In this case formulas simplify and the backreaction mechanism 
producing black hole modes at higher orders becomes more transparent
The resulting Ansatz for the internal master fields reads
\be \widehat \Phi'~=~\sum_{n\neq 0} \widetilde \nu_{n}{\cal P}_n \ ,\qquad \widehat V'_{\alpha}~=~\sum_{n\neq 0} {\cal P}_n\star\left( V_{n,\alpha}+\kappa_y\star \widetilde V_{n,\a}\right)\ ,\ee
\emph{idem} $\widehat{\bar{V}}'_{\ad}$. 
As already noted, even if the zero-form is expanded over only
projectors (and no twisted projectors), the internal connection
must be expanded over both projectors and twisted projectors.
At the linearized level, the presence of $\k_y$ on the 
right-hand side of \eq{prime2} requires the linearized internal 
connection to contain only twisted projectors, but 
the higher order corrections to the commutator on the left-hand side of \eq{prime2}
necessarily produce both types of projectors, as is 
evident from Eqs. \eq{genproj1}-\eq{genproj4}). 
More precisely, even- and odd-order 
corrections to the internal connection are expanded
over untwisted and twisted projectors, respectively.

Proceeding, it follows from $\widehat D'_\a \widehat \Phi'=0$ that
\be\widetilde \nu_{n}\pi(\widetilde V_{-n,\a})+\widetilde \nu_{-n}\widetilde V_{n,\a}~=~0\ ,\ee
\be\widetilde \nu_{n}\left(\pi(V_{-n,\a})+ V_{n,\a}\right)~=~0\ ,\ee
with the solution
\be\pi(V_{n,\a})~=~-V_{n,\a}\ ,\qquad \pi(\widetilde V_{n,\a})~=~-\widetilde V_{n,\a}\ ,\ee
\be V_{-n,\a}~=~V_{n,\a}\ ,\qquad \widetilde V_{-n,\a}~=~e^{-2i\varphi_n}\widetilde V_{n,\a}\ ,\label{minusn}\ee
where the phase factor $e^{2i\varphi_n}=\widetilde \nu_n/(\widetilde \nu_n)^\ast$.
Moreover, the condition $\widehat F'_{\a\ad}~=~0$ holds identically, 
provided that the internal connection is holomorphic, \emph{viz.}
\be \partial_{\ad} V_{n,\a}~=~0\ ,\qquad \partial_{\ad}\widetilde V_{n,\a}~=~0\ .\ee
The deformation term in Eq. \eq{ZcurvF1} now only has components along the twisted projectors, \emph{i.e.}, Eqs. \eq{defV} and \eq{defVt} reduce to
\be \partial_{[\a}V_{n,\b]}+V_{n,[\a}\star V_{n,\b]}+\widetilde V_{n,[\a}\star \widetilde V_{-n,\b]}~=~0\ ,\ee
\be \partial_{[\a}\widetilde V_{n,\b]}+V_{n,[\a}\star \widetilde V_{n,\b]}+\widetilde V_{n,[\a}\star V_{-n,\b]}~=~ -\frac{ib}4 \e_{\a\b}\widetilde\nu_n \kappa_z\ ,\ee
where, as already noted, quadratic terms in $\tV_{n,\a}$ source the $V_{n,\a}$.
Defining the reduced deformed oscillators
\be \Sigma^{[\pm]}_{n,\a}~:=~z_\a-2i\left(V_{n,\a}\pm e^{-i\varphi_n}\widetilde V_{n,\a}\right)\ , \label{sigmaV}\ee
one arrives at the decoupled deformed-oscillator equations for every $n$, \emph{viz.}
\be \left[\Sigma^{[\pm]}_{n,\a},\Sigma^{[\pm]}_{n,\b}\right]_{\star}~=~-2i\e_{\a\b}\left(1\mp b |\widetilde\nu_n|\kappa_z\right)\ ,\ee\be\pi(\Sigma^{[\pm]}_{n,\a})~=~-\Sigma^{[\pm]}_{n,\a}\ ,\quad \partial_{\ad}\Sigma^{[\pm]}_{n,\a}~=~0\ ,\ee
where the deformation parameter indeed corresponds to $M^{[\pm]}_n$ evaluated at $\n_n=0$. 
Solving these equations using the method based on integral
representations spelled out above, and taking $\l_k=0$, 
one obtains 
\bea V_{n,\a} & = & z_\a\,\frac{ib |\tn_n|}{2}\,\int_{-1}^1\frac{dt}{(t+1)^2}\,F^-\left[\frac{b|\tn_n|}{2}\log t^{2}\right] \,e^{ i\ft{t-1}{t+1} w_z} \ ,\label{Vprimen}\\[5pt]
\widetilde V_{n,\a} & = & - z_\a\,\frac{ib \tn_n}{2}\,\int_{-1}^1\frac{dt}{(t+1)^2}\,F^+\left[\frac{b|\tn_n|}{2}\log t^{2}\right] \,e^{ i\ft{t-1}{t+1} w_z} \ , \label{tVprimen}\eea
were we have defined the even and odd combinations
\be F^\pm [x] \ := \ \frac{1}{2}\left({}_1F_1\left[\frac{1}{2};2;x\right]\ \pm {}_1F_1\left[\frac{1}{2};2; -x\right]\ \right) \ . \ee
One can indeed see, as expected, that the black hole sector $V_{n,\a}$ is 
non-vanishing beyond the leading order; it contains contributions that are of
positive even orders in the deformation parameter $\tn_n$.


\scss{Embedding into an associative algebra}\label{Sec:closure}


Let us demonstrate that the enveloping algebra of
the internal deformed oscillator algebra forms
a subalgebra of an associative algebra with well-defined
star product\footnote{While the proof is more straightforward with the solutions cast in factorized form, we would like 
to stress that the latter is not necessary in any way, and every result, in particular the finiteness of star 
products among master fields in our solution space, can be proved 
equally well in factorized and in fully normal-ordered form.}$^,$\footnote{We remark that the fact that the connection on a non-commutative 
symplectic geometry can be shifted to a deformed oscillator
that belong to an adjoint section is important for constructing 
open Wilson lines, which are gauge invariant observables that 
have no analog in the commuting case.}.

To this end, we first summarize the internal master fields
as follows:
\bea \Phi' & = & \sum_{n=\pm1, \pm2,...}\left(\n_n \cP_n +\tn_n\tcP_n \right)\star\k_y\ , \\[5pt]
\wS'_{\a} & = & z_\a-2i\sum_{n=\pm1,\pm2,...} \left(V_{n,\a}\star\cP_n + \tV_{n,\a}\star\tcP_n\right)  \ ,\label{wSprime}\eea
with
\bea {\cal P}_{n}(E) & = &  2(-1)^{n-\ft{1+\ve}2}\,\oint_{C(\ve)} \frac{d\eta}{2\pi i}\,\left(\frac{\eta+1}{\eta-1}\right)^{n}\,e^{-4\eta E} \ ,\label{sumproj}\\[5pt]
\tcP_n(E) & := & \cP_n(E)\star\k_y \ = \ 4\pi (-)^{n-\ft{1+\ve}2}\,\oint_{C(\ve)} \frac{d\eta}{2\pi i}\,\left(\frac{\eta+1}{\eta-1}\right)^{n}\,\d^2(y-i\eta \s_0\yb) \ ,\label{sumtwproj}\eea
and
\bea V_{n,\a} & = & 2iz_\a\sum_{k=1}^\infty  {1/2 \choose k}\left(-\frac{b}{2}\right)^{k} \n^{(k)}_n \int_{-1}^1 \frac{dt}{(t+1)^2} \frac{(\log (1/t^2))^{k-1}}{(k-1)!}\,e^{i\frac{t-1}{t+1}w_z}  \ ,\label{Vna3}\\[5pt]
\tV_{n,\a} & = & 2iz_\a\sum_{k=1}^\infty  {1/2 \choose k}\left(-\frac{b}{2}\right)^{k} \tn^{(k)}_n \int_{-1}^1 \frac{dt}{(t+1)^2} \frac{(\log (1/t^2))^{k-1}}{(k-1)!}\,e^{i\frac{t-1}{t+1}w_z}  \ , \label{tVna3}\eea
where the deformation coefficients $\n^{(k)}_n$ and $\tn^{(k)}_n$ are given in \eq{nuk}-\eq{tnuk}. 
From the fact that the generalized projectors form a separate star product algebra,
as in Eqs. \eq{genproj1}--\eq{genproj4}, and using the self-replication formula 
\eq{selfrep} (see also \cite{us}), it follows that the associative algebra generated by the internal 
master fields is a subalgebra of 
\be \widehat {\cal A}'~:=~ \left\{\sum_{n} \cP_{n}(E)\star\sum_{i,\bar i,j,\bar j=0,1} V^{n}_{i,\bar i,j,\bar j}(z)\star \overline V^{n}_{i,\bar i,j,\bar j}(\bar z)\star(\kappa_y)^{\star i}\star (\bar\kappa_{\bar y})^{\star \bar i}\star(\kappa_z)^{\star j}\star (\bar\kappa_{\bar z})^{\star \bar j}\right\}\ ,\label{cA'}\ee
where (see \eq{WSansatz})
\be V^{n}_{i,\bar i,j,\bar j}~:=~ \int_{-1}^1 \frac{dt}{t+1} \Delta^{n}_{i,\bar i,j,\bar j}(t;\partial^{(\r)})\left. e^{\ft{i}{t+1} \left((t-1) z^+ z^-
+\rho^+ z^- -\rho^- z^+\right)}\right|_{\r=0}\ ,\ee
are defined in terms of 
\be \Delta^{n}_{i,\bar i,j,\bar j}(t;\partial^{(\r)})~:=~\sum_{p=0}^{p_0} f^{{n},\a_1\dots \a_p}_{i,\bar i,j,\bar j}(t) 
\partial^{(\r)}_{\a_1}\cdots \partial^{(\r)}_{\a_p}\ ,\ee
where $p_0$ is a finite interger, and $f^{{n},\a_1\dots \a_p}_{i,\bar i,j,\bar j}(t)$
belong to the ring product algebra ${\cal R}$ spanned by real functions on $[-1,1]$ of the form
\be (T f_m)(t):=\int_{-1}^1 ds_1\cdots \int_{-1}^1 ds_m f_m(s_1,\dots,s_m) \delta(t-s_1\cdots s_m)\ ,\ee
with $f_m(s_1,\dots,s_m)$ being a polynomial in $m$ variables.
Indeed, if $Tf_m, T g_n\in{\cal R}$, then 
\be (Tf_m) \circ (Tg_n) = T(f_m g_n)\ ,\ee
and ${\cal R}$ forms a linear space under addition and multiplication
by polynomials, \emph{viz.} $t^p (Tf_m)(t)=T ((s_1 \cdots s_m)^p f(s_1,...,s_m))(t)$.
The star product of two elements in $\widehat {\cal A}'$ 
thus involves holomorphic star products in ${\cal Z}$ of the form
\be V(z) \,\star\, V'(z)\hspace{15cm}\nn\ee\be=~ \int_{-1}^1 dt \int_{-1}^1  dt' \frac{1}{2(\tilde t+1)}  \Delta(t;\partial^{(\r)}) \Delta'(t';\partial^{(\r')})  \left. e^{\ft{i}{\tilde t+1}\left((\tilde t-1)z^+z^- +\tilde{\r}^+ z^ --\tilde{\r}^-z^+-\ft{1}2 \r^+\r^{\prime -}+\ft{1}2\r^{\prime +}\r^-\right)} \right|_{\r=0=\r'}\ ,\label{Apm^2} \ee
where $\tilde t=tt'$ and $\tilde \r^\pm$ are defined in \eq{tildet}. 
We can rearrange 
\be \ft12 \Delta(t;\partial^{(\r)}) \Delta'(t';\partial^{(\r')}) \left. e^{\ft{i}{\tilde t+1}\left(\tilde{\r}^+z^- -\tilde{\r}^-z^+-\frac{1}2 \r^+\r^{\prime -}+\ft{1}2\r^{\prime +}\r^-\right)}\right|_{\r=0=\r'} \hspace{9cm}\nn\ee\be ~=:~  \sum_{I} \Delta^I(t;\partial^{(\r)}) \Delta^{\prime I}(t';\rho,\partial^{(\r)}) \left. e^{\ft{i}{\tilde t+1} (\r^{+}z^--\r^{-}z^+ )}\right|_{\r=0}\ ,\ee
where $\Delta^{\prime I}(t';\rho,\partial^{(\r)})=\sum_{p'=0}^{p_0} \sum_{q'=0}^{p'_0}  f^{I,\a_1\dots \a_{p'},\b_1\dots \b_{q'}}(t) \rho_{\a_1}\cdots \rho_{\a_{p'}} \partial^{(\r)}_{\b_1}\cdots \partial^{(\r)}_{\b_{q'}}$, and  
$\left\{ f^{I,\a_1\dots\a_p}(t)\right\}$ and $\left\{f^{\prime I,\a_1\dots \a_{p'},\b_1\dots \b_{q'}}(t')\right\}$ are 
linear combinations of $\left\{f^{\a_1\dots\a_p}(t)\right\}$ and $\left\{f^{\prime\a_1\dots\a_{p'}}(t)\right\}$, respectively, with coefficients given by polynomials in $t$ and $t'$; 
hence the new functions remain elements of ${\cal R}$.
Thus
\be V(z) \star V'(z)~=~ \sum_I \int_{-1}^1 \frac{ dt}{t+1}  (\Delta^I(\partial^{(\r)})\circ \Delta^{\prime I}(\rho,\partial^{(\r)}))(t) \,\left.e^{\ft{i}{ t+1}\left((t-1)z^+z^- +\r^{+}z^--\r^{-}z^+  \right)}\right|_{\r=0} \ ,\ee
where 
\bea  &&\left(\Delta^I(\partial^{(\r)}) \circ \Delta^{\prime I}(\rho,\partial^{(\r)})\right)(t)(\cdot)\nn\\[5pt]
& :=~ &\sum_{p,q'=0}^{p_0} \sum_{p'=0}^{p'_0}  (f^{I,\a_1\dots \a_p}\circ f^{\prime I,\b_1\dots \b_{q'},\c_{1}\dots \c_{p'}})(t) \partial^{(\r)}_{\a_1}\cdots \partial^{(\r)}_{\a_p} \left(\rho_{\b_1}\cdots \rho_{\b_{q'}} \partial^{(\r)}_{\c_{1}}\cdots \partial^{(\r)}_{\c_{p'}}(\cdot)\right) \ ,\label{Deltaring}\eea
showing the closure of $\cA'$ under the star product. 
%


\scs{Spacetime Dependence of the Master Fields} \label{spacetime}


In this Section, we shall introduce different gauge 
functions $\widehat{g}(x,Y,Z)$ that activate the 
connection $\widehat U$ and the dependence of the master 
fields on ${\cal X}$, as to give rise to a 
nontrivial spacetime structure. 
As already stressed in Section 3, these functions implement 
large gauge transformations, altering the asymptotics of the fields, 
as opposed to small, or proper, gauge transformations, which represent 
redundancies in the local description of the dynamics, but that that 
can nonetheless be useful in order to remove unphysical singularities 
form the fields.
In what follows, we shall consider three gauge functions: i) the AdS
gauge function $L(x,Y)$, that takes the solution space to a gauge that 
we shall refer to as the \emph{$L$-gauge}; 
ii) the Kruskal-like gauge function $\widehat L(x,Y,Z)=L(x,Y)\star 
\widetilde L(x,Z)$, where $\widetilde L(x,Z)$ is a local Lorentz 
transformation that aligns the spin frame in ${\cal Z}$ with that of ${\cal Y}$ 
as to remove all singularities away from the center of the black hole solutions,
leading to a gauge reminiscent of the Kruskal coordinate system for the Schwarzschild solution;
and iii) the Vasiliev gauge function $\widehat G(x,Y,Z)=L(x,Y)\star \widehat H(x,Y,Z)$, that takes the
solution to the \emph{Vasiliev gauge} where Fronsdal fields arise 
asymptotically in $W$ (after Lorentz covariantization and setting $Z=0$).
We would like to stress that in the $L$-gauge, the internal connection 
becomes real-analytic in ${\cal Z}$ for generic points in ${\cal X}$,
which facilitates the perturbative construction of $\widehat H$, that we
shall undertake at the first order.
We also note that, due to the nature of the Ansatz in Section 4,
linearized Fronsdal fields with $s\geqslant 1$ only arise in 
the black hole sector; we shall discuss the prospect of switching
on these fields in the particle sector in the Conclusion.

\scss{Gauge functions}

\paragraph{$AdS_4$ vacuum solution.} The $AdS_4$ spacetime provides a vacuum configuration 
for Vasiliev's theory, given by
\be \widehat\Phi^{(0)}~=~ 0\ ,\qquad \widehat S^{(0)}_{\underline \a}~=~ Z_{\underline \a}\ ,\qquad
\widehat U^{(0)}~=~ \O\ ,\label{vacL0}\ee
where $\Omega$ is the $\msp(4;\Real)$-valued flat connection on ${\cal X}$ 
with invertible frame field.
The resulting vacuum values of $W$ and the canonical Lorentz connection
$\omega$ are given by $W^{(0)}=e^{(0)}$ and $\omega^{(0)}$, where
\be e^{(0)}_{\a\ad}~=~ \left.2i\l {\partial^2\over \partial y^\a \partial \yb^{\ad}} \O\right|_{Y=0}\ ,\qquad (\omega^{(0)}_{\a\b},\bar\omega^{(0)}_{\ad\bd})~=~2i \left.\left({\partial^2\over \partial y^\a \partial y^{\b}},{\partial^2\over \partial \yb^{\ad} \partial \yb^{\bd}}\right) \O\right|_{Y=0}\label{vac02}\ ,
\ee
introducing the inverse $AdS$ radius $\l$.
A vacuum gauge function \cite{Vasiliev:1990bu,Bolotin:1999fa} is a map
\be L:~{\cal R}\rightarrow Sp(4;\Real)/SL(2;\Comp)\ ,\ee
defined on a region ${\cal R}\subset {\cal X}$ where it obeys 
\be L^{-1}\star \,dL=\left.\Omega\right|_{\cal R}\ .\ee
The Killing symmetries, that is, the globally defined higher spin gauge transformations 
preserving the vacuum, have parameters given locally by
\be \left.\widehat \e^{(0)}\right|_{\cal R}~=~ L^{-1}\star \e'(Y)\star L\ ,\ee
where $\e'$ belongs to the bosonic higher spin algebra $\mhs_1(4)$ 
or its minimal subalgebra $\mhs(4)$. 
Writing 
\be f^L(Y)~:=~ L^{-1}(x,Y)\star f(Y)\star L(x,Y)\ ,\ee
and defining the matrix representation $L_{\underline{\a\b}}(x)$ of $L$ via
\be Y^{L}_{\underline\a}~:=~L^{-1}\star Y_{\underline\a}\star L~=~ L_{\underline\a}{}^{\underline\b} Y_{\underline\b}\ ,\ee
it follows that
\be f^L(Y)~=~f(Y^L)\ ,\ee
where Weyl ordering is assumed on both sides; in particular, we have  
$\widehat \e^{(0)}=\e^{\prime }(Y^L)$, which are globally defined
Killing parameters on $AdS_4$.

\paragraph{Vacuum gauge function in stereographic coordinates.} The metric can be given on manifestly Lorentz covariant form in
stereographic coordinates in units where $\l=1$ as follows:
\be ds^2_{(0)} \ = \ \frac{4dx^2}{(1-x^2)^{2}}\ ,\qquad x^a\in \Real^4\ ,\qquad  x^2\neq 1\ , \label{stereo}\ee
where $x^2:=x^a x^b\eta_{ab}$ and $dx^2:=dx^a dx^b\eta_{ab}$.
This coordinate system provides a global cover of $AdS_4$, provided that the 
surface $x^2=1$ is taken to be two-sided, after which it can be identified 
as the boundary; for relations to embedding coordinates 
and spherically symmetric global coordinates, see Appendix \ref{App:conv}.
Correspondingly, in the notation of Appendix \ref{App:conv}, we have
\be e^{(0)}_{\a\ad}= - h^{-2}(\s^a)_{\a\ad}dx_a\ ,\qquad 
\o^{(0)}_{\a\b}=- h^{-2} (\s^{ab})_{\a\b} dx_a x_b\ ,\qquad h~:=~\sqrt{1-x^2}\ .\ee
Defining  
\be  x^a=-\frac12 (\s^a)^{\a\ad}x_{\a\ad}\ ,\qquad x^{\a\ad}=(\s_a)^{\a\ad} x^a\ ,\ee
and
\be \xi~:=~(1-h^2)^{-\ft12}\tanh^{-1}\sqrt{\ft{1-h}{1+h}}\ , \ee
the vacuum connection can be integrated on 
\be {\cal R}=\left\{x^a\in \Real^4: 
x^2  ~<~  1\right\}\ ,\ee
and expressed in terms of the gauge function \cite{Bolotin:1999fa,Sezgin:2005pv,Iazeolla:2008ix}
\be  L~=~\exp_\star (4i\xi  x^a  P_a) \ =\ {2h\over 1+h} \exp {4ix^a P_a\over 1+h}\ .\ee
Its matrix representation reads
\be
L_{\underline{\a}}{}^{\underline{\b}} ~=~ \left(\ba{cc}\cosh(2\xi\,x)\,\d_\a{}^\b & \sinh(2\xi\,x)\frac{x_\a{}^{\bd}}{x} \\[5pt] \sinh(2\xi\,x)\frac{\bar x_{\ad}{}^{\b}}{x} & \cosh(2\xi\,x)\,\d_{\ad}{}^{\bd}\ea\right)\ .\label{3.20}\ee

\paragraph{Kruskal-like gauge function.} Following \cite{us}, we define
\be \widehat L(x,Y,Z)~:=~L(x|Y)\star \tilde L(x|Z)\ ,\qquad \tilde L:~{\cal R}_{4}\rightarrow SL(2;\Comp)/C_{SL(2;\Comp)}(E^{L})\ ,\label{gaugef}\ee
where $C_{SL(2;\Comp)}(M)$ denotes the subgroup of 
$SL(2;\Comp)$ that commutes with $M~\in~ \msp(4;\Comp)$, and
\be E^L= L^{-1}\star E\star L= -\ft18 Y^{\underline\a} Y^{\underline\b} E^{L}_{\underline{\a\b}}\ ,\qquad
E^{L}_{\underline{\a\b}}~=~L_{\underline\a}{}^{\underline\a'} L_{\underline\b}{}^{\underline\b'}(\Gamma_{0'0})_{\underline{\a'\b'}}\ .\label{killingmatrix}\ee
The role of $\tilde L$ is to align the spin-frame on $\cal Z$ 
with one adapted to the $Sp(4;\Real)$ generator $E^L$ used to construct the 
projectors $\cP^L_n$.
As shown in \cite{us}, and as we shall review below, 
this gauge choice removes all singularities from the 
the gauge fields away from the spatial origin of $AdS_4$.
In this gauge, the vacuum configuration is given by
\be \widehat\Phi^{(\widehat L)(0)}~=~ 0\ ,\qquad 
\widehat S^{(\widehat L)(0)}_{\underline \a}~=~ \tilde L^{-1}\star Z_{\underline \a}\star \tilde L\ ,\qquad 
\widehat U^{(\widehat L)(0)}~=~ \O+ \tilde L^{-1}\star \,d\tilde L\ ,\label{vac0}\ee
implying that still
\be W^{(\widehat L)(0)}=\Omega\ ,\qquad \omega^{(\widehat L)(0)}=\omega^{(0)}\ ,\ee
as $\tilde L^{-1}\star d\tilde L|_{Z=0}=0$ and $\widehat M^{(\widehat L)}_{\a\b}|_{Z=0}=y_{(\a}\star y_{\b)}$,
though the internal connection does no longer obey the Vasiliev gauge condition.

\paragraph{Vasiliev gauge function.} Provided that the internal connection 
is real analytic in ${\cal Z}$ for generic points on ${\cal X}\times {\cal Y}$,
the Vasiliev gauge function
\be \widehat G(x|Y,Z)~=~ L\star \widehat H\ ,\label{Gnew}\ee
where $\widehat H$ is a perturbatively defined field-dependent 
large gauge transformation defined by
\be Z^{\underline\a} S^{(\widehat G)}_{\underline \a} = 0 \ .\ee
We note that 
\be \widehat H=1+\sum_{n\geqslant 1} \widehat H^{(n)}\ ,\ee
and stress that $\widehat H^{(n)}$ may in general contain singularities 
in ${\cal C}$, as we shall exemplify further below at the linearized level 
in the particle sector.

\scss{Master fields in $L$-gauge and Kruskal-like gauge}

\scsss{Weyl zero-form }\label{sec:LWSWeyl}

From Eqs. \eq{Lrot} and \eq{Phiprproj}, and 
using $\pi(\widehat L)=\pi(L)\star \tilde L$,
it follows that 
\be \widehat \Phi^{(\widehat L)} \ = \ \tilde L^{-1}\star\widehat \Phi^{(L)} \star \tilde L 
\ = \ \widehat \Phi^{(L)} \ = \ \Phi_{\textrm{bh}}+ \Phi_{\textrm{pt}}\ ,\label{Phi}\ee
where we have use the fact that $\widehat \Phi^{(L)}$ is $Z$-independent, and 
defined
\be   \Phi_{\textrm{bh}} \ := \ \sum_{n=\pm 1,\pm2,\dots}\n_n\cP^L_n\,\star\, \kappa_y \ ,\qquad 
\Phi_{\textrm{pt}} \ := \ \sum_{n=\pm 1,\pm2,\dots} \widetilde \nu_{n} {\tcP}^L_n \,\star\, \kappa_y \ ,\label{starky}\ee
recalling that $\cP^L_n\equiv L^{-1}\star \cP_n\star L$ and 
$ {\tcP}^L_n\equiv L^{-1}\star \tcP_n\star L$.
We also define the generating functions 
\be C_{\textrm{bh}}(x,y):=\left.\Phi_{\textrm{bh}}\right|_{\yb=0}\ ,\qquad 
C_{\textrm{pt}}(x,y):=\left.\Phi_{\textrm{pt}}\right|_{\yb=0}\ ,\ee
for the dynamical scalar field and the (self-dual part of the) spin $s=1,2,\dots$
Weyl tensors in the black hole and particle sectors, respectively.
As we shall demonstrate next, all spins are activated in 
$C_{\textrm{bh}}(x,y)$, whose spin-2 component is the 
Schwarzschild black hole Weyl tensor \cite{us,Didenko:2009td},
while $\Phi_{\textrm{pt}}$ only consists of a
rotationally-invariant dynamical scalar field.

\paragraph{Black-hole sector.} Using the regular presentation \eq{sumproj}, we have
\be
\cP^L_n\star \k_y \ = \ 2(-1)^{n-\ft{1+\e}2}\oint_{C(\e)}\frac{d\eta}{2\pi i}\left(\frac{\eta+1}{\eta-1}\right)^{n} e^{-4\eta E^L}\star \k_y  \label{enhancedPhi} \ ,\ee
where 
\bea e^{-4\eta E^L} \ \equiv \  L^{-1}\star e^{-4\eta E}\star  L  
\ = \ e^{-\ft{\eta}2\left(y^\a y^\b \vark^L_{\a\b}+\bar y^{\ad} \bar y^{\bd} {\bar\vark}^L_{\ad\bd}+2y^a\bar y^{\bd} v^L_{\a\bd}\right)} \ .\label{EL}\eea
The matrix $v^L_{\a\bd}$ is a Killing vector with Killing two-form 
$(\vark^L_{\a\b},\bar{\vark}^L_{\ad\bd})$.
In terms of the global radial AdS coordinate $r$, one has
\be (\vark^L)^2 \ : = \ \ft12 (\vark^L)^{\a\b}(\vark^L)_{\a\b}\ = \ - r^2\ .\ee
By introducing an adapted spin-frame consisting of the
$x$-dependent eigenspinors $(u^{+}_{(E)\a} , u^{-}_{(E)\a})$
of $\vark^L_{\a\b }$, we can write
\be  \vark^L_{\a\b }~=~  r\,{\cal D}_{(E)\a\b}\ ,\qquad
 v^L_{\a\bd}~=~ \sqrt{1+r^2}\,{\cal T}_{(E)\a\bd} \ ,\label{kLE}\  \ee
where  
\be {\cal D}_{(E)\a\b}~:=~u^{+}_{(E)\a}  u^{-}_{(E)\b} +  u^{-}_{(E)\a} u^{+}_{(E)\b} \ , \qquad{\cal T}_{(E)\a\bd}~:=~ u^{+}_{(E)\a}\bar{ u}^{+}_{(E)\bd}+ u^{-}_{(E)\a}\bar{ u}^{-}_{(E)\bd}
\label{vLE} \ ;\ee
see \cite{us} for further details.
Performing the star product $e^{-4\eta E^L}\star \kappa_y$ in \eq{enhancedPhi} yield
\bea \Phi_{\textrm{bh}} & = &   \frac{2}{\sqrt{(
\vark^L)^2}}\sum_{n=\pm 1,\pm2,...} (-1)^{n-\ft{1+\e}2}\nu_{n}\oint_{C(\e)}\frac{d\eta}{2\pi i\eta}\left(\frac{\eta+1}{\eta-1}\right)^{n}  \ \times\nn\\[5pt]&\times&\exp\left\{\frac{1}{2\eta} \left[ y^\a
(\vark^L)^{-1}_{\a\b}y^\b+\yb^{\ad}(\bar{\vark}^L)^{-1}_{\ad\bd}\yb^{\bd}+2iy^\a\yb^{\bd}(\vark^L)^{-1}_{\a\b}(v^L)^\b{}_{\bd}\right]\right\}
\label{enhancedPhi2} \ . \eea
Since 
\be e^{-4\eta E^L}\star \k_y|_{\yb=0}= \ft1{\sqrt{(\eta \vark^L)^2}} \exp \ft1{2\eta} y^\a(\vark^L)^{-1}_{\a\b} y^\b\ ,\qquad ({\vark}^L)^{-1}_{\a\b}=\ft1{r} {\cal D}_{(E)\a\b}\ ,\ee
we have 
\bea C_{\textrm{bh}}(x,y) \ = \
\frac{2}{r}\sum_{n=1}^{\infty}i^{n-1}\oint_{C(1)}\frac{d\eta}{2\pi
i\eta}\left(\frac{\eta+1}{\eta-1}\right)^{n}\sum_{\e=\pm 1}
(-1)^{\ft{1+\e}2(n-1)}\mu_{\e n}\,e^{\ft{\e}{2\eta}\,
y^\a(\vark^L)^{-1}_{\a\b}y^\b}\ .\label{o3Weylnm}\eea
For a given $n$, the contribution from $\nu_n {\cal P}_n(E)$ to the
spin-$s$ sector is thus given, up to an $n$- and $s$-dependent real factor, by 
\bea C_{{\textrm{bh}},n,\a(2s)}\ \sim \ \frac{i^{n-1}\mu_n}{r^{s+1}}\,(u_{(E)}^+
 u_{(E)}^- )^s_{\a(2s)} \ ,\label{nsWeylspher}\eea
where we note that $i^{n-1}\mu_n$ is real and imaginary, respectively 
in the case of scalar (even $n$) and spinor (odd $n$) singleton projectors.
The Weyl tensors are thus of generalized Petrov type D \cite{us}, 
and the corresponding asymptotic charges are electric and magnetic,
respectively,
for scalar and spinor singletons in the A model, and vice versa in the
B model.
As first noted in \cite{Didenko:2009td} for the case $n=1$, this corresponds 
to an $AdS_4$ Schwarzschild black hole Weyl tensor together with its 
generalization to all integer spins.

\paragraph{Particle sector.} In order to compute $\Phi_{\textrm{pt}}$ in \eq{Phi}, 
we start from the regular presentation \eq{sumtwproj} of the twisted projector, 
which transforms under the adjoint action by $L$ into
\bea \widetilde{{\cal P}}^L_n(E) \ = \ 4\pi (-1)^{n-\ft{1+\ve}2}\,\oint_{C(\ve)} \frac{d\eta}{2\pi i}\,\left(\frac{\eta+1}{\eta-1}\right)^{n}\,\d^2(y^L-i\eta \s_0\yb^L)\ ,\label{tPEL} \eea
where, in stereographic coordinates,
\bea & \d^2(y^L-i\eta \s_0\yb^L) \ = \ \d^2\left(A(x,\eta)y + B(x,\eta)\yb\right) \ , \ \ \ \ \ \textrm{with} &\\[5pt]
&  A_\a{}^\b \ = \ \frac{1}{\sqrt{1-x^2}}\,\left(\d_\a{}^\b - i\eta (\s_0\bar{x})_{\a}{}^\b\right)  
\ , \qquad B_\a{}^{\bd} \ = \ \frac{1}{\sqrt{1-x^2}}\,\left(x_\a{}^{\bd} - i\eta (\s_0)_{\a}{}^{\bd}
\right) \ .&   \label{AB}\eea
Recalling the complex analyticity property \eq{deltay} of the delta function in non-commutative
twistor space, the star product in the second equation in \eq{starky} can be found to be  
\bea 2\pi \delta^2\left(y^L-i\eta\s_0\yb^L\right) \star \kappa_y \ = \ 
\frac{2\pi}{\det A}\,\delta^2(\widetilde y)\star\k_y \ = \ \frac{1}{\det A}\,e^{iy^\a M_\a{}^{\ad} \yb_{\ad}} \ ,\eea
where  
\be M_\a{}^{\bd} \ := \ A^{-1}_\a{}^\b B_\b{}^{\bd} \ = \ f_1(x,\eta)x_\a{}^{\bd}-if_2(x,\eta)(\s_0)_\a{}^{\bd} \ , \ee
 \be f_1 \ := \ \frac{1-2i\eta x_0+\eta^2}{1-2i\eta x_0 + \eta^2 x^2} \ , \qquad f_2 
 \ := \ \eta \,\frac{1-x^2}{1-2i\eta x_0 + \eta^2 x^2}   \label{A-1B}\ ,\ee
\be \det A \ = \ \frac{1-2i\eta x_0+\eta^2 x^2}{1-x^2} \ ,\ee
and we have introduced
\be  \widetilde y_\a \ := \ y_\a + M_\a{}^{\bd} (x,\eta)\yb_{\bd}\ , \label{modosc}\ee
which generate a (noncommutative) Weyl algebra for generic $x^a$ and $\eta$, with the notable
exception $\widetilde{y}_\a|_{x=0,\eta=\pm 1}= y_\a \mp i \left(\s_{0}\right)_\a{}^{\bd} \yb_{\bd}$,
which are abelian; see Footnote \ref{footnote2} after Eq. \eq{genproj4}.
Thus, we arrive at 
\bea \Phi_{\textrm{pt}} \ = \ 2(1-x^2) \sum_{n\neq 0} 
(-1)^{n-\ft{1+\e}2}\widetilde \nu_{n}\oint_{C(\e)}\frac{d\eta}{2\pi i} 
\left(\frac{\eta+1}{\eta-1}\right)^n 
\frac{e^{iy^\a M_\a{}^{\bd}(x|\eta) \yb_{\bd}}}{1-2i\eta x_0+\eta^2 x^2} \ .\label{Phipt}\eea
We note that the expansion in oscillators of this function only contains equal powers of 
$y_\a$ and $\yb_{\ad}$, \emph{i.e.}, all Weyl tensors of spin $1,2,3,...$ vanish and only scalar 
modes appear.
Moreover, the reality condition on $\wPhi$ implies that that each positive-energy
particle mode must be accompanied by a corresponding negative-energy anti-particle mode; 
see Eq. \eq{realonparameters2}. 
For example, if $\widetilde \nu_{n} = 0$ for $\forall \,n \neq \pm1$, then 
we recover the mode function of the ground state of the lowest-weight space
$\mD(1,0)$ accompanied by its negative-energy counterpart in $\mD(-1,0)$,
as \eq{Phipt} becomes
\bea \Phi_{\textrm{pt}}^{|n|=1} & = & 2(1-x^2) \left[\widetilde \nu_{1}\oint_{C(1)}\frac{d\eta}{2\pi i} \,\frac{\eta+1}{\eta-1}\,\frac{e^{iy^\a M_\a{}^{\bd}(x|\eta) \yb_{\bd}}}{1-2i\eta x_0+\eta^2 x^2}\right. \nn\\ & - & \left. \tn_{-1}\oint_{C(-1)}\frac{d\eta}{2\pi i}\, \frac{\eta-1}{\eta+1}\, \frac{e^{iy^\a M_\a{}^{\bd}(x|\eta) \yb_{\bd}}}{1-2i\eta x_0+\eta^2 x^2}\right]\ .\label{phi}\eea
The corresponding real physical scalar mode $\phi_{\textrm{pt}}^{|n|=1} $ 
is given by the coefficient of the unity, \emph{i.e.}
\be \phi_{\textrm{pt}}^{|n|=1} =4\left(\widetilde \nu_{1} \frac{1-x^2}{1-2i x_0+ x^2}+\tn_{-1} \frac{1-x^2}{1+2i x_0+ x^2}\right) \ = 4 \ \left(\widetilde \nu_{1}\, \frac{e^{it}}{(1+r^2)^{1/2}} + \tn_{1}^\ast\,\frac{e^{-it}}{(1+r^2)^{1/2}}\right)\ , \label{d10}\ee
where the last expression is given in the global spherically symmetric coordinates, and
we note that it is regular everywhere. 
Similarly, the scalar mode $\phi_{\textrm{pt}}^{|n|=2} $ from the ground state of $\mD(2,0)$
together with its negative energy counterpart in $\mD(-2,0)$
can be obtained from the ${\cal P}_2$ and ${\cal P}_{-2}$ projectors, \emph{i.e.} 
\be \phi_{\textrm{pt}}^{|n|=2}=8\left[\widetilde \nu_{2} \left(\frac{1-x^2}{1-2i x_0+ 
x^2}\right)^2 + \tn_{-2}\left(\frac{1-x^2}{1+2i x_0+ x^2}\right)^2\right]   \ = \ 
8\left[\widetilde \nu_{2}\, \frac{e^{2 it}}{1+r^2} + \tn_2^\ast\, \frac{e^{-2 it}}{1+r^2}\right] \ . \ee
The projectors with $|n| > 2 $ encode the rotationally-invariant massless scalar modes 
of energy $\pm n$. 
%

\scsss{Internal connection in $L$-gauge}\label{Sec:V}

Applying an adjoint $L$ transformation to the internal connection
\eq{wSprime}, and using \eq{enhancedPhi} and \eq{tPEL}, 
we find
\bea \widehat S^{(L)}_\a & = & L^{-1}\star \widehat S'_{\a}\star  L \ = \ z_{\a} -2i(\wV^{(L)}_{\textrm{bh},\a}+\wV^{(L)}_{\textrm{pt},\a}) \ , \label{Valpha}
\eea
where black hole sector is given by 
\bea \wV^{(L)}_{\textrm{bh},\a}&=&4i \sum_n (-)^{n-\ft{1+\ve}2}\,\oint_{C(\ve)} \frac{d\eta}{2\pi i}\,\left(\frac{\eta+1}{\eta-1}\right)^{n} \sum_{k=1}^\infty  {1/2 \choose k}\left(-\frac{b}{2}\right)^{k} \n^{(k)}_n  \nn\\[5pt] &&\times \int_{-1}^1 \frac{dt}{(t+1)^2} \frac{(\log (1/t^2))^{k-1}}{(k-1)!}\,e^{-4\eta E^L}\star 2iz_\a e^{i\frac{t-1}{t+1}w_z} \ ,
\label{VLbh}\eea
and the particle sector by
\bea
\wV^{(L)}_{\textrm{pt},\a}&=& 4i \sum_n (-)^{n-\ft{1+\ve}2}\,\oint_{C(\ve)} \frac{d\eta}{2\pi i}
\,\left(\frac{\eta+1}{\eta-1}\right)^{n}
\sum_{k=1}^\infty  {1/2 \choose k}\left(-\frac{b}{2}\right)^{k} \widetilde{\n}^{(k)}_n
\nn\\[5pt] && \times \int_{-1}^1 \frac{dt}{(t+1)^2} \frac{(\log (1/t^2))^{k-1}}{(k-1)!}\, \d^2(y^L-i\eta \s_0\yb^L) \star z_\a e^{i\frac{t-1}{t+1}w_z} \ ,\label{wVpt}\eea
whose real-analyticity properties in ${\cal C}$ will be spelled out next.

\paragraph{Black-hole sector.} The master field $\wV^{(L)}_{\textrm{bh},\a}$
was shown in \cite{us2} to be real-analytic in ${\cal T}$ except 
at the equatorial plane $\theta=\pi/2$ in the spherical global coordinates 
of $AdS_4$ defined in \eq{metricglob}, where singularities appear on a
plane in twistor space.
This can be seen by introducing a source $\rho^\a$ to write $V_{n,\a}$ 
as in \eq{WSansatz}, after which ${\cal P}^L_n \star V_{n,\a} $ can be computed
using the lemma
\bea  & e^{-\ft12\left(y^\a y^\b \vark^L_{\a\b}+\bar y^{\ad} \bar y^{\bd}{\bar\vark}^L_{\ad\bd}+2y^a\bar y^{\bd} v^L_{\a\bd}\right)}\,\star \, e^{\ft{i}{2(t+1)}\left((t-1)z^\a z^\b {\cal D}_{\a\b}+2\r^\a z_\a\right)} & \nn\\[5pt]
& \ = \ \frac{1}{\sqrt{ (\vark^{L})^2 \,G^2}}e^{-\ft12\bar y^{\ad} \bar y^{\bd}\left({\bar\vark}^L_{\ad\bd}-{\bar v}^L_{\ad}{}^\a(\vark^L)^{-1}_\a{}^\b  v^L_{\b\bd}\right)+\ft{i}{2(t+1)}\left((t-1)z^\a z^\b {\cal D}_{\a\b}+2\r^\a z_\a\right)+\ft12 b^\a b^\b G^{-1}_{\a\b} } \ ,&\label{lemmaS}\eea
where we recall that ${\cal D}_{\a\b}~:=~2u^-_{(\a} u^+_{\b)}$, and 
\bea G_{\a\b} &  := &  (\vark^L)^{-1}_{\a\b}+i\frac{ t-1}{t+1}\,{\cal D}_{\a\b}\ ,\label{Gab}\\[5pt] 
b^\a &:=& i\left[y^\a+\bar y^{\ad}{\bar v}^L_{\ad}{}^\b(\vark^L)^{-1}_\b{}^\a+\frac{1}{t+1}\left( (t-1)z^\b{\cal D}_\b{}^\a-\r^\a\right)\right]\ ,\label{ba}\eea
and we have used $G^{-1}_{\a\b}=-\ft{G_{\a\b}}{G^2}$ with 
\be G^2:=\ft12 G^{\a\b}G_{\a\b}= \frac{(t+1)^2-i (t^2-1) \vark^{L\a\b} {\cal D}_{\a\b}+(\vark^{L})^2 (t-1)^2}{(t+1)^2 (\vark^{L})^2} \ .\label{det}\ee
The crux of the matter is that, after the star product with the projectors, 
the singularities at $t=-1$ are moved to the zeros 
of $(t+1)^2-i(t^2-1) \vark^{L\a\b}{\cal D}_{\a\b}+(\vark^L)^2 (t-1)^2$, 
that have imaginary parts provided that $\vark^{L\a\b}{\cal D}_{\a\b} \neq 0$
(recall that $(\vark^{L})^2 = -r^2$ is real), that push them away from the 
integration domain $t\in[-1,1]$. 
The quantity $\vark^{L\a\b}{\cal D}_{\a\b}$, which is given by
the contraction of the $x$-dependent eigenspinors of $\vark^L_{\a\b}$ \eq{kLE}-\eq{vLE}
with the rigid spin-frame $(u^+_\a,u^-_\a)$, is proportional to $\cos \theta$; for details, see \cite{us2}.
Thus, the $t$-integral is convergent, and one is led to 
the conclusion stated above.  

\paragraph{Particle sector.} To analyze the singularity structure of 
$\wV^{(L)}_{\textrm{pt},\a}$, we need to compute the star product in \eq{wVpt}.
To this end, we introduce a source $\r_\a$ as in \eq{WSansatz}, and consider
\bea  && \d^2(y^L-i\eta \s_0\yb^L) \ \star \ e^{\ft{i}{2(t+1)}\left((t-1)z^\a z^\b {\cal D}_{\a\b}+2\r^\a z_\a\right)}  \nn\\[5pt]
& =& \   \frac{1}{2\pi}\,\frac{1-x^2}{1-2i\eta x_0 +\eta^2 x^2} \,\,\frac{t+1}{t-1}\, e^{\frac{i}{2} \frac{t+1}{t-1} \widetilde{y}^\a {\cal D}_{\a\b}\widetilde{y}^\b-i(z^\a-\frac{1}{t-1}\,\r^\b  {\cal D}_{\b}{}^{\a})\widetilde{y}_{\a}+ \frac{i}{2(t^2-1)}\,\r^\a \r^\b {\cal D}_{\a\b}  } \ , \label{tPstartV}
\eea
where the modified oscillators $\widetilde{y}_{\a}$ are defined in \eq{modosc},
and $(1-x^2)/(1-2i\eta x_0 +\eta^2 x^2)$ is regular, as shown explicitly in \eq{d10}.
The final form of $\widehat V^{(L)}_{{\rm pt},\a}$ is thus given by\footnote{
The solutions contain two sources of explicit 
Lorentz symmetry breaking: one due to the expansion over the generalized projectors,
and another one due to the introduction of $w_z$ in representing $\delta^2(z)$ as 
a delta sequence; of these two, the latter is responsible for the different signs in \eq{VLpt} in the decomposition
with respect to the spin frame.}
\bea \widehat V^{(L)\pm}_{{\rm pt}} &=& \mp 2i \sum_{n=\pm1,\pm2,\dots}
(-1)^{n-\ft{1+\ve}2}\,\oint_{C(\ve)} 
\frac{d\eta}{2\pi i}\,\left(\frac{\eta+1}{\eta-1}\right)^{n}\,
\sum_{k=1}^\infty {1/2 \choose k}\left(-\frac{b}{2}\right)^k
\frac{\tn^{(k)}_n}{(k-1)!} 
\nn\\[5pt]
&& \times \frac{1-x^2}{1-2i\eta x_0 +\eta^2 x^2} \,\widetilde{y}^\pm e^{i(\widetilde{y}^+z^- - \widetilde{y}^-z^+)}\int_{-1}^1\frac{dt}{(t-1)^2}
\left(\log\frac{1}{t^2}\right)^{k-1}
e^{i\frac{t+1}{t-1}\widetilde{y}^+\widetilde{y}^- }\,\ ,\label{VLpt}\eea
where $\widehat V^{(L)\pm}_{{\rm pt}}:=u^{\pm\a} \widehat V^{(L)}_{{\rm pt},\a}$.
Interestingly, the singularity in the $t$-integral has been moved from 
its position in the holomorphic gauge, namely at $t=-1$, to $t=+1$, unlike in the
case of the black hole sector where it was removed except at the
equatorial plane\footnote{\label{Footnote16}The formula \eq{tPstartV} can be obtained directly 
from \eq{lemmaS}-\eq{det} by realizing the delta function as 
$\d^2(y^L-i\eta \s_0\yb^L) \ = \ \frac{1}{\det A} \d^2(\ty) \ = \ \frac{1}{\det A} \lim_{\e\rightarrow 0^+} \frac{1}{\e}e^{-\frac{i}{\e}\ty^+\ty^-}$, which can be cast in the form of 
the first factor in \eq{det} with $\vark_{\a\b}=\frac{i}{\e}{\cal D}_{\a\b}$, $\bar{\vark}_{\ad\bd} = -\frac{i}{\e}(M^T{\cal D}M)_{\ad\bd}$, $v_{\a\bd}= \frac{i}{\e}({\cal D}M)_{\a\bd}$. 
Thus $\vark^{-1} \rightarrow  0$ in that limit, and the Gaussian determinant 
\eq{det} reduces to $\lim_{\e\rightarrow 0^+} (t+1)/(t-1-\e(t+1)) = (t+1)/ (t-1)$.}.
Taking into account the more detailed structure of the poles
at $t=+1$ in the exponential, it follows that $\widehat V^{(L)}_{{\rm pt},\a}$ is \emph{regular in 
${\cal Z}$}, while it has singularities on a plane in ${\cal Y}$,
given by a pole in the first order of the perturbative expansion (i.e. for $k=1$). 

\paragraph{Singularity-free twistor space curvature.} We remark that the singularities are cancelled, however, 
in the star commutator $[\wS^{(L)-},\wS^{(L)+}]_\star$, which 
yields the source term $-2i(1-b\widehat \Phi^{(L)}\star\k)$, which is regular
(and exact already at the first order, as the Weyl zero-form does not receive 
any non-linear corrections in the $L$-gauge). 
To exhibit this, we consider the linearized equation 
\bea \partial_+\wV^{(L)(1)+}_{\textrm{pt}} +\partial_-\wV^{(L)(1)-}_{\textrm{pt}} \ = \ \frac{b}{2i}\Phi^{(L)}_{\textrm{pt}}\star\kappa \ ,
\qquad \partial_\pm = \partial/\partial z^\pm\ ,\label{linV}\eea
in the special case where $\tn_n$ vanishes except for $|n|=1$, \emph{i.e.}
\bea \wV^{(L)(1)\pm}_{\textrm{pt}} & = & \pm 2ib(1-x^2) \left(\frac{\tn_1\,\widetilde{y}^\pm }{1-2i x_0 + x^2} 
e^{i\left(\widetilde{y}^+z^- - \widetilde{y}^-z^+\right)}
\int_{-1}^1\frac{dt}{(t-1)^2}\,e^{i\frac{t+1}{t-1}\widetilde{y}^+\widetilde{y}^-}\right|_{\eta=+1} \nn\\ 
&+& \!\!\! \left.\left.\frac{\tn_{-1}\,\widetilde{y}^\pm}{1+2i x_0 + x^2}  e^{i\left(\widetilde{y}^+z^- - \widetilde{y}^-z^+\right)}\int_{-1}^1\frac{dt}{(t-1)^2}\,e^{i\frac{t+1}{t-1}\widetilde{y}^+\widetilde{y}^- }\right|_{\eta=-1}  \right )\ ,\label{V1pt}\eea
where $\widetilde{y}_\a|_{\eta=\pm1}= \tilde y_\a + M_\a{}^{\bd} (x,\eta=\pm 1)\yb_{\bd}$.
The left-hand side of \eq{linV} thus reads
\be \partial_+\wV^{(L)(1)+}_{\textrm{pt}} +\partial_-\wV^{(L)(1)-}_{\textrm{pt}} = 4b(1-x^2)\left(
\frac{\tn_1 \,\widetilde{y}^+\widetilde{y}^-}{1-2i x_0 + x^2}  \,e^{i(\widetilde{y}^+z^- - \widetilde{y}^-z^+)}\int_{-1}^1\frac{dt}{(t-1)^2}\,e^{i\frac{t+1}{t-1}\widetilde{y}^+\widetilde{y}^- }\right|_{\eta=+1}\nn\ee 
\be +  \left.\left.\frac{\tn_{-1}\,\widetilde{y}^+\widetilde{y}^-}{1+2i x_0 + x^2}  \,e^{i\left(\widetilde{y}^+z^- - \widetilde{y}^-z^+\right)}\int_{-1}^1\frac{dt}{(t-1)^2}\,e^{i\frac{t+1}{t-1}\widetilde{y}^+\widetilde{y}^- }\right|_{\eta=-1}
\right)\ .\ee
As for the right-hand side of \eq{linV}, we use \eq{phi} to compute
\be \Phi_{\textrm{pt}}^{|n|=1}\star\kappa \ = \ 4(1-x^2)\left(\frac{\tn_1}{1-2i x_0 + x^2}\,\left.e^{iy^\a M_\a{}^{\bd} \yb_{\bd}} 
\right|_{\eta=+1}
+ \frac{\tn_{-1}}{1+2i x_0 + x^2} \,\left.e^{iy^\a M_\a{}^{\bd} \yb_{\bd}}\right|_{\eta=-1}\right) \star \kappa \nn\ee 
\be\hspace{1cm} \ = \ 4(1-x^2)\left(\frac{\tn_1}{1-2i x_0 + x^2}\,\left.e^{i(\widetilde{y}^+z^- - \widetilde{y}^-z^+)}\right|_{\eta=+1}
+\frac{\tn_{-1}}{1+2i x_0 + x^2} \,\left.e^{i(\widetilde{y}^+z^- - \widetilde{y}^-z^+)}\right|_{\eta=-1}\right)\ .\label{Phiptkappa}\ee
Thus, Eq. \eq{linV} is satisfied provided that 
\bea \widetilde{y}^+\widetilde{y}^-\int_{-1}^1\frac{dt}{(t-1)^2}\,e^{i\frac{t+1}{t-1}\widetilde{y}^+\widetilde{y}^- } \ = \ \frac{1}{2i} \ ,  \eea 
which holds on the grounds of the first of the properties of the distribution $I^\pm$
given in \eq{propI}.
This means, in particular, that $\wV^{(L)(1)\pm}_{\textrm{pt}} $ has 
a pole at $\widetilde{y}^\mp = 0$, and that, as anticipated, the latter 
is exactly cancelled by the commutator in \eq{INT2} at first order. 

\scsss{Internal connection in Kruskal-like gauge}

Applying an adjoint $\widehat L$ transformation to the internal connection
\eq{wSprime}, and using \eq{enhancedPhi} and \eq{tPEL} and $\widehat L=L\star \widetilde L$, we find
\bea \widehat S^{(\widehat L)}_\a & = & \widehat L^{-1}\star \widehat S'_{\a}\star  \widehat L \ = \ \tilde L_{\a}{}^{\b} z_{\b} -2i(\wV^{(\widehat L)}_{\textrm{bh},\a}+\wV^{(\widehat L)}_{\textrm{pt},\a}) \ , \label{Valpha2}
\eea
where the black hole and particle contributions 
\be \wV^{(\widehat L)}_{\textrm{bh},\a}=\widetilde L^{-1}\star \wV^{(L)}_{\textrm{bh},\a}\star \widetilde L\ ,\qquad
\wV^{(\widehat L)}_{\textrm{pt},\a}=\widetilde L^{-1}\star \wV^{(L)}_{\textrm{pt},\a}\star \widetilde L\ ,\ee
with $\wV^{(L)}_{\textrm{bh},\a}$ and $\wV^{(L)}_{\textrm{pt},\a}$ given in \eq{VLbh} and \eq{wVpt}, respectively.
Thus, more explicitly,
\bea \wV^{(\widehat L)}_{\textrm{bh},\a}&=&4i \sum_n (-)^{n-\ft{1+\ve}2}\,\oint_{C(\ve)} \frac{d\eta}{2\pi i}\,\left(\frac{\eta+1}{\eta-1}\right)^{n} \sum_{k=1}^\infty  {1/2 \choose k}\left(-\frac{b}{2}\right)^{k} \n^{(k)}_n  \nn\\[5pt] &&\times \int_{-1}^1 \frac{dt}{(t+1)^2} \frac{(\log (1/t^2))^{k-1}}{(k-1)!}\,e^{-4\eta E^L}\star 2iz^{\tilde L}_\a\, e^{i\frac{t-1}{t+1}w^{\tilde L}_z} \ ,\eea
and 
\bea
\wV^{(\widehat L)}_{\textrm{pt},\a}&=& 4i \sum_n (-)^{n-\ft{1+\ve}2}\,\oint_{C(\ve)} \frac{d\eta}{2\pi i}
\,\left(\frac{\eta+1}{\eta-1}\right)^{n}
\sum_{k=1}^\infty  {1/2 \choose k}\left(-\frac{b}{2}\right)^{k} \widetilde{\n}^{(k)}_n
\nn\\[5pt] && \times \int_{-1}^1 \frac{dt}{(t+1)^2} \frac{(\log (1/t^2))^{k-1}}{(k-1)!}\, \d^2(y^L-i\eta \s_0\yb^L) \star z^{\tilde L}_\a \,e^{i\frac{t-1}{t+1}w^{\tilde L}_z} \ ,\label{wVpt2}\eea
with
\be z^{\tilde L}_{\a} ~:=~ \tilde L_\a{}^\b z_\b \ , \qquad u^{\pm\b}\tilde L_\b{}^\a \ = \ u_{(E)}^{\pm \a}\ ,  \qquad  w^{\tilde L}_z:= \ft 12 z^\a z^\b {\cal D}_{(E)\a\b}\ ,\label{wtildeL}\ee
where these equalities follow from the definition made in \eq{vLE} (see Appendix E in \cite{us} for the explicit form of $\tilde L_\a{}^\b$).
In what follows, we shall first recall how the conjugation by $\tilde L$ removes 
unphysical singularities in ${\cal Z}$ in $\wV^{(L)}_{\textrm{bh}}$, as first found in \cite{us},
and then show that it preserves the real-analyticity property of 
$\wV^{(L)}_{\textrm{pt}}$ in ${\cal Z}$.

\paragraph{Black-hole sector.} As observed in \cite{us}, the $SL(2,\Comp)$ 
transformation induced by $\widetilde L$ aligns the spin-frame in ${\cal Z}$ 
with that of $(u^{+}_{(E)\a} , u^{-}_{(E)\a})$, as in \eq{wtildeL}, thereby
modiying the quantities in \eq{Gab} and \eq{ba}.
The resulting modification of the determinant in \eq{det} contains 
the factor $(t+1)^2-i(t^2-1) \vark^{L\a\b}{\cal D}_{(E)\a\b}+(\vark^L)^2 (t-1)^2$, the imaginary part of which only vanishes at $r=0$, which is thus the only singular point 
of $\wV^{(\widehat L)}_{\textrm{bh}\,\a}$.

As $r=0$ is the only singular point of the Weyl zero-form, we conclude 
that the singularities in the $L$-gauge at the equatorial plane away from $r=0$ 
are gauge artifacts.

\paragraph{Particle sector.} As explained before, the singularity structure in 
${\cal C}$ of $\wV^{(\widehat L)}_{\textrm{pt}}$ is related to the nonintegrable
divergencies of the measure of the $t$-integral.
To study the latter, we need the modified version of \eq{tPstartV} , \emph{viz.}
\bea  && \d^2(y^L-i\eta \s_0\yb^L) \ \star 
\ e^{\ft{i}{2(t+1)}\left((t-1)z^\a z^\b {\cal D}_{(E)\a\b}+2\r^\a z^{\tilde L}_\a\right)}  \nn\\[5pt]
& =& \   \frac{1}{2\pi}\,\frac{1-x^2}{1-2i\eta x_0 +\eta^2 x^2} \,\,\frac{t+1}{t-1}\, 
e^{\frac{i}{2} \frac{t+1}{t-1} \widetilde{y}^\a {\cal D}_{(E)\a\b}\widetilde{y}^\b-
i(z^\a-\frac{1}{t-1}\,\r^\c  {\cal D}_{\c}{}^{\b}\tilde{L}_\b{}^\a)\widetilde{y}_{\a}+
\frac{i}{2(t^2-1)}\,\r^\a \r^\b {\cal D}_{\a\b}  } \ . \label{tPstartVmod}
\eea
Thus, comparing to \eq{tPstartV}, we conclude that 
$\wV^{(\widehat L)}_{\textrm{pt},\a}$ is real-analytic in ${\cal Z}$
while it has singularities on a plane in ${\cal Y}$ depending
on ${\cal X}$ (that is a modification of the singular plane
in $L$-gauge).

We remark that as far as the real-analyticity 
property of $\wV^{(L)}_{\textrm{pt}}$ in ${\cal Z}$ is
concerned, one may argue as follows: Unlike in the black hole
sector, in the $L$-gauge, the star product in \eq{tPstartV} 
does not generate any imaginary part, 
or any other sort of contribution that pushes the singular points of the 
$t$-measure out of the integration domain, independently of whether the 
spin-frames in ${\cal Z}$ and ${\cal Y}$ are collinear or not (point-wise
over ${\cal X}$).
However, the details of the singular plane in ${\cal Y}$
requires the detailed calculation in \eq{tPstartVmod}.

\scss{Vasiliev gauge} 

Let us investigate the mechanism whereby the (large) gauge transformation,
with gauge function $\widehat G=L\star \widehat H$, that brings the solution
spaces from the holomorphic gauge to the Vasiliev gauge, defined by the condition 
\bea z^\a \wV^{(\widehat G)}_\a +\mbox{h.c.}\ = \ 
z^+\wV^{(\widehat G)-} - z^-\wV^{(\widehat G)+}  +\mbox{h.c.}\ = \ 0 \ , \label{VFgauge} \eea 
removes singularities in ${\cal T}$ in the internal connection
and restores the manifest Lorentz covariance, broken by the
introduction of the delta sequence \eq{deltasequence}.
\paragraph{Linearized analysis.} At linearized level, the gauge transformation reads
\bea  \wV_\a^{(\widehat G)(1)} \ = \ \wV^{(L)(1)}_\a +\partial_\a \widehat H^{(1)} \ . \label{VFtransf}\eea
Contracting by $z^\a$ and using that, by definition, $Z^{\underline\a} \wV^{(\widehat G)}_{\underline\a}=0$, one finds 
\bea \widehat H^{(1)} \ = \ -\left(\frac{1}{z^\b\partial_\b}z^\a \wV^{(L)(1)}_\a +\mbox{h.c.}\right)\ .\eea
Inserting $\widehat H^{(1)}$ into \eq{VFtransf} and decomposing using the spin-frame, yields
\bea  \wV^{(\widehat G)(1)\pm} \ = \ \wV^{(L)\pm}-\partial^\pm\frac{1}{z^+\partial_+
+z^-\partial_-}(z^+\wV^{(L)(1)-} - z^-\wV^{(L)(1)+}) \ . \label{VFpm}\eea
In the black hole sector, the resulting $t$-integral was analyzed in
\cite{Sundell:2016mxc}, and shown to be a real-analytic function 
on ${\cal C}$ for $r>0$, and to reproduce 
\be \wV^{(\widehat G)(1)}_{\textrm{bh},\a} \ = \ 
z_\a\int_0^1 dt \,t\,\Phi_{\textrm{bh}}(-tz,\bar{y}) \,e^{ity^\a z_\a} \ , \ee
that is, the linearized internal connection obtained by direct integration 
in the Vasiliev gauge, recalling that the linearized Weyl zero-form is the 
same in the $L$ and Vasiliev gauges; see, for example, \cite{more,Sezgin:2002ru}.
At $r=0$ the internal connection is instead singular, as expected. 
Examining the case $\n_n=0$, $\forall n\neq 1$, for concreteness, it reads
\bea \wV^{(\widehat G)(1)}_{\textrm{bh},\a}|_{r=0} \ = \
z_\a\int_0^1 dt \,t\,\Phi_{\textrm{bh}}(-tz,\bar{y})|_{r=0} \,e^{ity^\a z_\a} 
\ , \eea
and, inserting $\Phi_{\textrm{bh}}(-tz,\bar{y})|_{r=0} \ = \ \n_1 (\cP_1\star\k_y)(-tz,\bar{y})=\n_1\delta^2(-tz-i\s_0\yb)$, we get
\bea \wV^{(\widehat G)(1)}_{\textrm{bh},\a}|_{r=0} \ = \
-i\n_1(\s_0\yb)_\a \,e^{-4E}\int_0^1 dt \,\delta^2(tz+i\s_0\yb) 
\ ,\eea
which has singularites on a plane in twistor space.

Turning to the linearized particle sector, upon defining 
\be \widetilde u \ := \ \ty^\a z_\a \ = \ \ty^+z^- - \ty^- z^+\ ,\ee
we compute  
\bea \widehat H^{(1)}_{\textrm{pt}} \ = \ \left.-\frac{ib\tn_1}{4}\,\frac{1-x^2}{1-2i x_0 + x^2}\,
\frac{1}{\ty^+\ty^-}\,\frac{\ty^+z^- + \ty^- z^+}{\widetilde u}\, 
\left(e^{i\widetilde u}-1\right)\right|_{\eta=+1} + \mbox{idem}|_{\eta=-1}\ ,\eea 
which is regular in ${\cal Z}$ but has a pole in ${\cal Y}$.
The appearance of a pole in ${\cal Y}$ in $\widehat H^{(1)}$ can be 
traced back to the fact that the curvature deformation $\Psi$ that 
builds up the internal connection is a $\d$-function in the particle 
sector \eq{tPE}.
It follows that 
\bea \wV^{(\widehat G)(1)\pm} \ = \ -\frac{b\tn_1}{2}\,\frac{1-x^2}{1-2i x_0 + x^2}\,\frac{z^\pm}{\widetilde u}\left[e^{i\widetilde u}-\frac{e^{i\widetilde u}-1}{i\widetilde u}\right]+ 
\mbox{idem}|_{\eta=-1}\ , \eea
which is indeed real-analytic \emph{everywhere} on ${\cal C}$.
Moreover, in the Vasiliev gauge the manifest Lorentz covariance
is restored, and we can write
\bea \wV^{(\widehat G)}_\a &=& -\frac{b\tn_1}{2}\,\frac{1-x^2}{1-2i x_0 + x^2}\,\frac{z_\a}{\widetilde u}\left[e^{i\widetilde u}-\frac{e^{i\widetilde u}-1}{i\widetilde u}\right] 
+ \mbox{idem}|_{\eta=-1} \nn\\[5pt]
&=& z_\a\int_0^1 dt \,t\,\Phi(-tz,\bar{y}) \,e^{ity^\a z_\a} \ , \eea
in agreement with direct integration in the Vasiliev gauge,
just as in the black hole case.

\paragraph{Vasiliev gauge beyond the linearized approximation} 

So far, we have set up a perturbative scheme that provides
solutions in Vasiliev gauge up to first order.
We will first discuss the prospects of
extending the solutions to the Vasiliev gauge to
all order.
Then we will propose a correspondence to an
alternative perturbative scheme based on
normal order on the quasi-local branch of the
theory, and that these two equivalent schemes are reproducing 
the deformed Fronsdal theory on-shell.

The singular nature of $\widehat H^{(1)}_{\textrm{pt}}$ in ${\cal Y}$ 
raises the issue of whether the master fields in the Vasiliev gauge, 
which are elements in $\Omega({\cal T}')$, can be mapped to $\Omega({\cal T})$,
as required by the assumptions made in Section 2 in order
to ensure the existence of invariants based on integrals
over ${\cal B}$ and traces over the extended Weyl algebra ${\cal W}$. 
Two related issues are whether the gauge function $\widehat G$
is large, that is, affects the values of invariants,
and whether it induces redefinitions of the inital 
data for the Weyl zero-form and spacetime one-form in the
$L$-gauge, which may be related to those recently proposed by 
Vasiliev in order to obtain a quasi-local perturbation 
theory in terms of Fronsdal fields \cite{Vasiliev:2016xui}. 

Let us outline the pending steps in somewhat more detail.
We thus start from the initial family $\widehat f'_{\nu,\tilde \nu}$ 
of exact solutions in holomorphic gauge, built from generalized Fock space 
operators using parameters 
$\nu$ and $\tilde \nu$ corrresponding to black hole modes and massless 
particle modes, respectively, which can be given equally
well in Weyl and normal order.
Letting $\widehat S$ denote Kontsevich gauge transformations,
which act on horizontal sections of ${\cal C}$,
and $\widehat f$ the families of master fields making up the
exact solution spaces, we would like to establish the following
sequence of \emph{large Kontsevich gauge transformations}:
\be \widehat f'_{\nu,\tilde\nu} \stackrel{S_0\circ S_L}{\longrightarrow} \widehat f^{(L)}_{\nu,\tilde\nu}
\stackrel{S_{H_\mu}}{\longrightarrow} \widehat f^{(\widehat G)}_{\nu,\tilde\nu;\mu}|_{\Omega({\cal T}')}\stackrel{(S_0)^{-1}}{\longrightarrow} 
 \widehat f^{(\widehat G)}_{\nu,\tilde\nu;\mu}|_{\Omega({\cal T})}
\ ,\ee
where
\begin{itemize}
\item[] $S_L$ implements the gauge transformation with
gauge function $L:{\cal X}\rightarrow G$;
\item[] $S_0$ denotes the map from Weyl to normal order;
\item[] $S_{H_\mu}$ implements the gauge transformation 
that bring the solution to Vasiliev's gauge, which fixes 
the gauge function $H_\mu$ up to a homogenous solution 
parametrized by a large gauge function $M_\mu:{\cal X}\rightarrow G$.
\end{itemize}
If the map exists, the final configuration $\widehat f^{(\widehat G)}_{\nu,\tilde\nu;\mu}|_{\Omega({\cal T})}$
is thus a set of forms on ${\cal B}$ valued in ${\cal W}$, for which
we can compute an on-shell action as a functional of asymptotic 
data and invariant quantities.
Thus, denoting the on-shell action by $S[\widehat f]=\int_{\cal B}{\rm Tr}_{\cal W} \widehat{\cal L}(\widehat f)$, 
where $\widehat f$ collectively denotes the master fields, we have
\be S[\widehat f^{(\widehat G)}_{\nu,\tilde\nu;\mu}|_{\Omega({\cal T})}]=
\int_{\cal B}{\rm Tr}_{\cal W} (\widehat {\cal L}\circ (S_0)^{-1}\circ S_{H_\mu}
\circ S_0\circ S_L)(\widehat f'_{\nu,\tilde\nu})\ .\ee
We note that the moduli (boundary states) entering via
$M_\mu$ can be transferred from $H_\mu$ to $L$ by defining
$\check H_\mu=(M_\mu)^{-1}\star H_\mu$ and $L_\mu:=L\star M_\mu$,
which is thus a gauge function including a boundary state.

As for the existence of the map, a nontrivial compatibility 
condition arises already at the second order as follows: 
$\widehat H^{(1)}$ induces a second order correction 
$\widehat \Phi^{(2)}$ to the Weyl zero-form, whose
$Z$-dependent piece cannot be corrected using any 
second-order initial data for $\Phi'$.
Thus, the $Z$-dependent part of $(S_0)^{-1}(\widehat \Phi^{(\widehat G)(2)})$
must be an element of $\Omega({\cal T})$, that is, upon 
expanding it in the basis of ${\cal W}$, the resulting
component fields should be elements of $L^1({\cal Z})$.
The simplest realization of this condition would be a 
Weyl zero-form that approaches a constant value at 
the point at infinity of ${\cal Z}$, but more generally
it may turn out to be necessary to allow integrable
divergencies.
Proceeding to higher orders, it would also be interesting to examine
to what extent the compatibility may actually fix the precise form of
the Vasiliev gauge condition beyond leading order; if so, then this 
would provide an intrinsic method for fixing this apparent ambiguity 
currently plaguing the theory.


\scs{Conclusions and Outlook}


In what follows, we first summarize our results, after
which we turn to omitted details, physical interpretations
and prospects for future research.

Extending the methods of \cite{us,us2}, in this paper we have obtained 
a solution space to Vasiliev's four-dimensional 
higher spin gravity by superposing spherically-symmetric black hole
and scalar particle modes.
Interestingly enough, in the gauges we use we can observe that the
scalar modes give rise to a backreaction in the form of black holes modes, that arises
already at the second of classical perturbation theory.
This effect, unusual from the point of view of ordinary gravitational theories, 
can be interpreted as a consequence of the non-locality of the vertices extracted 
from Vasiliev's equations at every perturbative order.

Our solution method combines large gauge functions 
with the Ansatz in Eqs. \eq{ansatz1}--\eq{ansatz3}, 
based on separation of twistor space variables and 
auxiliary integral presentations of intial data. More precisely, our construction involves:
\begin{itemize}
\item[i)] The large gauge function $\widehat G=L\star \widehat H$,
where $L$ creates the asymptotic anti-de Sitter region, 
and $\widehat H$ restores Vasiliev's gauge, in which the 
asymptotic field configurations consist of unfolded (free) 
Fronsdal fields.
We have determined the linearized contribution 
$\widehat H^{(1)}$, which is regular on twistor 
space away from the origin for black hole modes, 
and singular in the twistor fiber space for particle modes. 
We have also studied the field configurations to all orders 
in the $L$-gauge, reached from the holomorphic gauge via $L$ 
alone, and in a Kruskal-like gauge, 
reached via a further local $SL(2,\Comp)$-rotation of the $Z$ oscillators. 
In such gauges the Weyl zero-form is first-order exact, and the
interpretation of its content in terms of black-hole and particle
modes, barring all the subtleties mentioned in the Introduction, more transparent. 
\item[ii)] Prior to switching on the gauge function, 
the field configurations are localized to twistor space
in a holomorphic gauge, where they consist of an undeformed 
Weyl zero-form and a deformed twistor space connection,
obeying a deformed oscillator algebra whose 
enveloping algebra is a subalgebra of the star 
product algebra $\widehat {\cal A}'$ in \eq{cA'}.
\item[iii)] The algebra 
$\widehat {\cal A}'$ is expanded over an algebra of projectors 
and twisted projectors in supersingleton spaces realized 
as functions on ${\cal Y}$ using a regular presentation 
that implements the normalization of states in compact weight spaces.
\item[iv)] In the holomorphic gauge, the Weyl zero-form is 
constant on ${\cal Z}$, while the deformed oscillator is a 
distribution on ${\cal Z}$ presented via a generalized 
Laplace transformation, whose exterior derivative 
reproduces the inner Klein operator in the two-form curvature. 
\end{itemize}
The construction above remains to be completed in three
main respects:
\begin{itemize}
\item[a)] Extension as to include general particle 
and black hole modes in the deformed oscillator algebra,
to which we see no obstacles.
\item[b)] Computation of the gauge fields to first order,
and verification of the central on-mass-shell theorem. 
\item[c)] Computation of $\widehat H$ and the spacetime
gauge fields to higher orders, which requires an extension
of Vasiliev's gauge, and the study of whether there exists
a quasi-local branch of moduli space in which the spacetime 
gauge fields remain smooth at the origin for nontrivial 
particle modes. 
\end{itemize}
More generally, we would like to apply the Ansatz 
to more general initial data, such as for example
the generalized projectors built via the different 
inequivalent choices of the Cartan generators of 
$\msp(4;\Comp)$ listed in \cite{us}, which
will give rise to more general Type-D solution
spaces with fluctuations. 

\vspace{0.7cm}

Concerning the interpretations of our results,
and future directions, a number of remarks are in order:

\paragraph{Black hole microstates.}
It was observed in 
\cite{Iazeolla:2008ix} that the Flato--Fronsdal theorem,
which states that the massless particle spectrum
is contained in the direct product of two
supersingletons, has a natural generalization to the
direct product of a supersingleton and an anti-supersingleton,
which decomposes under $\mso(2,3)$ into compact
weights filling up the adjoint representation
of the higher spin algebra.
We would like to associate the latter states
to the black hole sector.
To this end, we begin by recalling from 
\cite{Iazeolla:2008ix} that reflection of 
one of the two spaces in the direct product 
to the dual space, turns these two representations 
into operators on which the higher spin algebra 
acts in the twisted adjoint fashion, that is,
one should think of them as being terms in a 
twisted-adjoint zero-form.
Moreover, 
both of these representations admit two dual 
pairs of bases, with Lorentz covariant and
compact bases, connected by means of harmonic expansion.
In the former basis, all basis elements are operators 
that are polynomial in ${\cal Y}$. 
Thus, the aforementioned association requires that 
in the free theory limit, the generalized Type D 
sector of a Fronsdal field of Lorentz spin $s$ 
contains the (complexified) $\mso(2,3)$ irrep with 
highest weight $(s-1,s-1)$;  see \cite{Didenko:2008va}.
Turning to the compact bases, indeed, the massless 
spectrum remains spanned by basis vectors that are 
real-analytic, with reference state given by the 
Gaussian element ${\cal P}_1$, while the reference 
state of the adjoint representation is given by the 
twisted projector 
$\widetilde{\cal P}_1\sim \delta^2(y-i\s_0 \bar y)$.

We would like to stress that the particle and black
hole spaces are isomorphic as complex vector spaces; 
indeed, the isomorphism is implemented by means of 
one-sided star product multiplication with the inner 
Klein operator $\kappa_y$, that is, by a twistor space 
Fourier transform.
However, as far as the real structure as higher spin
modules is concerned, there is a dichotomy, as it must 
be imposed using the twisted-adjoint reality condition.
It follows that particle states are realized in terms of 
complex operators, whose complex conjugates are
the realizations of anti-particles, while the 
black hole states form real vector spaces;
this can be made more manifest by looking at the
adjoint $\Psi$ fields, which are hermitian in the 
black hole sector and twisted-hermitian in the
particle sector. 
Thus, the particle sector is a complex vector space with
a positive definite sesquilinear form (which one can realize
using a suitable extension of the supertrace
operation to the extended Weyl algebra \cite{FCS}), 
\emph{i.e.} a Hilbert space, while the black hole sector
is real vector space with positive definite
bilinear form.
The correspondence between star product algebras 
and free conformal fields \cite{Colombo:2012jx,DidenkoSkvortsov,MishaOlga}
suggests that the $\mathbb Z_2$ symmetry between the 
modules embedded into the oscillator algebra has
a holographic counterpart, whereby the free conformal 
field theory at the boundary of anti-de Sitter spacetime, 
with its Hilbert space structure, can be mapped
by a $\mathbb Z_2$ transformation to another theory 
that we propose to associate to the black holes (or, rather, to the black-hole microstates, as discussed in the Introduction), 
with its Euclidean structure.

Having obtained a classical moduli space consisting of
particle and black hole modes, it would be desirable
to provide it with a free energy (or on-shell action) 
expressible as a higher-spin gauge-invariant functional 
of the particle and black-hole deformation parameters.
From it, a number of physically interesting quantities 
can be obtained, such as holographic correlation
functions obtained from multi-particle states, and 
derivation of thermodynamical relations 
for black holes from various ensembles constructed 
by including off-diagonal operators into $\Phi_{\rm bh}$ 
and multi-black hole solutions (for which we have reason 
to expect that there exist a ``dilute gas'' approximation 
at large distances \cite{Colombo:2010fu,us}).
We recall that, as noted above, the gauge invariant characterization 
of single black hole microstates is in terms of the parameters
$\nu_n$, \emph{i.e.} the eigenvalues
of $\Psi_{\rm bh}$, which are indeed independent variables.
On the other hand, the asymptotic charges ${\cal M}_s$
only form independent variables in the free theory,
\emph{i.e.} the physical characterizations of black hole 
microstates in terms of individual (linearized) Weyl 
tensors is reliable only asymptotically.

\paragraph{Large nature of $\widehat H$.}

In terms of the gauge field equations, the role of $\widehat H^{(1)}_{\rm bh}$ is essentially
to rotate the spin-$s$ fields from the $L$-gauge to the Fronsdal frame. At the level of the
Fronsdal fields, this amounts to a generalized Weyl rescaling that mixes Fronsdal tensors
of different ranks and gradients of the physical scalar.
In other words, in the black hole sector, we can interpret both the $L$-gauge
and the Vasiliev gauge as two different frames for Fronsdal fields, of which the
Vasiliev frame is preferred as it is diagonalized.

In the particle sector, on the other hand, one simply cannot expand the master fields 
around $Y=0$, since there are inverse powers of $Y$ in the internal connection
and hence in the generating function for the gauge fields.
In other words, in the particle sector the linearized transformation from
$L$-gauge to Vasiliev gauge has no interpretation in terms of the original
Fronsdal field content.
Instead, it would be interesting to introduce an enlarged class of functions on $Y$-space,
and interpret the equations of motion in $L$-gauge directly in terms of the
corresponding enlarged set of component fields, thus containing the original 
set of Fronsdal fields as a subset.

The identification of the asymptotic charges requires the large
transformation connecting the holomorphic gauge to the standard 
Vasiliev gauge.
At first order, we have found these transformations in this paper.
Interestingly, in the particle sector, the gauge function 
$\widehat H^{(1)}$ inherits a singularity in ${\cal Y}$
from the twistor space connection, while the master fields 
in Vasiliev's gauge are completely regular. 
The origin of the singularity in $\widehat H^{(1)}$ is intimately 
related to the holomorphic gauge we start from, in which the two-form curvature
has a $\delta$-function source in twistor space. 
It is expected that, extending Vasiliev's gauge to higher orders, $\widehat H$ 
induce a redefinition of the initial data 
for the zero-form and for the spacetime one-form, to be compared 
with those recently proposed by Vasiliev in order to rewrite 
the non-linearities encoded in the equations in terms of 
current interactions \cite{Vasiliev:2016xui} (see also \cite{Sezgin:2017jgm,Didenko:2017lsn} for recent holographic tests of the local second-order interaction terms found in \cite{Vasiliev:2016xui}). 
We plan to assess how large the gauge transformation 
$\widehat H$ is, \emph{i.e.}, how far the $L$- and Vasiliev gauge are from
each other, by examining its effect on observables.
As a first step, it is natural to examine zero-form
charges, that is, decorated open Wilson lines in ${\cal Z}$ \cite{Bonezzi:2017vha},
which are known to contain higher spin amplitudes in their
leading order.

\paragraph{Why the Vasiliev gauge?}

We have obtained different solution spaces to classical equation of
motion in different gauges that
are presumably far from each other.
Each solution space may have its own set of classical observables 
(finite higher spin invariants);
for example, the holomorphic gauge supports zero-form charges, but 
gives trivial results for observables that involve spacetime gauge fields.
Thinking of the classical theory as a saddle point of a path integral
adds one more condition, namely the finiteness of a classical action
on-shell, which can then be interpreted as a free energy given as
a function of the various classical observables in question.
If the quantum theory in question is gravitational, then we
have reason to expect on physical grounds that there exists
a holographic dual in the form of a conformal field theory
that has been deformed by operators with sources dual to the
boundary conditions of the bulk theory.
Thus, if the perturbative expansion of the conformal theory 
in these sources is an expansion around a free theory with
higher spin symmetries, then the sources may correspond
to Fronsdal fields in the bulk.
Therefore, the existence of such a free energy functional would
single out the Vasiliev gauge, as it is required for Fronsdal
fields asymptotically.
As for the higher spin black holes, it is within this version
of the theory that we envisage a possible interpretation of them 
as gravitational fuzzballs, as discussed in the Introduction. 
We hope to make this proposal more concrete in a future publication. 

As for the $L$-gauge and other gauges that one may surely think of 
as well, they may describe field configurations on correspondence 
space geometries that are no longer having the topology of asymptotically 
anti-de Sitter spacetime (as is the case for the Kruskal-like gauge),
and/or whose component field descriptions go beyond Fronsdal theory 
(as for the $L$-gauge, for instance), in the sense discussed above.
Moreover, they may admit their proper set of classical observables.
It is reasonable to expect that there may also exist suitable
action principle and related free energy functionals, lending these
solution spaces their physical interpretations.
Finally, as for the holomorphic gauge, its physical interpretation
may instead be in terms of various deformations of symplectic 
manifolds.
We shall comment on generalized correspondence space geometries
and related extensions of Vasiliev's equations towards the end.

\paragraph{Dual boundary conditions in Weyl and normal order.}

The boundary value formulation that we are trying to implement
involves dual boundary conditions, connecting field configurations with finite star product and traces defined in Weyl order with corresponding ones that are real analytic at the origin of twistor space in normal order.
A key check is whether the second order correction to the Weyl
zero-form can be mapped back to Weyl order and expanded over the
assumed class of fiber functions and zero-forms on the base manifold.
This is a form of litmus test, since there is not ambiguity in the
Vasiliev gauge at leading order, which means that there is no ambiguity
in the $Z$-dependent piece of $\widehat \Phi^{(2)}$ in the Vasiliev gauge.
Upon going to Weyl order, the latter should belong to $L^1({\cal Z})$, 
but it need not be real analytic on $\cal Z$.

\paragraph{Comparison with quasi-local Fronsdal branch.}

In a similar fashion, starting from our solutions, it would be very 
interesting to see whether one could use the gauge freedom 
to eat up the black hole modes at every order in perturbation theory, 
while imposing the dual boundary conditions in Weyl and normal order,
thereby being left 
with fully non-linear field configurations expressed only 
in terms of Fronsdal fields, that could be more directly 
compared with the non-linear results on higher spin gravity 
derived from holography \cite{Bekaert:2014cea,Bekaert:2015tva,Sleight:2016dba}.

\paragraph{Alternative perturbative scheme in normal order.}
We have presented so far a hybrid perturbative approach,
combining Weyl and normal order and relying on the trace operation
on the extended Weyl algebra.
However, there also exists a perturbative approach 
formulated on-shell entirely within normal order \cite{Vasiliev:2015wma}
and relying on a different type of (super)trace 
operations for computing invariants \cite{Vasiliev:2015mka}.
The main differences between the two approachs are as follows:
a) Concerning homotopy integrals, these arise in the hybrid 
approach in obtaning perturvative expansion for $\widehat H$,
while they arise in the normal-ordered approach in obtaining
the perturbative expansion of the master fields; 
b) Concerning the treatment of star products of operators in the
Fock and anti-Fock spaces, these need to be regularized using the
regular prescription following the hybrid approach, whereas it has
been proposed in \cite{Vasiliev:2015mka} that they are regularized 
by the homotopy integrals following the normal-ordered approach.
We expect that the two approaches lead to master fields that look
different as distributions in twistor space, but that they may
nonetheless agree at the level of amplitides, as they rely
on different trace operations in twistor space.

\paragraph{Comparison of Weyl- and normal-ordered approaches.}

Our method starts from a factorized internal Ansatz (which is tantamount to starting from total Weyl ordering in twistor space) in order to highlight the generalized projector algebras of special physical meaning that lie at the heart of our construction, and arrives at normal-ordered quantities after taking the star product explicitly, thereby mixing the $Y$ and $Z$ dependence. 
This procedure can be contrasted with the aforementioned 
normal-ordered approach to solving Vasiliev's equations in the sector of particle
and black hole modes (see for example \cite{Didenko:2009td} for the latter), which is already large
enough to provide a physically nontrival 
comparison.
The precise relation between these two approaches -- which in general also needs to take into account that
the gauge-function method introduces other non-polynomial 
functions of the oscillators from the start -- is also related 
to the problem of determining a proper restriction on the allowed 
class of functions, gauge transformations and field redefinitions 
\cite{Vasiliev:2015wma,Vasiliev:2015mka, Vasiliev:2016xui,Boulanger:2015ova,Skvortsov:2015lja}.
We expect there to be a correspondence between the two
approaches only at the level of dual observables,
including the on-shell action, but not necessarily 
at the level of the master fields --- similarly to what happens evolving
classical fields along either equal time slices
or light-fronts.

\paragraph{Generalized higher spin geometries.}

The higher spin black holes are constructed on correspondence spaces 
with base manifolds of a definite topology using definite classes of 
fiber functions given by operators in Fock spaces (the Hilbert space
of the harmonic oscillator) including the inner Klein operators introduced
via the (non-dynamical) two-form $\widehat J$.
As we have seen, there is an intricate interplay between the fiber algebra
and the singularity structure on the base manifold.
This suggests a higher spin landscape based on master field equations 
containing a dynamical two-form that can be formulated on correspondence
spaces of more general topology and with more general noncommutative
structure.
Dynamical two-forms have been introduced using three-graded
internal Frobenius algebras in \cite{FCS}, leading to a more
general class of Frobenius--Chern--Simons
theories \cite{Bonezzi:2016ttk}, that in their turn can be embedded into 
a broader class of models based on homotopy associative algebras.
As for the noncommutative structure of the base manifold, it
arises naturally from the quantization of two-dimensional topological
sigma models, with target spaces being differential Poisson manifolds that in their
turn are special instances of homotopy Poisson manifolds. 
In particular, in \cite{Arias:2016agc} a class of differential
Poisson sigma models with extended supersymmetries generated by the
Kiling vectors spanning the fibers of the correspondence space
has been constructed and proposed as a first quantized description 
of the Frobenius--Chern--Simons.
The fermionic zero-modes of the sigma model generates a closed and central
top-form on the fiber space that can be used to provide a more general
definition of master fields useful for a globally defined formulation
of Frobenius--Chern--Simons theory.
We would like to stress that in the landscape of model that we
envisage we expect new models whose Weyl zero-forms lend themselves
to physical interpretations as density matrices of various quantum mechanical 
systems obeying nonlinear equtions of motion that one may propose as
natural deformations of the von Neumann--Liouville equations derived
from the Schr\"odinger equation.
It would be interesting to make this relation clearer and
to investigate whether there exists a more general framework for
nonlinear quantum mechanics based on underlying topological
field theories.

\vspace{2cm}

\paragraph{Acknowledgments}

We are grateful to R. Aros, R. Bonezzi, N. Boulanger, D. De Filippi, V.E. Didenko, J. Raeymaekers, E. Sezgin, 
E.D. Skvortsov, C. Sleight, M. Taronna, M.A. Vasiliev and Y. Yin for stimulating discussions. 
The research of C.~I. was supported by the funding from 
Grant Agency of Czech Republic, project number
P201-12-G028, and by the Russian Science Foundation grant 
14-42-00047 in association with the Lebedev Physical Institute 
in Moscow, that C.~I. wishes to thank for kind hospitality 
at various stages of this work. 
The work of P. S. is supported by Fondecyt iRegular grant N$^{\rm o}$ 
1140296, Conicyt grant DPI 20140115 and UNAB internal grant DI-1382-16/R.

\vspace{2cm}

\vspace{2cm}

\begin{appendix}


\scs{Spinor conventions and $AdS_4$ Background}\label{App:conv}


We use the conventions of \cite{Iazeolla:2008ix} in which $SO(2,3)$ generators $M_{AB}$ with $A,B=0,1,2,3,0'$ obey
\be [M_{AB},M_{CD}]\ =\ 4i\y_{[C|[B}M_{A]|D]}\ ,\qquad
(M_{AB})^\dagger\ =\ M_{AB}\ ,\label{sogena}\ee
which can be decomposed using $\eta_{AB}~=~(\eta_{ab};-1)$ with $a,b=0,1,2,3$ as
\be [M_{ab},M_{cd}]_\star\ =\ 4i\y_{[c|[b}M_{a]|d]}\ ,\qquad
[M_{ab},P_c]_\star\ =\ 2i\y_{c[b}P_{a]}\ ,\qquad [P_a,P_b]_\star\ =\
i\lambda^2 M_{ab}\ ,\label{sogenb}\ee
where $M_{ab}$ generate the Lorentz subalgebra $\mso(1,3)$, and $P_a=\l M_{0'a}$ with $\l$ being the inverse $AdS_4$ radius related to the cosmological constant via $\L=-3\l^2$. Decomposing further under the maximal compact subalgebra, the $AdS_4$ energy generator $E=P_0=\l M_{0'0}$ and the spatial $\mso(3)$ rotations are generated by $M_{rs}$ with $r,s=1,2,3$.
In terms of the oscillators $Y_{\underline\a}=(y_\a,\yb_{\ad})$ defined in \eq{oscillators}, their realization is taken to be
\be M_{AB}~=~ -\ft18  (\C_{AB})_{\underline{\a\b}}\,Y^{\underline\a}\star Y^{\underline\b}\ ,\label{MAB}\ee
 \be
 M_{ab}\ =\ -\frac18 \left[~ (\s_{ab})^{\a\b}y_\a\star y_\b+
 (\sb_{ab})^{\ad\bd}\bar y_{\ad}\star \yb_{\bd}~\right]\ ,\qquad P_{a}\ =\
 \frac{\l}4 (\s_a)^{\a\bd}y_\a \star \yb_{\bd}\ ,\label{mab}
 \ee
using Dirac matrices obeying $(\C_A)_{\underline\a}{}^{\underline\b}(\C_B C)_{\underline{\b\c}}=
\eta_{AB}C_{\underline{\a\c}}+(\C_{AB} C)_{\underline{\a\c}}$, and van der Waerden symbols obeying
 \be
  (\s^{a})_{\a}{}^{\ad}(\sb^{b})_{\ad}{}^{\b}~=~ \y^{ab}\d_{\a}^{\b}\
 +\ (\s^{ab})_{\a}{}^{\b} \ ,\qquad
 (\sb^{a})_{\ad}{}^{\a}(\s^{b})_{\a}{}^{\bd}~=~\y^{ab}\d^{\bd}_{\ad}\
 +\ (\sb^{ab})_{\ad}{}^{\bd} \ ,\label{so4a}\ee\be
 \ft12 \e_{abcd}(\s^{cd})_{\a\b}~=~ i (\s_{ab})_{\a\b}\ ,\qquad \ft12
 \e_{abcd}(\sb^{cd})_{\ad\bd}~=~ -i (\sb_{ab})_{\ad\bd}\ ,\label{so4b}
\ee
\be ((\s^a)_{\a\bd})^\dagger~=~
(\sb^a)_{\ad\b} ~=~ (\s^a)_{\b\ad} \ , \qquad ((\s^{ab})_{\a\b})^\dagger\ =\ (\sb^{ab})_{\ad\bd} \ .\ee
and raising and lowering spinor indices according to the
conventions $A^\a=\epsilon^{\a\b}A_\b$ and $A_\a=A^\b\epsilon_{\b\a}$ where
\be \e^{\a\b}\e_{\c\d} \ = \ 2 \d^{\a\b}_{\c\d} \ , \qquad
\e^{\a\b}\e_{\a\c} \ = \ \d^\b_\c \ ,\qquad (\e_{\a\b})^\dagger \ = \ \e_{\ad\bd} \ .\ee
The $\mso(2,3)$-valued connection
 \be
  \O~:=~-i \left(\frac12 \omega^{ab} M_{ab}+e^a P_a\right) ~:=~ \frac1{2i}
 \left(\frac12 \omega^{\a\b}~y_\a \star y_\b
 +  e^{\a\dot\b}~y_\a \star {\bar y}_{\dot\b}+\frac12 \bar{\omega}^{\dot\a\dot\b}~{\bar y}_{\dot\a}\star {\bar y}_{\dot\b}\right)\
 ,\label{Omega}
 \ee
  \be
 \o^{\a\b}~=~ -\ft14(\s_{ab})^{\a\b}~\o^{ab}\ , \qquad \omega_{ab}~=~\ft12\left( (\s_{ab})^{\a\b} \o_{\a\b}+(\bar\s_{ab})^{\ad\bd} \bar\o_{\ad\bd}\right)\ ,\ee
 \be e^{\a\dot\a}~=~ \ft{\lambda}2(\s_{a})^{\a \dot\a}~e^{a}\ , \qquad e_a~=~ -\l^{-1} (\s_a)^{\a\ad} e_{\a\ad}\ ,\label{convert}\ee
and field strength
\be {\cal R}~:=~ d\O+\O\star \O~:=~-i \left(\frac12 {\cal R}^{ab}M_{ab}+{\cal R}^a P_a\right) ~:=~ \frac1{2i}
 \left(\frac12 {\cal R}^{\a\b}~y_\a \star y_\b
 +  {\cal R}^{\a\dot\b}~y_\a \star {\bar y}_{\dot\b}+\frac12 \bar{\cal R}^{\dot\a\dot\b}~{\bar y}_{\dot\a}\star {\bar y}_{\dot\b}\right)\
 ,\label{calRdef}\ee
\be
 {\cal R}^{\a\b}\ =\ -\ft14(\s_{ab})^{\a\b}~{\cal R}^{ab}\ ,
 \qquad {\cal R}_{ab}~=~\ft12\left( (\s_{ab})^{\a\b} {\cal R}_{\a\b}+(\bar\s_{ab})^{\ad\bd} \bar{\cal R}_{\ad\bd}\right)\ ,\ee
 \be
 {\cal R}^{\a\dot\a}\ =\ \ft{\lambda}2(\s_{a})^{\a \dot\a}~{\cal R}^{a}\ ,
 \qquad {\cal R}_a~=~ -\l^{-1} (\s_a)^{\a\ad} {\cal R}_{\a\ad}\ .\ee
In these conventions, it follows that
 \be
 {\cal R}_{\a\b}~=~ d\o_{\a\b} -\o_{\a}^{\c}\o_{\c\b}-
 e_{\a}^{\cd}\bar e_{\cd\b}\ ,\qquad
 {\cal R}_{\a\dot\b}~=~  de_{\a\bd}+ \o_{\a\c}\wedge
 e^{\c}{}_{\bd}+\bar{\o}_{\bd\dd}\wedge e_{\a}{}^{\dd}\
 ,\ee\be
 {\cal R}^{ab}~=~ R_{ab}+\lambda^2
 e^a\wedge e^b\ ,\qquad R_{ab}~:=~d\o^{ab}+\o^a{}_c\wedge\o^{cb}\ ,\ee\be
 {\cal R}^a~=~ T^a ~:=~d e^a+\o^a{}_b\wedge e^b\ ,
 \label{curvcomp} \ee
where $R_{ab}:=\frac12 e^c e^d R_{cd,ab}$ and $T_a:=e^b e^c T^a_{bc}$ are the Riemann and torsion two-forms.
The metric $g_{\mu\nu}:=e^a_\mu e^b_{\nu}\eta_{ab}$. The $AdS_4$ vacuum solution $\O_{(0)}=e_{(0)}+\o_{(0)}$ obeying $d\O_{(0)}+\O_{(0)}\star\O_{(0)}=0$, with Riemann tensor $ R_{(0)\m\n,\r\s}=
 -\lambda^2 \left( g_{(0)\mu\rho} g_{(0)\nu\sigma}-
  g_{(0)\nu\rho} g_{(0)\mu\sigma} \right)$ and vanishing torsion, can be expressed as $\O_{(0)}=L^{-1}\star dL$ where the gauge function $L\in SO(2,3)/SO(1,3)$. The stereographic coordinates $x^\mu$ of Eq. \eq{stereo}, are related to the coordinates $X^A$ of the five-dimensional embedding space with metric
$ds^2  =dX^A dX^B\eta_{AB}$,
in which $AdS_4$ is embedded as the hyperboloid
$X^A X^B \eta_{AB}=  -\ft1{\l^2} $,
as
\bea x^\m \ = \ \frac{X^\m}{1+\sqrt{1+\l^2 X^\m X_\m}} \ , \qquad X^\m \ = \ \frac{2x^\m}{1-\l^2 x^2}\ , \qquad \m \ = \ 0,1,2,3 \ .\label{A.15}\eea
%
The global spherical coordinates $(t,r,\theta,\phi)$ in which the metric reads
\bea  ds^2 \ = \ -(1+\l^2r^2)dt^2+\frac{dr^2}{1+\l^2 r^2}+r^2(d\theta^2+\sin^2\theta d\phi^2) \ ,\label{metricglob}\eea
are related locally to the embedding coordinates by
\bea & X_0 \ = \ \sqrt{\l^{-2}+r^2}\sin t \ , \qquad X_{0'} \ = \ \sqrt{\l^{-2}+r^2}\cos t \ , & \nn\\[5pt]
& X_1 \ = \ r\sin\theta\cos\phi \ , \quad  X_2 \ = \ r\sin\theta\sin\phi \ , \quad X_3 \ = \ r\cos\theta \ ,& \label{AdSspherical}\eea
providing a one-to-one map if $t\in [0,2\pi)$, $r\in[0,\infty)$, $\theta\in[0,\pi]$ and $\phi\in[0,2\pi)$ defining the single cover of $AdS_4$. This manifold can be covered by two sets of stereographic coordinates, $x^\mu_{(i)}$, $i=N,S$, related by the inversion $x^\m_N = -x^\m_S/(\l x_S)^2$ in the overlap region $\l^2 (x_N)^2, \l^2 (x_S)^2  <  0$, and the transition function $T_N^S=(L_N)^{-1}\star L_S\in SO(1,3)$. The map $x^\mu \rightarrow -x^\m/(\l x)^2$ leaves the metric invariant, maps the future and past time-like cones into themselves and exchanges the two space-like regions $0<\l^2 x^2< 1$ and $\l^2 x^2 > 1$ while leaving the boundary $\l^2 x^2 =1$ fixed. It follows that the single cover of $AdS_4$ is formally covered by taking $x^\mu\in \Real^{1,3}$.

Petrov's invariant classification of spin-2 Weyl tensors \cite{Petrov:2000bs,PenroseRindler} is based on their algebraic properties at any spacetime point. Generalized to the higher spin context and by making use of spinor language, it amounts to study the roots of the degree-$2s$ polynomial $\O(\zeta):=C_{\a(2s)}\z^{\a_1}\ldots\z^{\a_{2s}}$, where $C_{\a(2s)}\equiv C_{\a_1\a_2\ldots\a_{2s}}=C_{(\a_1\a_2\ldots\a_{2s})}$ is the self-dual part of the Weyl tensor and $\z^{\a}$ an arbitrary non-vanishing two-component spinor. Factorizing the polynomial in terms of its roots defines a set of $2s$ spinors which one refers to as \emph{principal spinors},  \emph{viz.} $\O(\zeta)=u^1_{\a_1}\z^{\a_1}\ldots u^{2s}_{\a_{2s}}\z^{\a_{2s}}$, so $C_{\a(2s)}=u^1_{(\a_1}\ldots u^{2s}_{\a_{2s})}$. If $\O(\z)$ has multiple roots, the corresponding principal spinors are collinear. The classification then amounts to distinguish how many different roots $\O(\z)$ has,\emph{i.e.}, how many non-collinear principal spinors enter the factorization of the spin-$s$ Weyl tensor. Clearly, this classification can be given in terms of the partitions $\{p_1,...,p_k\}$ ($k\leq 2s$) of $2s$ in integers obeying $p_1+p_2+...+p_k=2s$ and $p_i\geqslant p_{i+1}$. In the spin-2 case, this singles out the familiar six different possibilities: $\{1,1,1,1\}$ (type I in Petrov's original terminology); $\{2,1,1\}$ (type II); $\{2,2\}$ (type D); $\{3,1\}$ (type III); and $\{4\}$ (type N) plus the trivial case of a vanishing Weyl tensor (type O). The Type D case is related to gravitational field configurations surrounding isolated massive objects; for arbitrary spin-$s$, we refer to the type $\{s,s\}$ as \emph{generalized type D}.


\scs{Properties of inner Klein operators}\label{App:A}


Working with the chiral integration domain ${\cal R}_{\mathbb R}$, 
it makes sense to define the following complex analytic delta functions 
($M_{\a}^{\b}\in GL(2;\mathbb C)$):
\bea \d^2(y)&:=&  \d(y_1)\d(y_2)\ ,\qquad \d^2(My)\ =\ \frac{1}{\det M} \d^2(y) \label{deltay}\\[5pt]
\d^2(z)&:=&  \d(z_1)\d(z_2)\ ,\qquad \d^2(Mz)\ =\ \frac{1}{\det M} \d^2(z)\ .\label{deltaz}\eea
Their hermitian conjugates are defined by
$\d^2(\bar y)= (\d^2(y))^\dagger$ and $\d^2(\bar z)=
(\d^2(z))^\dagger$.
By splitting $y_\a$ and $z_\a$ into a complexified Heisenberg algebras
\bea [\eta^-,\eta^+]_\star&=&1\ ,\qquad \eta^\pm~=~v^{\pm\a}  y_\a\ ,\qquad v^{+\a} v^-_\a~=~-\frac{i}2\ ,\\[5pt]
[\zeta^-,\zeta^+]_\star&=&1\ ,\qquad \zeta^\pm~=~v^{\pm\a} z_\a\ ,\eea
one can define idempotent inner Kleinian operators
\bea \k_y&:=&(-1)_\star ^{N_y}\ ,\qquad N_y~:=~ \eta^+\star \eta^-\ ,\\[5pt]
\k_z&:=&(-1)_\star ^{N_z}\ ,\qquad N_z~:=~ \zeta^+\star \zeta^-\ ,\eea
using the notation ($c\in\Comp$)
\bea c_\star^{\widehat P}&=&\exp_\star (\widehat P\log c)\ ,\qquad \exp_\star \widehat P\ =\ \sum_{n=0}^\infty \frac1{n!} \widehat P^{\star n}\ ,\qquad \widehat P^{\star n}\ =\ \underbrace{\widehat P\star\cdots\star\widehat P}_{\tiny \mbox{$n$ times}}\ ,\eea
and representing $(-1)_\star ^{N_y}$ as
\bea (-1)_\star ^{N_y}&=& \lim_{\e\rightarrow 0} \exp_\star(i(\pi+\e)N_y)
\ ,\eea
\emph{idem} $(-1)_\star ^{N_z}$. 
The broken $SL(2,\Comp)$-invariance is restored in the limit $\e\rightarrow 0$ 
in Weyl order, 
\emph{viz.}
\bea \k_y&=& 2\pi\delta^2(y)\ ,\qquad \k_z~=~2\pi\delta^2(z)\ .\eea
We also define
\bea \bar\k_{\yb}&:=&(\k_y)^\dagger\ =\ (-1)^{\bar N_\yb}_\star~=~ 
2\pi\delta^2(\yb)\ ,\\[5pt]
\bar\k_{\zb}&:=&(\k_z)^\dagger\ =\ (-1)^{\bar N_\zb}_\star~=~ 2\pi\delta^2(\zb)
\ ,\eea
using
\bea \bar N_{\yb}&:=& (N_y)^\dagger~=~\bar\eta^+\star \bar\eta^-\ ,\qquad \bar\eta^\pm~:=~(\eta^\mp)^\dagger~=~\bar v^{\pm\ad} \yb_{\ad}\ ,\qquad \bar v^\pm_{\ad}~:=~(v^\mp_\a)^\dagger\ ,\\[5pt]
\bar N_{\zb}&:=& (N_z)^\dagger~=~\bar\zeta^+\star \bar\zeta^-\ ,\qquad \bar\zeta^\pm~:=~(\zeta^\mp)^\dagger~=~\bar v^{\pm\ad}\zb_{\ad}\ ,
\ ,\eea
such that $[\bar\eta^-,\bar\eta^+]_\star=[\bar\zeta^-,\bar\zeta^+]_\star=1$.
The inner Kleinian elements generate the involutive automorphisms
\bea \pi_y(\widehat F)&:=& \k_y\star\widehat F\star \k_y\ ,\qquad \pi_z(\widehat F)\ :=\ \k_z\star\widehat F\star \k_z\ ,\\[5pt]
\bar\pi_{\yb}(\widehat F)&:=& \bar\k_{\yb}\star\widehat F\star \bar\k_{\yb}\ ,\qquad \pi_z(\widehat F)\ :=\ \bar\k_{\zb}\star\widehat F\star \bar\k_{\zb}\ ,\eea
that act locally on symbols defined in Weyl order, but not 
on symbols defined in normal order that depends non-trivially 
on both $Y$ and $Z$.
The inner automorphisms $\pi= \pi_y\pi_z$ and 
$\bar\pi=\bar\pi_{\bar y}\bar\pi_{\bar z}$, however,
act locally on symbols defined both in Weyl and 
normal order, \emph{viz.}
\bea \pi(\widehat f(y,\bar y;z,\bar z))&=&\widehat f(-y,\bar y;-z,\bar z)\ .\eea
This action is generated by conjugation by the elements
\bea
\widehat\kappa&=& \k_y\star \k_z\ ,\qquad \widehat{\bar\kappa}\ =\ \k_{\bar y}\star \k_{\bar z}\ .\eea
Their Weyl-ordered symbols can be read off from
\bea {\cal O}_{\rm Normal}(\widehat\kappa)&=& {\cal O}_{\rm Weyl}(
(2\pi)^2\d^2(y)\d^2(z))\ ,\qquad 
{\cal O}_{\rm Normal}(\widehat{\bar\kappa})\ =\ 
{\cal O}_{\rm Weyl}((2\pi)^2 \d^2(\yb)\d^2(\zb))\ ,\eea
providing an example of the fact that one and the same operator 
can be factorized in one order and completely entangled in another 
order.


\scs{Deformed oscillators with delta function Klein operator}\label{App:defosc}


In this Appendix, we recall the main steps of the solution of Eqs. 
\eq{piV}-\eq{ZcurvF2}, that determine the connection $V'_\a$ on
${\cal Z}$ within the Ansatz \eq{ansatz1}-\eq{ansatz3}. 
Note that the deformed oscillator problem with a distributional deformation term can be reduced to an ordinary one, with a regular Gaussian source, upon changing ordering prescription, from the Weyl ordering used here to normal ordering on $\cal Z$. It is then possible to move back the so-obtained solution to Weyl ordering and get the same result that we shall review here (see Appendix G in \cite{us} for a detailed proof).

\scss{Problem setting }

As a consequence of the choice \eq{ansatz1}, and of the separation of $Y$ and $Z$ variables 
in \eq{ansatz2}-\eq{ansatz3}, the deformed oscillators obey
\bea [\wS'_\a,\wS'_\b]_\star & = & -2i\e_{\a\b}\left(1-b\Psi\star\kappa_z\right) \ ,\label{intS}\\[5pt]
[\widehat{\bar S}'_{\ad},\widehat{\bar S}'_{\bd}]_\star & = & -2i\e_{\ad\bd}\left(1-\bar{b}\bar{\Psi}\star\kappa_z\right) \ .\label{intSb}
 \eea
where the right-hand sides have a distributional deformation term.
This problem was solved in \cite{us} (and, in a different gauge, in \cite{Didenko:2009td}) 
for a constant deformation parameter.
We can use the same solution method by replacing the constant
deformation parameter by $\Psi$, as we shall describe next.
%

\scss{Integral equation}

The deformed oscillators $(\wS'_\a,\widehat{\bar{S}}'_{\ad})$ can 
be obtained explicitly by employing the $\circ$-product method of 
\cite{Prokushkin:1998bq}, later refined in \cite{Sezgin:2005pv} 
(see also \cite{Iazeolla:2007wt}) and adapted to a distributional 
deformation term in \cite{us}. 
The method can be articulated in the following two steps:

\begin{itemize}
\item[i)] We introduce a spin-frame $u^\pm_\a$ to split ($u^{\a+}u^-_\a=1$)
\be \wS'_\a(Y|z)~=~ u^-_\a \wS'^+(Y,z)-u^+_\a \wS'^-(Y,z)\ ,\qquad [\wS'^-,\wS'^+] _\star~=-2i(1-b\Psi\star \k_z)\ ,\label{defoscpm}\ee
and represent $(\wS'^\pm,\widehat{\bar{S}}'^\pm)$ as
($z^\pm:=u^{\pm\a}z_\a$, $w_z:=z^+z^-$, $[z^-,z^+]_\star = -2i $)
\bea \wS'^\pm &\equiv& u^{\pm\,\a}\wS'_\a \ = 4\int_{-1}^1 u^{\pm\,\a} \frac{dt}{(t+1)^2}\,f^{\pm}(\Psi|t) \star z_\a \,e^{i\ft{t-1}{t+1} w_z}\nn \\[5pt]&=&  \ 4\int_{-1}^1 \frac{dt}{(t+1)^2}\,f^{\pm}(\Psi|t) \star z^{\pm} \,e^{i\ft{t-1}{t+1} w_z}\nn \\[5pt]&=& -4iu^{\pm\,\a}\frac{\partial}{\partial\r^{\a}}  \int_{-1}^1 \frac{dt}{t+1} \,f^{\pm}(\Psi|t) \star\, \left. e^{\ft{i}{t+1} \left((t-1)w_z +\r^\b z_\b \right)}\right|_{\r^\a=0}\nn \\[5pt]&=&   -4i\frac{\partial}{\partial\r_{\pm}}  \int_{-1}^1 \frac{dt}{t+1} \,f^{\pm}(\Psi|t) \star\, \left. e^{\ft{i}{t+1} \left((t-1)w_z +\r^+z^- -\r^-z^+\right)}\right|_{\r^\pm=0}\ ,\label{WSansatz}\eea
where $\r_\a$ are classical sources, $\e_{\a\b} = u^-_\a u^+_\b-u^-_\b u^+_\a$ and 
$f^{\pm}(\Psi|t)$ are star-functions of $\Psi$.
The virtue of these generalized Laplace transforms is that
they are closed under star-product, as can be seen from
\bea & \frac{f^{\pm}(\Psi|t) }{t+1}\,\star\,e^{\ft{i}{t+1} \left( (t-1) w_z +\r^+z^- -\r^-z^+\right)}\,\star \,\frac{f^{\pm}(\Psi|t') }{t'+1}\,\star\, e^{\ft{i}{t'+1} \left( (t'-1)w_z +\r^{\prime+}z^- -\r^{\prime-}z^+\right)}  &\nn\\[5pt]  &  \ = \  \frac{f^{\pm}(\Psi|t)\star f^{\pm}(\Psi|t')}{2(\tilde{t}+1)} \,\star\,e^{\ft{i}{\tilde{t}+1}\left((\tilde{t}-1)w_z +\tilde{\r}^+z^- -\tilde{\r}^-z^+-\ft{1}2\r^+\r^{\prime-}+\ft{1}2\r^-\r^{\prime+}\right)} \ ,&  \label{selfrep} \eea
\be \tilde{t}~:=~tt'\ ,\qquad \tilde{\r}^+~:=~\r^{\prime +} +t'\r^+\ ,\quad \tilde{\r}^-~:=~\r^-+t\r^{\prime -}\ ,\label{tildet}\ee
where the induced map $(t,t')\mapsto tt'$ sends $[-1,1]\times[-1,1]$ to $[-1,1]$,
while the $Y$-dependent pieces behave as spectators. 
In particular,
 \be \left[\frac{f^{-}(\Psi|t)}{(t+1)^{2}}\star z^-\,e^{i\ft{t-1}{t+1}w_z}\, ,\, \frac{f^{+}(\Psi|t')}{(t'+1)^{2}}\star z^+\,e^{i\ft{t'-1}{t'+1}  w_z}\right]_\star  \ee
  \be = \  -\frac{i f^{-}(\Psi|t)\star f^{+}(\Psi|t')}{2(\tilde{t}+1)^2}\,\star\left(1+i \frac{\tilde{t}-1}{\tilde{t}+1}w_z\right)\, e^{i \ft{\tilde{t}-1}{\tilde{t}+1}w_z}  \ . \ee
Thus, upon inserting \eq{WSansatz} into \eq{defoscpm}, the latter 
turns into the integral equations
\bea 4\int_{-1}^1 dt\int_{-1}^1 dt'\,\frac{f^{-}(\Psi|t)\star f^{+}(\Psi|t')}{(tt'+1)^2}\,\star\,\left[1+i\frac{tt'-1}{tt'+1}w_z \right]\,e^{i\ft{tt'-1}{tt'+1}w_z} ~=~1-b\Psi\star \k_z\ ,\label{stepi}\eea
where we recall that $\kappa_z=2\pi \delta^2(z)$;
\item[ii)] Inserting $1=\int_{-1}^1 du\,\d(u-tt')$ ($t,t'\in[-1,1]$) into the 
left-hand side of \eq{stepi}, changing order of integration,
and defining 
\bea  (h_{1} \circ h_{2})(\Psi|u) ~:=~ \int_{-1}^1 dt\int_{-1}^1 dt'\,h_{1}(\Psi|t)\,\star\,h_{2}(\Psi|t')\,\d(u-tt')\ ,\label{ringo2}\eea
which is a commutative and associative product\footnote{For this generalization 
of the $\circ$-product composition to accommodate functions of oscillators, 
it is of course crucial that the $Y$-dependence comes through one and the 
same function, $\Psi$ in this case.} on the space of functions on the unit interval, 
one arrives at 
\bea 4\int_{-1}^1\frac{du}{(u+1)^2}\,h(\Psi|u)\star \left[1+i\frac{u-1}{u+1}w_z\right]\,e^{i\ft{u-1}{u+1}w_z} ~=~1-2\pi b \Psi \star \delta^2(z)\ ,\label{penult}\eea
where $h(\Psi|u) : = (f^-\circ f^+)(\Psi|u)$.
Next, using the delta function sequence
\be \lim_{\varepsilon\rightarrow 0}\frac{1}{\varepsilon}e^{-i\s\ft1\varepsilon w_z}~=~\s \k_z\ , \label{deltalimit} \ee
we find the unique solution
\be h(\Psi|t)~=~\d(t-1)-\ft{b \Psi}2\ .\ee
Thus, Eq. \eq{stepi} is equivalent to the $\circ$-product 
equation 
\be  (f^{-}\circ f^{+})(\Psi|u) ~=~ \d(u-1)-\frac{b\Psi}{2} \ ,\label{ringeq2}\ee
with the following solution space, as we shall show in the next subsection:
\be f^\pm~=~ g^{\circ(\pm1)} \circ f\ ,\qquad f(\Psi|t)~=~\delta(t-1)+j(\Psi|t)\ ,\label{j2bis}\ee
\be j(\Psi|t)~=~q(\Psi|t)+\sum_{k=0}^\infty \l_{k}(\Psi) p_k(t)  \ ,\qquad q(\Psi|t)~=~-\frac{b\Psi}{4}\,\star\,{}_1F_{1}\left[\frac{1}{2};2;-\frac{b\Psi}{2}\log t^{2}\right]\ ,\label{j3}\ee
where $g$ is a gauge artifact (and we use the notation $g^{\circ(+1)}=g$ and $g^{\circ(-1)}\circ g=1$); $p_k(t):=\ft{(-1)^k}{k!}\,\d^{(k)}(t)$ act as projectors in the $\circ$-product algebra; and $\l_{k}$ 
are given by \eq{lambdak} and \eq{Lkm}.
\end{itemize}
The presentation \eq{deltalimit} is compatible with 
$\kappa_z\star f(z)=f(-z)\star \kappa_z$, 
$\kappa_z\star\kappa_z=1$, $\tau(\kappa_z)=-\kappa_z$ 
and $\kappa_y\star \kappa_z=\widehat \kappa$.
The fact that $g$ contains gauge artifacts follows 
by using holomorphic gauge parameters
of the form
\be \e(Y,z)~=~\int_{-1}^1 \frac{dt}{1-t^2}\breve{\e}(\Psi|t) \star e^{i\ft{t-1}{t+1} w_z}\ ,\ee
which induce
\be \delta_{\e} f^\pm~=~\pm \ft{1}{2} \breve{\e}\circ f^\pm\ .\ee

\scss{Solution using symbol calculus}

In order to solve Eq. \eq{ringeq2}, \emph{i.e.}
\bea  (f^-\circ f^+)(\Psi|u) \ = \ \d(u-1)-\frac{b\Psi}{2} \ ,\label{ringprobl}\eea
one begins by observing that the $\circ$-product algebra decomposes into
even and odd functions on the interval $[-1,1]$, \emph{viz.}
\bea f^{(\pi)}\circ g^{(\pi')} ~=~ \d_{\pi\pi'} f^{(\pi)}\circ g^{(\pi')} \ , \qquad f^{(\pi)}(-s) \ = \ \pi f^{(\pi)}(s) \ , \qquad \pi,\pi' \ = \ \pm 1 \ .\eea
Therefore \eq{ringprobl}  separates into the following two independent equations:
\bea  (f^{-(+)}\circ f^{+(+)})(\Psi|u) & = & I_0^{(+)}(u)-\frac{b\Psi}{2} \ ,\label{ringprobl+}\\[5pt]
(f^{-(-)}\circ f^{+(-)})(\Psi|u) & = & I_0^{(-)}(u) \ ,\label{ringprobl-}\eea
where
\bea I_0^{(\pm)} \ := \ \frac{1}{2}\left[\d(u-1)\pm\d(u+1)\right] \eea
acts as the identity in the even and odd $\circ$-product subalgebras.
Equations \eq{ringprobl+} and \eq{ringprobl-} can be cast into algebraic 
equations by expanding ($t\in[-1,1]$)
\bea f^{\pm(\pi)}(\Psi|t)&=& m^{\pm(\pi)}(\Psi|t)+\sum_{k=0}^\infty \l^\pm_{k}(\Psi)\,p^{(\pi)}_{k}(t) \
,\\[5pt]
m^{\pm(\pi)}(\Psi|t) &=& \sum_{k=0}^\infty \mu^{\pm}_{k}\,I^{(\pi)}_k(t) \Psi^{\star k}\ ,\label{mexp}\eea
in terms of ($k\geqslant 1$)
 \bea
 I^{(\pi)}_k(u)&:=&\left[{\rm sign}(u)\right]^{\frac12(1-\pi
 )}~\int_{-1}^1 ds_1 \cdots \int_{-1}^1 ds_k~\delta(u-s_1\cdots
 s_k)\nn\\[5pt]
 &=&\left[{\rm sign}(u)\right]^{\frac12(1-\pi)}{\left(\log
 \frac1{u^2}\right)^{k-1}\over (k-1)!}\ ,
 \eea
which obey ($k,l\geqslant 0$)
 \be
 I^{(\pi)}_k\circ I^{(\pi)}_l\ =\ I^{(\pi)}_{k+l}\ ,
 \label{ring}
 \ee
and of the $\circ$-product projectors ($k\geqslant 0$)
\bea p^{(\pi)}_k(s)&:=& {(-1)^k\over k!} \d^{(k)}(s)\ ,\qquad \pi =\
(-1)^k\ ,\label{pk}\eea
obeying
\bea p^{(\pi)}_k\circ f&=& L_k[f] p^{(\pi)}_k\ ,\qquad L_k[f](\Psi)\ =\ \int_{-1}^1
ds~ s^k f(\Psi|s)\ .\label{proj1}\eea
The property \eq{ring} implies that upon defining the symbols ($\xi\in\Comp$)
\bea \widetilde{m}^{\pm(\pi)}(\Psi|\xi)\ := \ \sum_{k=0}^\infty \mu^{\pm(\pi)}_{k}(\Psi)\,\xi^k \ , \eea
the $\circ$-product $m^{(\pi)}_{-}\circ m^{(\pi)}_{+}$ is mapped to 
the algebraic product $\widetilde{m}^{(\pi)}_-\star \widetilde{m}^{(\pi)}_+$.
Thus, substituting \eq{mexp} into \eq{ringprobl+} and \eq{ringprobl-}, 
and using \eq{ring} and \eq{proj1}, one is left with the algebraic equations
\bea  \widetilde{m}^{-(+)} (\Psi|\xi)\star \widetilde{m}^{+(+)} (\Psi|\xi) &=& 1-\frac{b\Psi}{2}\,\xi \ ,
\label{meq2a}\\[5pt]
\widetilde{m}^{-(-)} (\Psi|\xi)\star \widetilde{m}^{+(-)} (\Psi|\xi) &=&1 \ ,\label{meq2}\eea
and the following condition on the coefficients of the projectors in the expansion \eq{mexp}:
\bea \l^{-(\pi)}_{k}\star  L_{n}[m^{+(\pi)}]+ \l^{+(\pi)}_{k}\star L_{n}[m^{-(\pi)}]
+\l^{-(\pi)}_{k}\star\l^{+(\pi)}_{k}  \ = \ 0 \ .\label{proj2}\eea
The solution space to Eqs. \eq{meq2a} and \eq{meq2} is parameterized by an 
undetermined function $\widetilde{g}_\s^{(\pi)}$ as follows\footnote{
Note that, differently from the Lorentz-covariant solutions in 
\cite{Prokushkin:1998bq,Sezgin:2005pv,Iazeolla:2007wt}, the 
algebraic equations  involve the product of two different functions 
rather than the square of a single one.}:
\be \widetilde{m}^{\pm(+)}(\Psi|\xi)~=~ (\widetilde{g}^{(+)}(\Psi|\xi))^{\star(\pm 1)}\star 
\sqrt{1-\frac{b\Psi}{2}\,\xi}\ ,\qquad\widetilde{m}^{\pm(-)}(\Psi|\xi)~=~ (\widetilde{g}^{(-)}(\Psi|\xi))^{\star(\pm 1)}\ .\ee
Likewise, the solution space to \eq{proj2} contains an undetermined 
set of coefficients, say $\l^{+(\pi)}_{k}$. 
One can show that the undetermined quantities are gauge artifacts. 
A natural gauge choice is to work with the symmetric solutions
\be f^{\pm}~=~f\quad \Rightarrow\quad \mu^{\pm(\pi)}_{k}~=~\mu^{(\pi)}_{k}\ ,\qquad \l^{\pm(\pi)}_{k}~=~\l^{(\pi)}_{k}\ ;\label{symgauge}\ee
henceforth we shall drop the $\pm$ referring to the spin-frame. 
Thus
\be \widetilde m^{(+)}(\Psi|\xi)~=~\widetilde\varepsilon^{(+)}(\xi)\star\sqrt{1-\ft{b\Psi}{2}\xi}\ ,\qquad 
\widetilde{m}^{(-)}(\Psi|\xi)~=~\widetilde\varepsilon^{(-)}(\xi)\ ,\ee
where $(\widetilde\varepsilon^{(\pm)}(\xi))^{\star 2}=1$, and 
\bea \l^{(\pi)}_{k}\star\,\left(\l^{(\pi)}_{k}+2L_k[m^{(\pi)}]\right) \ = \ 0 \ .\eea
It follows that
\bea m^{(+)}~=~\ \varepsilon^{(+)} \circ (I^{(+)}_0+q^{(+)}))
\ , \qquad m^{(-)} ~=~ \varepsilon^{(-)} \circ I^{(-)}_0\ ,\eea
where $q^{(+)}(\Psi|s)$ has the symbol
\be \widetilde q^{(+)}(\Psi|\xi)~ =~\sum_{k=1}^{\infty}\m^{(\pi)}_{k} \xi^k~=~\sqrt{1-\ft{b\Psi}2\,\xi}-1\ ,\ee
corresponding to the confluent hypergeometric function
\bea q^{(+)}(\Psi|s) \ = \ \sum_{k=1}^\infty {\ft12 \choose k}\left(-\frac{b}{2}\right)^{k}(\Psi)^{\star k}\frac{\left(\log\ft1{s^2}\right)^{k-1}}{(k-1)!} \ = \ -\frac{b\Psi}{4}\,\star\,{}_1F_{1}\left[\frac{1}{2};2;\frac{b\Psi}{2}\log\frac{1}{s^2}\right] \ .\label{1F1}\eea
As for the coefficients of the projector $p_k$, we have
\bea \l^{(\pi)}_{k} \ = \ -2\th_{k} L_k[m^{(\pi)}] \ , 
\qquad \th_{k} \ \in \ \{0,1\} \ . \label{lambdak}\eea
Requiring that $\wS'_\a=z_\a$ for $\Psi=0$ and $\th_{k}=0$, that is 
\be f^\pm (0|s)|_{\theta_{k}=0} \ = \ \d(s-1) \ = \ I^{(+)}_0(s)+I^{(-)}_0(s)\ ,\ee
implies that
\be \varepsilon^{(\pm)}~=~1\ ,\qquad \mu^{(\pi)}_{0}~=~1\ ,\qquad \ .\ee
From \eq{lambdak} and 
\bea L_k[m]&=&L_k\left[\d(s-1)+q^{(+)}\right]~=~1+L_k[q^{(+)}]\ ,\label{Lkm}\\[5pt]
 L_k[q^{(+)}]&=& -{1+(-1)^k\over
2}\left[1-\left(1-{b\Psi\over 1+k}\right)^{\star 1/2}\right]\ ,\label{Lkq}\eea
it follows that $\l^{(\pi)}_{k}$ are $\Psi$-dependent only for even $k$.

\scss{Non-trivial vacuum connections on ${\cal Z}$}

For $\Psi=0$, even though the deformation is absent from \eq{intS}-\eq{intSb}, 
there still exist non-trivial vacuum solutions, as the  
coefficients $\l_{k\star}$ of $\circ$-product projectors $p_k$
need not vanish, but rather they reduce to $\l_{k\star}|_{\Psi=0} \ = \ -2\th_k$,
giving rise to a non-trivial flat connection on ${\cal Z}$ \cite{us};
see also \cite{Iazeolla:2007wt} for similar, $SO(p,4-p)$-invariant, 
non-trivial vacua. 
In this paper, for simplicity, we shall take $\th_k = 0$
for all $k$, deferring the study of the non-trivial 
vacua to a future work. 
With this choice, insertion of \eq{1F1} and \eq{mexp} into \eq{WSansatz}
in the symmetric gauge \eq{symgauge}, yields \eq{wV'}.

\end{appendix}


\end{document}